%% file: ms_revised.tex
\documentclass[12pt,preprint]{aastex}
\usepackage{natbib}
\usepackage{rjwmath}
\shorttitle{The Distribution of YSOs in the Orion Molecular Clouds}
\shortauthors{Megeath et al.}
\bibpunct{(}{)}{,}{a}{}{;}
\bibpunct{(}{)}{,}{a}{}{;}

\begin{document}

%% LaTeX will automatically break titles if they run longer than
%% one line. However, you may use \\ to force a line break if
%% you desire.

\title{The {\it Spitzer} Space Telescope Survey of the Orion A \& B Molecular Clouds - Part I: A Census of Dusty Young Stellar Objects and a Study of their Mid-IR Variability}

%% Use \author, \affil, and the \and command to format
%% author and affiliation information.
%% Note that \email has replaced the old \authoremail command
%% from AASTeX v4.0. You can use \email to mark an email address
%% anywhere in the paper, not just in the front matter.
%% As in the title, use \\ to force line breaks.

\author{S. T. Megeath\altaffilmark{1}, R. Gutermuth\altaffilmark{2}, J. Muzerolle\altaffilmark{3}, E. Kryukova\altaffilmark{1}, K. Flaherty\altaffilmark{6}, J. L. Hora\altaffilmark{4}, L. E. Allen\altaffilmark{5},  L. Hartmann\altaffilmark{7},  P. C. Myers\altaffilmark{4}, J. L. Pipher\altaffilmark{8},  J. Stauffer\altaffilmark{9}, E. T. Young\altaffilmark{10},  G.G. Fazio\altaffilmark{4}}

\altaffiltext{1}{Department of Physics and Astronomy, University of Toledo, Toledo, OH 43560 (megeath@physics.utoledo.edu)}

\altaffiltext{2}{Department of Astronomy, University of Massachusetts, Amherst, MA 01003, USA}

\altaffiltext{3}{Space Telescope Science Institute, Baltimore, MD 21218, USA}

\altaffiltext{4}{Harvard Smithsonian Center for Astrophysics,  Cambridge, MA 02138, USA}

\altaffiltext{5}{National Optical Astronomical Observatory, Tucson, AZ 85719, USA}

\altaffiltext{6}{Steward Observatory, University of Arizona, Tucson, AZ 85721, USA }

\altaffiltext{7}{Department of Astronomy, University of Michigan, Ann Arbor, MI 48109, USA}

\altaffiltext{8}{Department of Physics and Astronomy, University of  Rochester, Rochester NY 14627, USA}

\altaffiltext{9}{Spitzer Science Center, California Institute of Technology,  Pasadena, CA 91125, USA}

\altaffiltext{10}{SOFIA-Universities Space Research Association, NASA Ames Research Center, Moffett Field, CA 94035, USA}

%% Notice that each of these authors has alternate affiliations, which
%% are identified by the \altaffilmark after each name.  Specify alternate
%% affiliation information with \altaffiltext, with one command per each
%% affiliation.

%% Mark off your abstract in the ``abstract'' environment. In the manuscript
%% style, abstract will output a Received/Accepted line after the
%% title and affiliation information. No date will appear since the author
%% does not have this information. The dates will be filled in by the
%% editorial office after submission.

\begin{abstract}

We present a survey of the Orion A and B molecular clouds undertaken
with the IRAC and MIPS instruments onboard Spitzer.  In total, five
distinct fields were mapped covering 9~deg$^2$ in five mid-IR bands
spanning 3-24~$\mu$m. The survey includes the Orion Nebula Cluster, the 
Lynds~1641, 1630 and 1622 dark clouds, and the NGC~2023, 2024, 2068 and 
2071 nebulae.  These data are merged with the 
2MASS point source catalog to generate a catalog of eight band photometry.  
We identify 3479 dusty young stellar objects (YSOs) in the Orion molecular clouds by
searching for point sources with mid-IR colors indicative of reprocessed light from
dusty disks or infalling envelopes. The YSOs are subsequently classified on the basis of their mid-IR colors
 and their spatial distributions are presented.  We classify 2991 of the YSOs as pre-main sequence
stars with disks and 488 as likely protostars.  
Most of the sources were observed with IRAC in 2-3 epochs over 6 months; we search for 
variability between the epochs by looking for correlated
variability in the 3.6 and 4.5~$\mu$m bands.  We find that 50\% of the 
dusty YSOs show variability.  The variations are typically small  ($\sim 0.2$~mag.) with the protostars showing a higher incidence of variability and larger variations.  The observed correlations between
the 3.6, 4.5, 5.8 and 8~$\mu$m variability suggests that 
we are observing variations in the heating of the inner disk due to changes in the
accretion luminosity or rotating accretion hot spots.

\end{abstract}


%% From the front matter, we move on to the body of the paper.
%% In the first two sections, notice the use of the natbib \citep
%% and \citet commands to identify citations.  The citations are
%% tied to the reference list via symbolic KEYs. The KEY corresponds
%% to the KEY in the \bibitem in the reference list below. We have
%% chosen the first three characters of the first author's name plus
%% the last two numeral of the year of publication as our KEY for
%% each reference.

\section{Introduction}

Strewn across the constellation of Orion are the most massive and active molecular clouds
within 500 pc of the Sun.  These clouds host the nearest sites of high mass star formation
as well as thousands of young low mass stars.  They are important laboratories for
studying star and planet formation in a range of environments, from near isolation,
to small groups, and finally to massive clusters.  The large angular (and spatial) size of the Orion molecular
clouds, which extend 12$^o$ (90~pc) on the sky, necessitates the use of wide field surveys to observe the clouds
in their entirety.  We present here the first wide field, $2''-5''$ resolution, mid-IR survey
of the clouds using the  the InfraRed Array Camera (IRAC) and the Multiband 
Imaging Photometer for Spitzer (MIPS) onboard the  {\it Spitzer} Space Telescope.
With these data we have undertaken a spatially unbiased survey for dusty young stellar objects associated
with the Orion clouds.  

The cloud complex is composed of two distinct molecular clouds, the Orion~A and
Orion~B clouds, with a total mass in excess of $2 \times 10^5$~M$_{\odot}$ \citep{2005A&A...430..523W}.  The
Orion~A cloud contains the Orion Nebula and the first known embedded cluster, the Orion
Nebula Cluster \citep[hereafter: ONC,][]{1931PASP...43..255T,1937ApJ....86..119B, 1993prpl.conf..429Z,2008hsf1.book..544O}.
The ONC is one of the two richest young stellar cluster currently known within 1 kpc of the Sun
\citep{2003AJ....126.1916P,2003ARA&A..41...57L,2012ApJ...750..125A}.  This cluster has emerged as an important laboratory for
understanding the initial mass function
\citep{1997AJ....113.1733H,2000ApJ...540.1016L,2002ApJ...573..366M}, the evolution of disks around young
stars \citep{1998AJ....116.1816H,2000AJ....120.3162L,2004AJ....128.1254L}, and the
structure and kinematics of young clusters \citep{1998ApJ...492..540H,2008ApJ...676.1109F,2009ApJ...697.1103T}.  
The OMC-2/3 star forming region \citep[which may be considered part of the ONC cluster, ][]{2008hsf1.book..590P} and the L1641 molecular cloud \citep{2008hsf1.book..621A} are also part of the Orion A cloud. 
The Orion~B cloud contains the NGC 2024 HII region as well as the NGC~2023, 2068, 2071 reflection nebulae. All
of these regions also contain clusters or groups of young stars 
\citep{1991ApJ...371..171L,1994ApJS...90..149C,2008hsf1.book..621A,2008hsf1.book..590P,2008hsf1.book..662M,2008AJ....135..966F}.  Directly to the north of
the Orion B cloud is the Lynds~1622 cloud. Since this clouds appears to be at the same distance as the Orion~B
cloud, we include Lynds~1622 as part of Orion~B \citep{2008hsf1.book..782R,2009AJ....137.3843B}.
VLBI parallax measurements of stars and masers in the Orion Nebula give an average distance 
of 420~pc \citep{2007ApJ...667.1161S,2007A&A...474..515M,2007PASJ...59..897H}.
Although there is evidence that the distances to the various clouds in Orion may vary by as much
as 100 pc \citep{2005A&A...430..523W}; we adopt 420 pc as the distance to the entire cloud 
complex in our analysis.

%The ONC is found at the northern end of Orion~A, a 30 pc long,
%fialmentary molecular cloud. Directly to the south of the ONC is
%L~1641, a section of the Orion~A cloud containing several small groups
%of stars and a significant distributed population
%\citep{strom1993,chen1994,allen1996}.  

Stellar photometry with the IRAC and MIPS instruments can be used to detect excess infrared emission
due to starlight reprocessed by dust grains in disks and envelopes 
\citep{2004ApJS..154..367M,2004ApJS..154..363A, 2004ApJS..154..379M,2004ApJS..154..315W, 
2005ApJ...629..881H,2007prpl.conf..361A}.  Many of the
dusty young stellar objects (YSOs) identified by their mid-IR ($\ge 3.6$~$\mu$m) emission do 
not exhibit substantial excess  emission in the near-IR (1-2.2~$\mu$m) bands, which are often dominated by 
the photospheric emission from the star \citep{2004ApJS..154..374G}. In contrast, at mid-IR
wavelengths the emission from disks and envelopes exceeds that from the star \citep{2003ApJ...598.1079W,
2006ApJ...638..314D}. In this paper, we identify young stars with disks and protostars using a suite of 
color-color diagrams and color-magnitude diagrams constructed from the merged  3-24~$\mu$m 
photometry of the IRAC and MIPS instruments of {\it Spitzer} and the near-IR photometry
from the 2MASS point source catalog \citep{2009ApJS..184...18G,2012AJ....144...31K}.  
With this merged photometry, we map the  spatial distribution of dusty YSOs in a 14~deg$^2$ 
survey field covering the  Orion A and B molecular clouds.   

By identifying such objects through their infrared-excesses, we can study for the first time the extended spatial distribution of protostars and young pre-main
sequence stars with disks in the entire cloud complex.  In contrast to previous papers which mapped the distribution of young stellar objects using the surface density of near-IR sources \citep{1991ApJ...371..171L,2000AJ....120.3139C},  we can identify individual YSOs irrespective of whether they are found in clusters or in isolation. The resulting census of dusty YSOs provides a more representative portrait of the full range of star forming environments present in the Orion clouds. It also provides the means to detect variability in the emission  from dusty disks and envelopes between the 2-3 epochs of IRAC photometry. When compared to an independent estimate of the number of pre-main sequence stars without disks, the census can be used to estimate the fraction of pre-main sequence stars that have circumstellar disks and the fraction of  young stars in their protostellar phase. 

This contribution is the first of two papers presenting the results of the {\it Spitzer} Orion Survey.  In this paper, we
describe the construction of the catalog of all point sources in the surveyed fields and the identification of dusty young stellar objects in the catalog.  We then show the distribution of both pre-main sequence stars with disks
and protostars in the Orion clouds.  Since much of the IRAC data were taken in 2-3 epochs over an 8 month period, we also use
these data to examine the variability of the YSOs. This work complements the warm mission YSOVAR variability survey of the ONC by searching for variability in a larger sample of YSOs and by measuring the variability in all four IRAC wavelength bands \citep{2011ApJ...733...50M}.  In addition, we present a new means for characterizing the spatially varying completeness of point sources in nebulous fields.

In a second paper, we will correct for the spatially varying incompleteness found in the YSO
catalog and map the density of dusty YSOs across the clouds.  Using the corrected distribution
of YSOs, we will examine the  the demographics of protostars and young stars with disks in the 
diverse environments found within Orion molecular clouds, compare the Orion clouds to other nearby 
molecular clouds, study the structure of embedded clusters, and estimate the fraction of young stars with disks within the clouds.  

In addition, this paper has two companion papers which use the YSO catalog extracted
from the {\it Spitzer} Orion Survey: a study  of the relationship between stellar and gas surface densities in molecular clouds  \citep{2011ApJ...739...84G}  and a comparative study of protostellar luminosity functions in nine nearby molecular clouds  \citep{2012AJ....144...31K}.  Together, these papers comprise an analysis of many facets of the {\it Spitzer} Orion Survey including the mechanisms for YSO variability, the demographics of star formation, the properties of embedded clusters, and the influence of environment
on star and planet formation.

\section{The Spitzer Orion Point Source Catalog}
\label{sec:psc}

The Orion clouds were surveyed with the IRAC and MIPS instruments onboard the {\it Spitzer} 
Space Telescope \citep{2004ApJS..154...10F,2004ApJS..154...25R}.  This paper describes 
the extraction of point source photometry from the survey in the four IRAC bands and the 
MIPS 24~$\mu$m band and the initial results derived from that photometry. In this section, we 
discuss the observations, the data reduction, the identification of point sources,  and the determination of photometric magnitudes in the {\it Spitzer} 3.6, 4.5, 5.8, 8 and 24~$\mu$m bands.  We then overview the  basic properties of the resulting point source catalog.

\subsection{IRAC Observations}

\begin{table}[h]
\begin{center}
\caption{IRAC Observations}
\begin{tabular}{lccccccc}
\tableline\tableline
Field Name & Frame & No.$^a$ of  & Total Int. & Field$^{b}$ & 1st Epoch & 2nd Epoch & PID \\
       & Time (s)  & Dithers & Time (s) & (deg$^2$) & & & \\
\tableline
ONC                     & 12 HDR   & 4 & 41.6 & 1.33 & 2004 Mar 9 & 2004 Oct 12  & 43 \\
L1641                   & 12 HDR   & 4 & 41.6 & 2.71 & 2004 Feb 16-18 & 2004 Oct 8, 27$^{c}$ & 43 \\
Ori~A ext.          & 12 HDR   & 4 & 41.6 &  1.91 &  2007 Oct 21 & 2008 Apr 9-10& 30641 \\
$\kappa$ Ori ext.$^d$      & 12 HDR   & 4 & 41.6 & 1.21 & 2009 Mar 21-22 & - & 50070 \\
ONC center$^d$         & 2  &  24     & 28.8 & 0.08   & 2004 Oct 27 & - & 50 \\
NGC2024/23              & 12 HDR  & 4  & 41.6 & 1.20 & 2004 Feb 17 & 2004 Oct 27 & 43 \\
NGC 2024$^d$           & 2 &   24       & 28.8 & 0.04  & 2004 Oct 8 & - & 50 \\
NGC2024 ext.            & 12 HDR  & 4  & 41.6 & 0.07 & 2007 Mar 30 & 2007 Oct 16 & 30641 \\
NGC~2068/71             & 12 HDR & 4  & 41.6 & 1.10 & 2004 Mar 7 & 2004 Oct  28 & 43 \\
L1622                   & 12 HDR   & 4 & 41.6 & 0.23 & 2005 Oct 27  & 2006  Mar 25 & 43 \\
Ori B fil.           & 12 HDR  & 4  & 41.6 & 0.36 & 2007 Mar 29-30 & 2007 Oct 16-17 & 30641 \\
Reference 1$^d$         & 12 HDR   & 4 & 41.6 & 0.13 & 2005 Oct 26-29 &  & 43 \\
Reference 2$^a$           & 12 HDR   & 4 & 41.6 & 0.16 & 2007 Oct 16 & & 30641 \\
Orion A total           & 12 HDR   & 4 & 41.6 & 5.86 & - & - & - \\
Orion B total           & 12 HDR   & 4 & 41.6 & 2.96 & - & - & - \\
\tableline
\tablenotetext{a} {Number of pointings per map positions for both epochs combined.}
\tablenotetext{b} {Area with complete 4-band coverage except for the 2~s exposures,
where we state the single-band coverage fields.}
\tablenotetext{c} {The two dithers of the 2nd epoch observation were observed on separate dates.}
\tablenotetext{d} {Only one epoch was observed.}
\end{tabular}
\end{center}
\label{tab:irac_coverage}
\end{table}

%IRAC produces simultaneous images in four photometry
%bands: 3.6~$\mu$m, 4.5~$\mu$m, 5.8~$\mu$m and 8.0~$\mu$m each covering
%a $5.1' \times 5.1'$ FOV.  All four images are not coincident, but
%instead are grouped into two fields of view separated by $6'$; one
%field of view for the 3.6~$\mu$m and 5.8~$\mu$m channels and one for
%the 4.5~$\mu$m and 8.0~$\mu$m channels.  

The IRAC observations were taken as part of the Guaranteed Time
Observation (GTO) programs PID~43, 50, 30641 and 50070.  Table~1
contains a complete summary of the IRAC observations. The maps were
made using a triangular grid of pointing centers in celestial
coordinates.  Since this grid was fixed to the celestial sphere, the
map could be adjusted to follow the contours of the molecular cloud.
The triangular grid was designed so that the observations
could be executed at the full range of possible spacecraft orientations without holes
in the coverage.   Each field was observed in two epochs scheduled during
consecutive visibility periods;  this ensured that the detector arrays
were rotated by as much as 180~degrees between the epochs.  The
rotation between the two epochs effectively swapped the positions
of the two spatially offset IRAC field of views (i.e. the 3.6/5.8~$\mu$m and 
the 4.5/8~$\mu$m fields of views),  thus providing full four-band coverage 
over most of the survey area.  The rotation also facilitated the removal of 
bright source artifacts in the data.

The 12 second high dynamic range (HDR) mode was used; this produced two frames
with integration times of 0.4~sec and 10.4~sec (corresponding to frame times
of 0.6 and 12~sec).  At every map position,
two dithered observations were obtained during each epoch.  Thus, for
most positions, the total integration time is at least 41.6 seconds per band for the
long HDR frames and 1.6 seconds for the short HDR frames.  However,
due to the rotation of the two field of views between the two epochs,
many positions at the edges of the map have an integration time of
20.8 seconds per band for the long HDR frames and 0.8 seconds for the short HDR
frames.  

\begin{figure}
\epsscale{1.1}
\plottwo{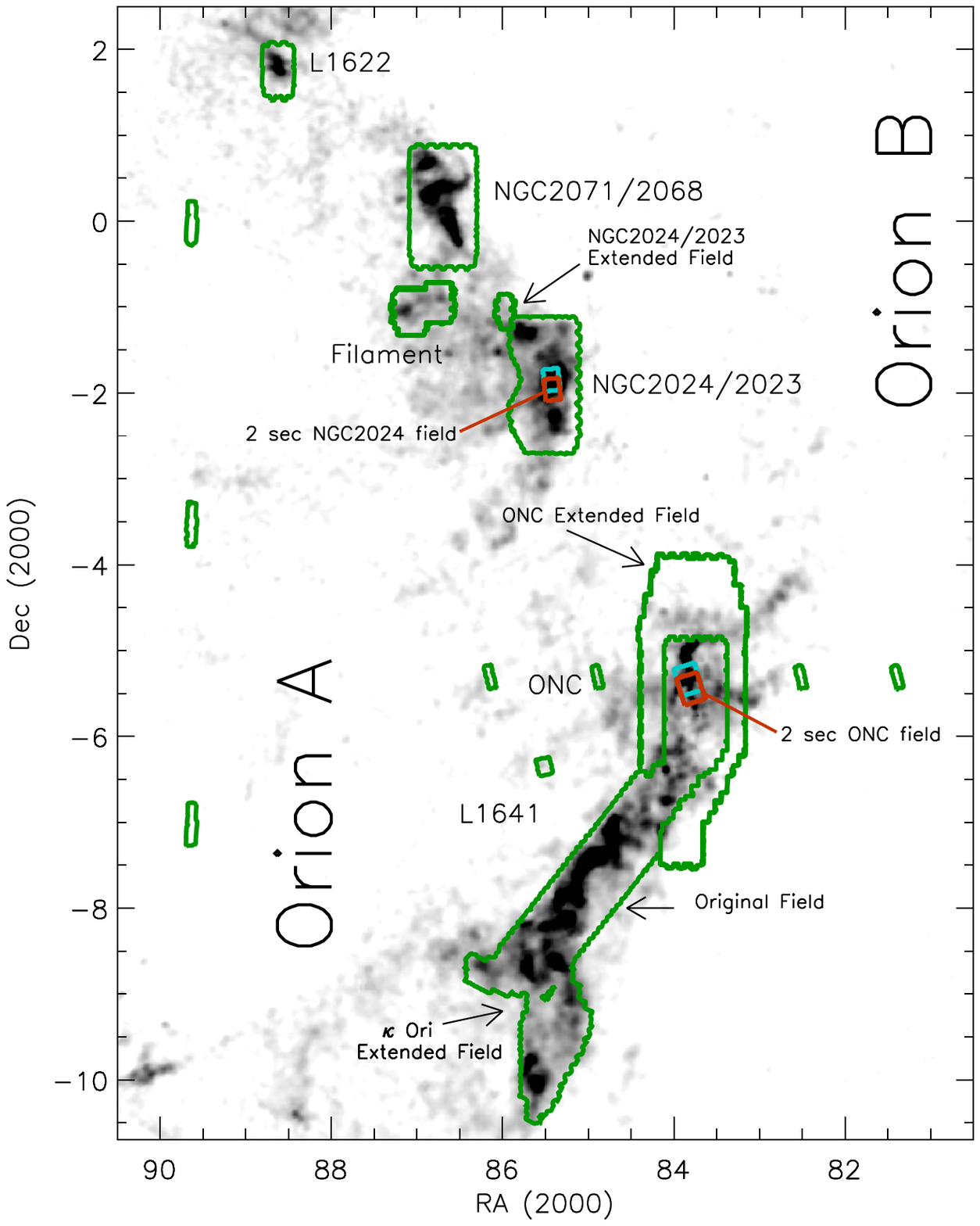}{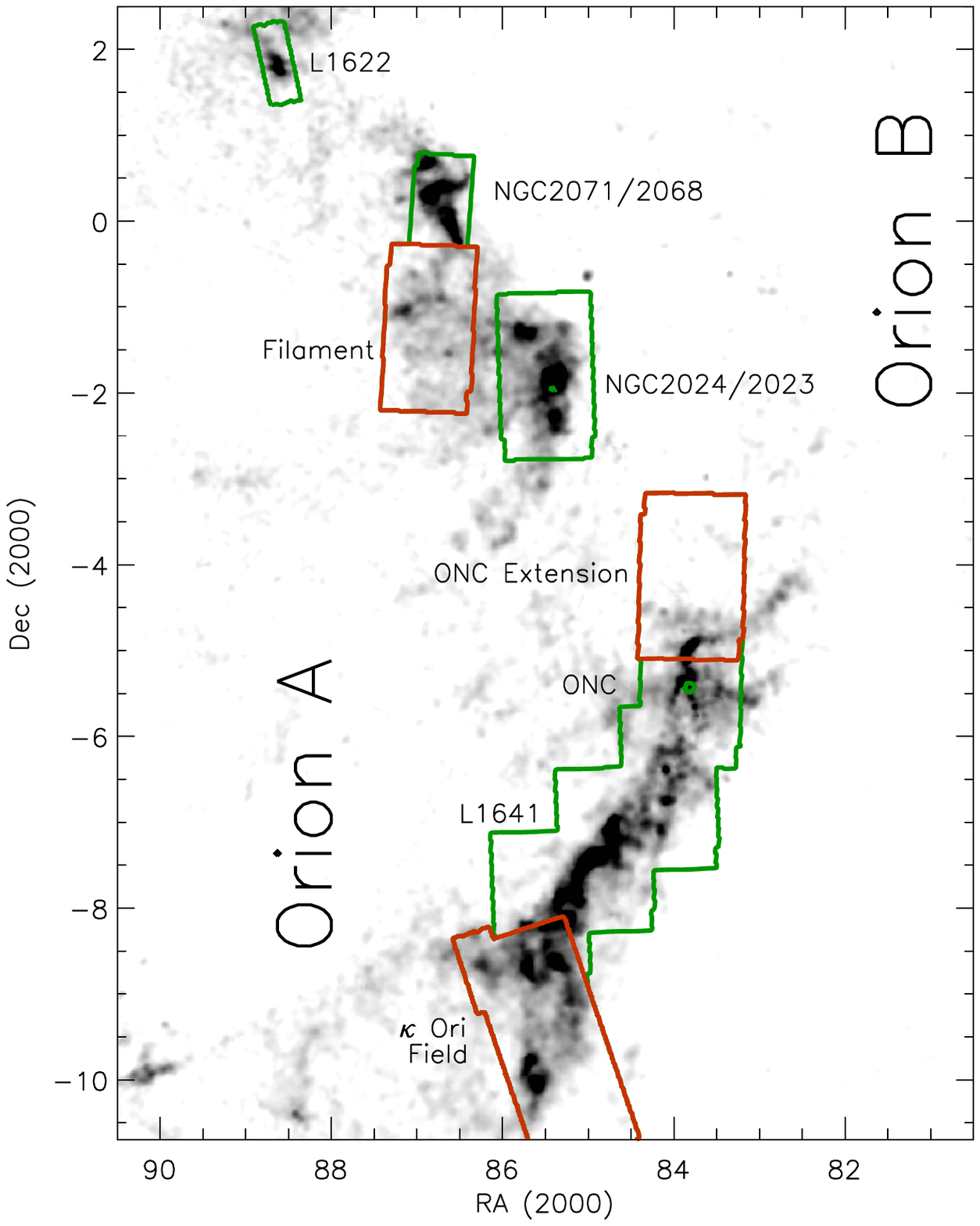}
\caption{The coverage of the {\it Spitzer} Orion cloud survey.  The
greyscale image is an extinction map of the Orion region made from the
2MASS PSC  \citep{2011ApJ...739...84G}.  {\bf
Left:} the fields surveyed with IRAC.  The fields are labeled by the designations
used in the text, these labels are drawn from their constituent star forming regions, 
their designation in the Lynds dark nebulae catalog, or their apparent proximity to a 
prominent star ($\kappa$ Ori).  In the case of Orion A, these fields merge into
one contiguous field.   The extensions of the NGC~2024, ONC
and L1641 fields are also identified. To extend the completeness of our
YSO sample in the Orion~B cloud, we also mapped a small, clumpy
filament separate from the major star forming regions (denoted "Filament") . 
Flanking the Orion A cloud are five fields placed to measure the density of young
stellar objects and contaminating objects outside the cloud, but
within the Orion OB1 association. To the east are three additional
reference fields; these are designed to measure density of
contaminating objects outside the Orion OB1 association. The red/cyan boxes
outline the mosaics obtained in the two IRAC fields of view using the 2 second frame time.
{\bf Right:} the fields surveyed with MIPS.  The five major fields are indicated;
the Orion B filament and NGC~2068/2071 fields overlap. The initial fields
are outlined in green, the extension fields are outlined in red. The saturated regions of
the MIPS data in the Orion nebula and NGC~2024 are outlined in green.}
\label{fig:coverage}
\end{figure}

In Fig.~\ref{fig:coverage}, the surveyed fields of view are shown overlaid on an extinction map of the 
Orion clouds from \citet{2011ApJ...739...84G}.
The Orion cloud complex consists of two distinct giant molecular
clouds (GMCs), Orion A and Orion B, as shown in $^{12}$CO and
$^{13}$CO surveys of this region \citep{1987ApJ...312L..45B,1994ApJ...429..645M,2005A&A...430..523W}.  The
Orion A cloud was surveyed in one contiguous 4.04~deg$^2$ map extending
from NGC 1977 down through the L1641 dark cloud. The survey field was
constructed to follow the distribution of $^{13}$CO emission in the
Bell Labs survey of Orion~A~\citep{1987ApJ...312L..45B}. The survey was later
extended to fully map the Orion nebula cluster (ONC). Subsequent
extinction maps of the Orion clouds showed the Orion~A cloud extending
to a declination of $- 10^{\circ}$  \citep[$\kappa$ Ori extension in Fig.~\ref{fig:coverage}, ][]{2000AJ....120.3139C}; the survey was expanded again to include this
southern extension of the cloud.  In addition to the L1641 dark cloud, the 5.9~deg$^2$ map of Orion~A also contains the
L1640 dark cloud (part of the ONC region) and the L1647 dark cloud (which we include as part of the L1641
region).   The Orion~B map
was broken into three fields centered on the regions of strong $^{13}$CO emission \citep{1994ApJ...429..645M}: 
a 1.27~deg$^2$ field covering the NGC~2024 and NGC~2023 nebulae in L1630, a 1.1~deg$^2$ field covering
the NGC~2068 and NGC~2071 nebulae in L1630, and a 0.36~deg$^2$ field
covering a region of bright $^{13}$CO emission in between the NGC~2068/2071 and NGC~2024/2023
fields (hereafter: Orion B filament).  In addition, a 0.23~deg$^2$ field covering the dark cloud L1622 was 
observed; in this paper, we consider L1622 to be part of the Orion~B cloud \citep{2009AJ....137.3843B}.

%The ONC portion of the Orion A
%cloud was first surveyed in two epochs on 09 March 2004 and 12 October
%2004.  The L1641 protion of Orion A was surveyed on 18-19 February
%2004 (first two dithers), 8 October 2004 (3rd dither) and 27 October
%2004.  The map of Orion A was extended during two epochs on 21-22
%October 2007 and 10 April 2008; the extended fields are delineated in
%Fig.~1.  The NGC~2024/2023 sub-cloud was mapped on 17 February 2004
%and 27-28 October 2004, and was extended on 30 March 2007 and 16
%October 2007 (Fig.~1). The NGC~2068/2071 sub-cloud was mapped 8 March
%2004 and 28 October 2004.  During two epochs on 27 October 2005 and 26
%March 2006, the Lynds 1622 field was obtained. The Orion B filament
%was observed on 29 March 2007 and 17 October 2008.
%The observations of
%the Orion Nebula and NGC~2024 were obtained on 27 October 2008 and 9
%October 2008, respectively.

In program PID 50, the NGC~2024 nebula and Orion Nebula  were
observed with 1.2~sec integration time frames (corresponding to a frame time of 2~sec) 
to obtain high signal to noise in
regions toward which the 10.4 second frames were saturated by bright
nebulosity.  Toward the Orion Nebula, the map consisted of a $3 \times
4$ rectangular grid in spacecraft coordinates with $270''$ offsets
between pointing centers.  Toward NGC~2024, a $2 \times 3$ rectangular
grid in spacecraft coordinates was made with the same offsets.  In
both maps, 24 dithered 2 second frames were obtained per map position
for total integration times of 28.8~sec per band.

A total of eight reference fields were also observed; these covered a total of 0.41~deg$^2$ in all four bands 
(Fig.~\ref{fig:coverage}).  For each field, a single epoch observation with 4 dither positions was
executed; the typical integration time was 41.6 seconds per band.  Five of the fields
were observed in October 2005 (PID 43); these consist of four fields
on either side of the Orion Nebula Cluster and one field targeting a
small cloud east of Orion A detected by \citet{1987ApJ...312L..45B}.  The
remaining fields, all of which are found to the east of the cloud
complex, were observed in October 2007 (PID 30641).  

\begin{table}[h]
%\begin{center}
\caption{MIPS 24~$\mu$m Observations}
\vskip 0.2 in
\begin{tabular}{lccccccc}
\tableline\tableline
Field Name & Scan Rate & Cross Scan & No. of & Total Int.  & Field & Epoch & PID \\
           &         & Offset ('') & Frames &  Time (s) & (deg$^2$) & &  \\
\tableline
ONC        & fast    & 160'' & 6 & 30 & 2.18 &  2004 Mar 20  & 58 \\
ONC Ext.   & fast    & 160'' & 6 & 30 & 2.28 & 2008 Apr 19 & 30641 \\
L1641$^{a}$ & fast   & 148'' & 6 & 30 & 5.35 &  2005 Apr 2 & 47 \\
$\kappa$ Ori$^{b}$  & fast/medium    & 148''/302" & 6/20 & 30/40 & 3.77 & 2008 Nov 26-27 & 50070 \\
NGC2024/23 & fast    & 148'' & 6 & 30 & 2.04 &  2005 Apr 3   & 47 \\
NGC2068/27 & medium  & 160'' &  20 & 80 & 0.97 &  2004 Mar 15  & 58 \\
L1622      & medium  & 302'' & 20 & 40 & 0.31 &  2005 Nov 4  & 43 \\ 
Ori B fil. & fast    & 160'' & 6 & 30 &1.94 &  2008 Apr 16  & 30641 \\
Orion A total &  &  & 11.43 & & \\
Orion B total &  &  &  4.93 & & \\
\tableline
\tablenotetext{a} {This field was mapped in three separate scan maps.}
\tablenotetext{b} {This field was combined from slow and medium rate scan maps.}
\end{tabular}
%\end{center}
\end{table}

\subsection{MIPS Observations}

The MIPS observations were obtained from GTO programs PID 43, 47, 58, 30641 and 50070.  The observations are 
tabulated in Table~2 and displayed in Fig.~\ref{fig:coverage}.  The Orion A cloud was mapped in six distinct fast 
scan maps: one covering the Orion Nebula Cluster (ONC in Table~2), one extending the
ONC map to the north, and three covering L1641 and one covering the part of the
cloud adjacent  in the sky to the star $\kappa$ Ori. The cross scan steps
were 148'' or 160'', resulting in a typical integration time of 30
seconds per pixel.  The combined map covered 11.43~deg$^2$. The
Orion B cloud was mapped using four individual scan maps.  The
NGC~2068/2071 region was mapped using medium rate scans and 160''
cross scan offsets, giving a typical integration time of 80 seconds per
pixel.  The NGC~2024/2023 region and Orion B filament were mapped with fast
scans and either 148'' or 160'' cross scan steps; the resulting integration time
was 30 seconds.  The result was two contiguous maps: a 2.58 deg$^2$ map
including NGC~2068/2071 and the filament, and a 2.04 deg$^2$ map of
NGC2024/2023. Finally, a field covering 0.31~deg$^2$ toward L1622 cloud was mapped using medium rate
scans and 302'' cross scan offsets with a typical integration time of 40
seconds.

\subsection{IRAC Data Reduction}

The data were mosaicked using the custom IDL package Cluster Grinder
\citep{2009ApJS..184...18G}.  The input images were BCD data processed
with the Spitzer Science Pipeline.  The BCD data pipeline versions included
s14, s17 and s18.  Cluster Grinder first prepares the BCD images for mosaicking
by matching the background of adjacent images, executing  cosmic ray masking using overlapping
images, and  ameliorating artifacts due to column pull-down, banding, muxbleed and residual bandwidth effects (see 
the chapter on IRAC in the Spitzer Observation Manual for a discussion of these
artifacts).  

The cosmic ray rejection is done by comparing overlapping pixels in the four or more frames covering a given mosaic pixel. Pixels which differ by more than 10  median absolute deviations from the median
value are masked out of the mosaic.  On the outskirts of the mosaics there are sometimes only two overlapping pixels due to the FOV rotation between epochs; in these cases we increase the redundancy by including multiple pixels from
each overlapping image.  Specifically, we  include pixels that are in the same row or column and
are directly adjacent  to each overlapping pixel resulting in 10 pixels used in the cosmic ray rejection for each mosaic pixel.  Pixels that deviate by more than 20 median absolute deviations from the median value of the 10 pixels are then masked.  For the 1.2~sec imaging of the ONC and NGC~2024, the high level of redundancy will translate into improved cosmic ray rejection compared to the longer integration time mosaics.

The data units were converted from the native BCD units of
MJy/sr$^{-1}$ into DN by dividing by the conversion factor given in
the BCD header (``FLUXCONV'') and then multiplying
by the integration time. Cluster Grinder then uses the refined WCS information in the BCD
image headers  to register the images to the mosaic grids and interpolate
the BCD  images onto the mosaic grid, which had an identical
pixel scale to the BCD data.  The mosaic pixels were given the value
of the average interpolated BCD images weighted by a bad pixel mask.

For the data taken in HDR mode, two mosaics were produced for each
wavelength band: one from the 10.4~sec integration time data and one from the 
0.4~sec integration time 
data.  These are then combined into a single mosaic by replacing all
saturated pixels in the 10.4~s mosaic with the pixel values from the
0.4~s mosaic scaled to a 10.4~s integration time. For the 1.2~sec data,
a single mosaic was produced for each wavelength band.

\subsection{The Generation of the 2MASS and IRAC Point Source Catalog}

The point sources on the HDR combined mosaics were identified using PhotVis v1.10
\citep{2008ApJ...674..336G}.  PhotVis convolved the mosaics with a sunken
Gaussian with a FWHM similar to that of a point source (2 pixels).  It then identified all sources in the convolved
image where the peak was 7 times the RMS of the pixel values in a $10 \times 10$ region
centered on the peak.  Since regions with bright structured nebulosity have higher RMS values,
this approach dramatically reduced the number of nebular knots misidentified as stars.
The point sources identified on each mosaic were then inspected by eye using PhotVis and all obvious
artifacts misidentified as point sources remaining in the data were
removed at this stage.  In addition, a small number of sources which
were missed by PhotVis due to confusion with nebulosity were added.  Photometric
magnitudes were extracted with PhotVis using a 2~pixel radius
aperture and an annulus from  2 to 6~pixels to estimate the sky
contribution.  The zero points were 19.6642, 18.9276, 16.8468, 17.3909
(for units of DN/s) and aperture corrections were 1.213, 1.234, 1.379,
and 1.584 for the 3.6, 4.5, 5.8 and 8~$\mu$m bands, respectively
\citep{2005PASP..117..978R}. The uncertainties for each of the four
IRAC bands were calculated by the IDL Astronomy Library routine
'aper.pro' by combining the  estimated shot noise from the source, the 
uncertainty in the background estimation and the  measured standard 
deviation in the sky annulus \citep[PhotVis incorporates routines
from the IDL Astronomy Library of ][]{1993ASPC...52..246L}.

At this point, an initial point source catalog was generated using the
bandmerging routine within Cluster Grinder.  The routine employs the
following procedure.  First, each 3.6~$\mu$m point source was
associated with the nearest point source in the longer
wavelength bands as long as they were separated by 
$\le 1$\arcsec.  If the nearest point source was beyond 1\arcsec, no association
was made and the source would have a detection only in the 3.6~$\mu$m
band.  This procedure was repeated in turn for the 4.5, 5.8 and
8~$\mu$m point sources that were not previously merged with a shorter
wavelength band.  The point sources were then associated with the
nearest 2MASS point source as long as the 2MASS source was within
a separation of 
1.2\arcsec.  The final R.A. and Dec. pair chosen was that given by the
first wavelength band with an uncertainty $\le 0.25$~mag in the following
progression: 4.5~$\mu$m, 3.6~$\mu$m, 8~$\mu$m, 5.8~$\mu$m and 2MASS.
The end product was a catalog of point sources with R.A.,Dec., photometric
magnitudes for all seven bands, and the uncertainties in all seven bands.

%The photometry from
%the four IRAC bands were then merged by comparing coordinates and
%identifying the closest source with 1 arcsecond.  In addition, the
%sources were merged with sources from the 2MASS point source catalogs,
%resulting in up to seven photometric bands: $J$, $H$, $Ks$, $[3.6]$,
%$[4.5]$, $[5.8]$ and $[8]$.

Photometric magnitudes of stars observed at multiple positions on
flat-fielded BCD images show variations of $5\%$ (see IRAC chapter
of Spitzer Observing Manual).  
To correct for these variation and to search for
photometric variability, final photometric magnitudes were then
derived from the individual BCD images using the coordinates in the point source catalog derived from the mosaics.  These final magnitudes were
obtained for all point sources identified in the mosaics with
uncertainties $\le 0.25$~mag in one of the four IRAC bands.  The steps
of this process were the following.  First, each frame containing a
given source was identified, divided by the conversion 
factor of MJy/sr$^{-1}$ over DN/s (i.e. ``FLUXCONV''),
and then multiplied by the integration time to convert the
units into DN.  These frames were then corrected for column pull-down, banding, muxbleed
and residual bandwith effects using custom IDL programs developed for
Cluster Grinder.  Finally, the R.A. and Dec. pair of each source, as determined
from the mosaics, was converted to pixel
coordinates in the BCD images using the refined WCS information in the BCD headers.
Aperture photometry was extracted at these positions using ``aper.pro''
with an aperture of 2~pixels and a sky annulus of 2 to 6~pixels.  The
zero points and aperture corrections were identical to those used for
the mosaics. The photometry was then further corrected using the array
location dependent photometric corrections given by the Spitzer
science center\footnote{http://ssc.spitzer.caltech.edu/irac/locationcolor/}; these
corrections account for the 5\% variation in point source photometry observed
across the BCD images.  We note that point response function (hereafter: PRF) fitting was not used due to the difficulty
in taking into account the variations of the IRAC PRF across the focal plane. (The PRF is the
point spread function convolved with the pixels response function.)

The photometry was extracted independently from both the 10.4~sec and
0.4~sec frames. Thus, each point sources was typically measured 
eight times (4 dithers and 2 integration times), although in many
cases the sources were not detected at the shorter integration time.
For the ONC and NGC~2024 observations with 1.2~sec frames, each point
source was measured independently for all 24 dithers.  The photometry
of each point source was collated as a function of Julian date for
all three integration times.  Magnitudes were calculated for all 
sources that were detected in two or more images of the same integration time.  For each combination
of source, wavelength band and integration time,
the median magnitude was tabulated using the average of the middle two
magnitudes when an even number of photometry values were present (if there
were only two photometry values, the average of those two values was used). The use of the median value 
eliminates the impact of cosmic ray strikes as long as there were three or more dithers and only one dither 
out of the three was affected.  Each
of the median magnitudes was assigned an uncertainty equal to $(\sum_{i=0}^{N}\sigma_i^2/N^2)^{1/2}$ where
$\sigma_i$ was the uncertainty of the $ith$ magnitude and $N$ was the total number of magnitudes
for a given source, band and frame time combination.  To construct the final table of photometry, the
10.4, 1.2 and 0.4~sec data were compared, and the highest signal to
noise photometry achieved without saturation was adopted.  The integration 
time at which a given point source would saturate was determined from 
the peak pixel value of that point source in the HDR combined mosaics.

Comparison of the photometry determined directly from the mosaics with
that measured from the individual BCD frames showed a systematic offset.  The median values of the differences of 
the magnitudes, $m_{mosaic} - m_{BCD}$, were 0.02, 0.03 and 0.03~mag 
for the 3.6, 4.5 and 8~$\mu$m bands, but was -0.05~mag 
for the 5.8~$\mu$m band.  This resulted in slightly red $[5.8]-[8]$
colors even for pure, Vega-like photospheres.  To remove this offset, we corrected the magnitudes
in the 5.8~$\mu$m band using the average difference between the 5.8~$\mu$m band offset
and the offsets in the other bands. The correction was applied with the equation $m(5.8~\mu m) = m_{BCD}(5.8~\mu m)-0.077$~mag.

A final photometric catalog of IRAC and 2MASS photometry was then
created by replacing the photometry from the mosaics with the median
value of the individual frame photometry.  Note that a photometric magnitude in a given band would be replaced only
if the uncertainty for the mosaic photometry was $\le 0.25$~mag {\it in that band}.  Otherwise,
the magnitude for that band was set to $-100$ and the uncertainty to $10$.  Although the
0.25~mag limit might be considered only a 4$\sigma$ detection, all of the sources
were previously found by PhotVis by showing a peak pixel value 7$\sigma$ above
the noise.   Furthermore, we apply more stringent limits to the photometry when
we search for young stellar objects.

\subsection{The Reduction of the MIPS~24~$\mu$m data and the Generation of the 24~$\mu$m Point Source Catalog}
\label{sec:mipsred}

%[this needs to be revised using Erin's paper - should I shrink this]
The MIPS 24~$\mu$m data were reduced, calibrated and mosaicked using
the MIPS intrument team's Data Analysis Tool
\citep{2005PASP..117..503G}.  The resulting MIPS mosaics have a pixel
size of 1\farcs 245 and units of DN~s$^{-1}$~det~pix$^{-1}$ (where det pix is a detector
pixel).  Point source
identification and aperture photometry was first performed with
PhotVis.  As was done for the IRAC data, PhotVis convolved the mosaics with a sunken
Gaussian having a 4.5 mosaic pixel FWHM and then identified all sources in the convolved
image where the peak pixel value was 7 times the RMS of the pixel values in a $10 \times 10$ region
centered on the peak. An aperture of 5 pixels (6\farcs 225) and a sky annulus
extending from 12 (14\farcs 94) to 15 (18\farcs 675) pixels was used.  We adopted
a zero point magnitude of 16.05; this was calculated using a mosaic pixel
that was 1/2 the diameter of a detector pixel, a calibration factor
of $6.4 \times 10^{-6}$~Jy per DN~s$^{-1}$~det~pix$^{-1}$ (again, pix is
a detector pixel), and a zero magnitude flux in the MIPS 24~$\mu$m band of 7.17~Jy.  
In addition, an aperture correction was included in the zero point magnitude which
consisted of two first parts.  First, we multiplied the signal in the aperture by 1.146; 
this is the correction from a 12 pixel aperture to infinity \citep{2007PASP..119..994E}.  
Furthermore, we applied the aperture correction between the 5 pixel aperture used in 
our photometry and the 12 pixel aperture. This was determined by comparing the 
photometry in 5 and 12 pixel apertures for a sample of bright, isolated and unsaturated 
stars in our sample; the resulting correction was -0.428~mag.  

%[this needs to be revised using Erin's paper]
Given that that the MIPS point spread function (hereafter: PSF) has a FWHM more than twice 
that of the IRAC PRF, it was necessary to revise the photometry using PSF-fitting
photometry to minimize the effect of confusion with other sources and nebulosity.  
The extraction of the MIPS photometry using PSF-fitting is described
in \citet{2012AJ....144...31K}.  The PSF photometry was calibrated
using the aperture photometry obtained for the stars used to construct
the reference PSF. Thus, it will have the same calibration as the aperture
photometry described above.  

%PSF fitting photometry was then performed directly on the mosaics using the
%implementation of DAOPHOT in the IDL Astronomy Users's Library
%\citep{1993ASPC...52..246L}.  A point spread function (PSF) was
%constructed idependently for each mosaic using 10 isolated, relatively
%bright stars.  The calibration of the PSF fitting photometry was
%determined by the aperture photometry of these 10 stars.  The
%resulting PSFs were fit to all the point sources identified in the
%mosaics using a 2 pixel fitting radius (i.e. the fit included only
%pixels in the inner 2 pixel radius). The PSF was then subtracted and
%the above procedure was repeated to identify and photometer sources
%hidden in the wings of brighter sources.

There are two regions of spatially extended saturation in the MIPS 24~$\mu$m survey
where point sources could not be detected.  The largest region of saturation is found in the 
Orion Nebula at the center of the Orion Nebula Cluster.  The saturation
encompasses the Trapezium as well as the region surrounding BN/KL. This region
has been recently mapped at 6-37~$\mu$m with FORCAST on SOFIA by \citet{2012ApJ...749L..22S}.  
The other  region of extended saturation is located in the NGC~2024 nebula and contains
the exciting star of NGC~2024, IRS 2b \citep{2003A&A...404..249B}.

Finally, the MIPS point sources were merged with the combined 2MASS and IRAC  point
source catalog.  The R.A. and Dec. of point sources in the IRAC point source catalog were
compared to the positions of the MIPS 24~$\mu$m point sources.  The MIPS
magnitudes and uncertainties were then assigned to the 
nearest IRAC point sources if they were separated by $\le 2.5''$.  If there was
no IRAC photometry within this limit, the MIPS coordinates, magnitudes and
uncertainties were added to the point source catalog  as a new source.

\subsection{The Spitzer Point Source Catalog and its Completeness}
\label{sec:pscdetail}

\begin{figure}
\epsscale{.9}
\plotone{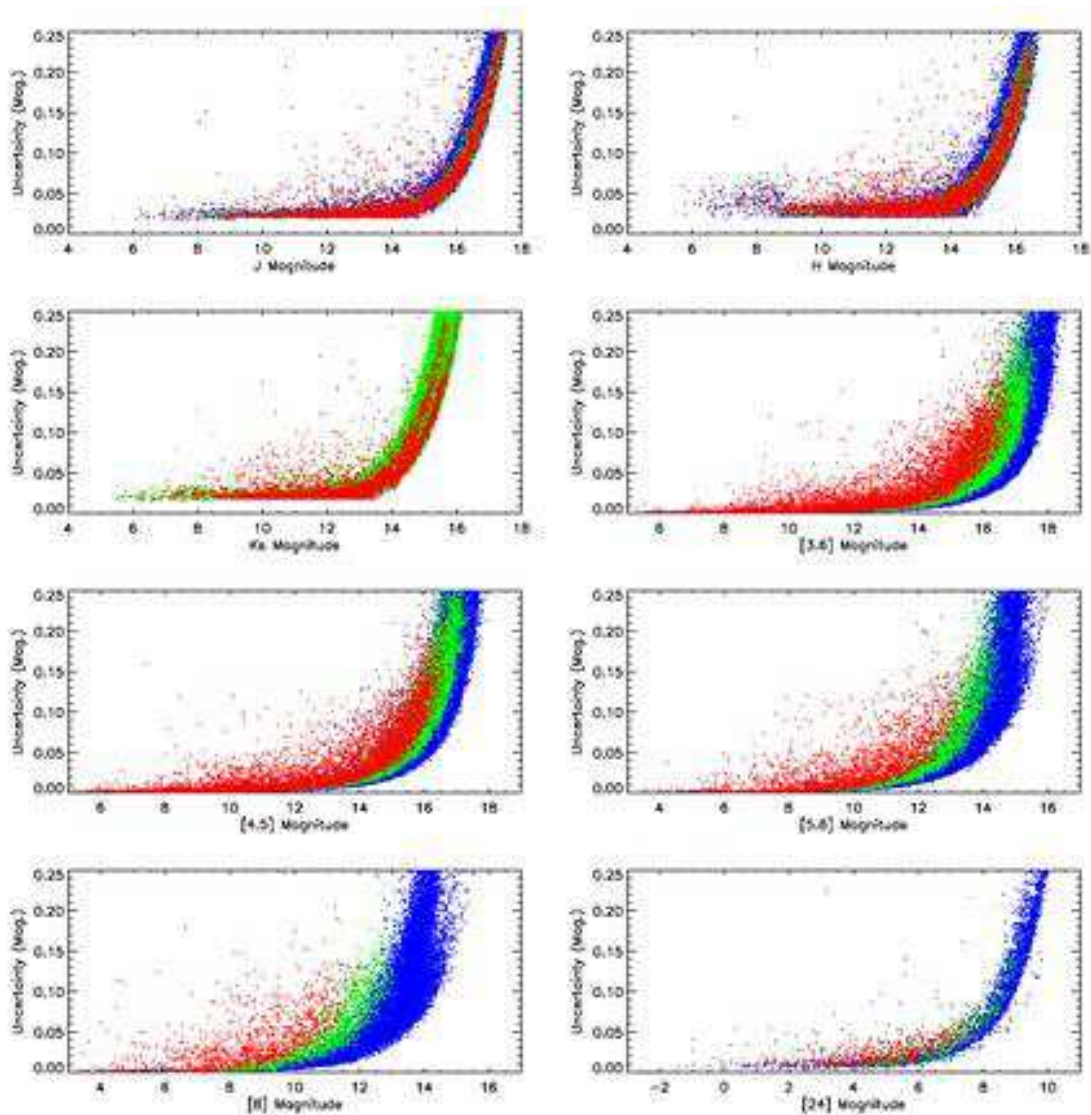}
\caption{Uncertainties as a function of magnitude for the final
adopted photometry in the three 2MASS bands, the four IRAC bands
and the MIPS 24~$\mu$m band.  The colors indicate the RMEDSQ
for a given source in the 8~$\mu$m band (the 8~$\mu$m band exhibits the
brightest nebulosity of the four IRAC bands): blue indicates $RMEDSQ < 30$,
green indicates $30 \ge RMEDSQ < 100$ and red indicates $RMEDSQ \ge 100$.
In the identification of YSOs, only magnitudes with uncertainties less than 0.1~mag 
are used for the $J$, $H$, $Ks$, [3.6] \& [4.5]  bands and with uncertainties
less than 0.15~mag in the [5.8] \& [8] bands.}
\label{fig:unc}
\end{figure}

The final point source catalog contains 306162 sources which are detected in at least
one of the IRAC bands or in the MIPS 24~$\mu$m bands. 
The magnitude uncertainties for the point source photometry are displayed in Fig.~\ref{fig:unc}. 
The uncertainties in all eight bands follow the expected upward sloping curve with the uncertainties increasing with increasing magnitude; 
however, the uncertainties in the IRAC bands show a large scatter for a given magnitude.  The reason for this spread
is that the uncertainties depend not only on the magnitude of the source, 
but also their location. Sources located toward bright,
structured nebulosity or in regions crowded with other point sources
show higher uncertainties than sources with equal magnitudes in regions
without bright nebulosity or crowding.  

To assess the dependence of the measured uncertainties on the presence of nebulosity
and nearby point sources, we determine the root median square 
 deviation (hereafter: RMEDSQ) of the pixel values in an annulus centered on the point source. 
 The RMEDSQ  for a given  star is given by 

\begin{equation}
RMEDSQ(i_0,j_0) = \sqrt{median[(S_{ij}-median[S_{ij}])^2]}
\label{eqn:remedsq}
\end{equation}

\noindent
where $i_0$ and $j_0$ are the pixel coordinates of the source, and
$i$, $j$ are the pixels which are found in an annulus centered on the
source.  For the IRAC photometry, the annulus typically extends from 6 to 11 pixels (7\farcs 2 to
13\farcs 2).  For bright stars, fluctuations in the PRF may dominate the RMEDSQ in this annulus.
To ensure that the PRF itself does not contribute to the RMEDSQ, the radius of the annulus is extended to
larger radii until variations in the PRF make a negligible contribution to the RMEDSQ. The maximum
allowed size of an annulus is from 45 to 50 pixels (54\arcsec to
60\arcsec).  Since the units of our mosaics are DN, the resulting
RMEDSQ are in units of DN.  For the MIPS photometry, the annulus also 
extends from  6 to 11 pixels (7\farcs 47 to
13\farcs 695).  Again,  the radius of the annulus is extended for bright stars to
larger radii until variations in the PSF are negligible with a maximum
allowed size of 45 to 50 pixels (56\farcs 025 to
62\farcs 25).  The units for the MIPS data are DN~s$^{-1}$~det~pix$^{-1}$.
%are these the same used by MIPS  

The resulting RMEDSQ is a measurement of the spatial varying signal
surrounding each point source. These spatial variations are due to 
highly structured nebulosity, nearby point sources, and background photon
noise.  To simplify the analysis, we do not distinguish between the source of
the variations and include any variation in the signal in the RMEDSQ regardless
of the cause.    The  RMEDSQ  is similar to a standard deviation except that it uses  
median values instead of mean values.  Consequently, RMEDSQ is less sensitive to pixels with 
extreme values than the standard deviation
calculated from the same pixels.   This moderates the influence of nearby 
stars, which  only affect a small region of the annulus surrounding the star.  Note that the
RMEDSQ gives the same value as the median absolute deviation.  

In Fig.~\ref{fig:unc}, we show the uncertainties of the point sources
segregated by their RMEDSQ values in the 8~$\mu$m band.  The 8~$\mu$m band is 
used since the spatially structured nebulosity apparent in all four of the IRAC bands is strongest 
in that band.  By measuring the RMEDSQ in the 8~$\mu$m band, we are examining the dependence of the uncertainties on the nebulosity.  In all four of the displayed IRAC bands, there is a clear progression to higher uncertainties with increasing RMEDSQ.  The same progression is seen in the MIPS 24~$\mu$m band, although the spread in uncertainties for a given magnitude is much lower than that observed in the IRAC bands.  In contrast, the 2MASS bands do
not show the same progression.  The reason is that the bright
nebulosity that typically dominates the RMEDSQ values at wavelengths $>
3$~$\mu$m is much fainter relative to stellar sources in the 1-2.5~$\mu$m
bands. Thus, the near-IR $J$, $H$ and $Ks$-bands are not strongly affected by nebulosity.  

The catalog exhibits spatially varying completeness due to confusion 
with nebulosity and, to a lesser extent, the crowding of point sources in dense clusters and groups.
An assessment of the completeness of the point source catalog must take into account
this spatially varying background. Instead of measuring completeness for each position in the 
IRAC maps, we determine  completeness as a function of RMEDSQ.
The completeness for a given source magnitude at any position in the Orion survey can then be 
determined by measuring the RMEDSQ at that position.  

As described in Appendix~A, we added artificial stars to the field containing 
the ONC and then recovered these stars using the same techniques 
described in the previous sections.  For the IRAC bands and the 
MIPS 24~$\mu$m band,  we determined the fraction of 
artificial stars recovered as a function of the stellar magnitude and
the RMEDSQ  in the selected wavelength band.  
The fraction of recovered stars vs RMEDSQ are shown for different source magnitudes  in 
Fig.~\ref{fig:medvfrac} (displaying the four IRAC bands) and Fig.~\ref{fig:medvfrac24}
(showing the MIPS 24~$\mu$m band).   These show a strong dependence of completeness
on both magnitude and the RMEDSQ.

\subsection{The Identification and Classification of Dusty YSOs}
\label{sec:ysoid}

The capability of {\it Spitzer} to identify young stellar objects
(YSOs) with infrared excesses from dusty disks and envelopes is now
well established in the literature
\citep[e.g.][]{2004ApJS..154..363A,2004ApJS..154..367M,
  2004ApJS..154..374G,2004ApJS..154..379M,2004ApJS..154..315W,2005ApJ...629..881H,2007prpl.conf..361A,
  2007ApJ...669..493W,2007AJ....134..411M,2007ApJS..169..328R,2008ApJ...674..336G}.  
 We have employed the methodology developed in \citet{2009ApJS..184...18G} whereby
YSOs exhibiting infrared excesses are identified and classified using
color-color diagrams constructed from the combined 2MASS, IRAC and
MIPS photometry.  We have augmented this approach by using criteria for
identifying and classifying protostars first developed in
\citet{2009AJ....137.4072M} and further refined in \citet{2012AJ....144...31K}.
We have also adopted new criteria to identify a small number of
YSOs that were missed in the criteria of \citet{2009ApJS..184...18G}
and  \citet{2012AJ....144...31K}.

To identify and classify YSOs, we utilized all photometric magnitudes
with uncertainties less than 0.1~mag for the $J$, $H$, $K_s$, 3.6 and
4.5~$\mu$m bands and 0.15~mag for the 5.8, 8 bands.  We did not apply 
an uncertainty cutoff for the 24~$\mu$m band, which in cases of saturated
sources that are fit to the wings of PSFs, may exceed 0.2~mag \citep{2012AJ....144...31K}. 
The sources satisfying these criteria are displayed in
the color-color and color-magnitude diagrams shown in
Figs.~\ref{fig:cc_select}, \ref{fig:cc_classic} \& \ref{fig:cm_mips}.

\subsubsection{The Reddening Law}

The identification of YSOs requires the adoption of a standard
extinction law.  We used a weighted average of the extinction law
determined for three regions in the Orion molecular clouds: NGC2024/NGC2023, NGC2068/NGC2071 and Orion~A  \citep[Table~3 of][]{2007ApJ...663.1069F}.
We adopted values of
$A_J/A_{K_s} = 2.65$,  $A_H/A_{K_s} = 1.55$, $A_{[3.6]}/A_{K_s} = 0.634$, 
$A_{[4.5]}/A_{K_s} = 0.537$, $A_{[5.8]}/A_{K_s} = 0.5$ and
$A_{[8]}/A_{K_s} = 0.504$.
The value of $A_J/A_{K_s}$ has been increased from the value of 2.5 reported in \citet{2005ApJ...619..931I}
to 2.65 for two reasons. 
First, the measured values of $E(H-K_s)/E(K_s-J)$ in \citet{2007ApJ...663.1069F}  and 
their adopted value of $A_H/A_{K_s}=1.55$ requires a value $A_J/A_{K_s}=2.65$.
Furthermore, the reddening vectors in the $H-[4.5]$ vs $J-H$ diagram
more closely follow the loci of the reddened photospheres if
$A_J/A_{K_s} = 2.65$ (see Fig~\ref{fig:cc_classic}). 

\subsubsection{YSO Identification}

\begin{figure}
\plotone{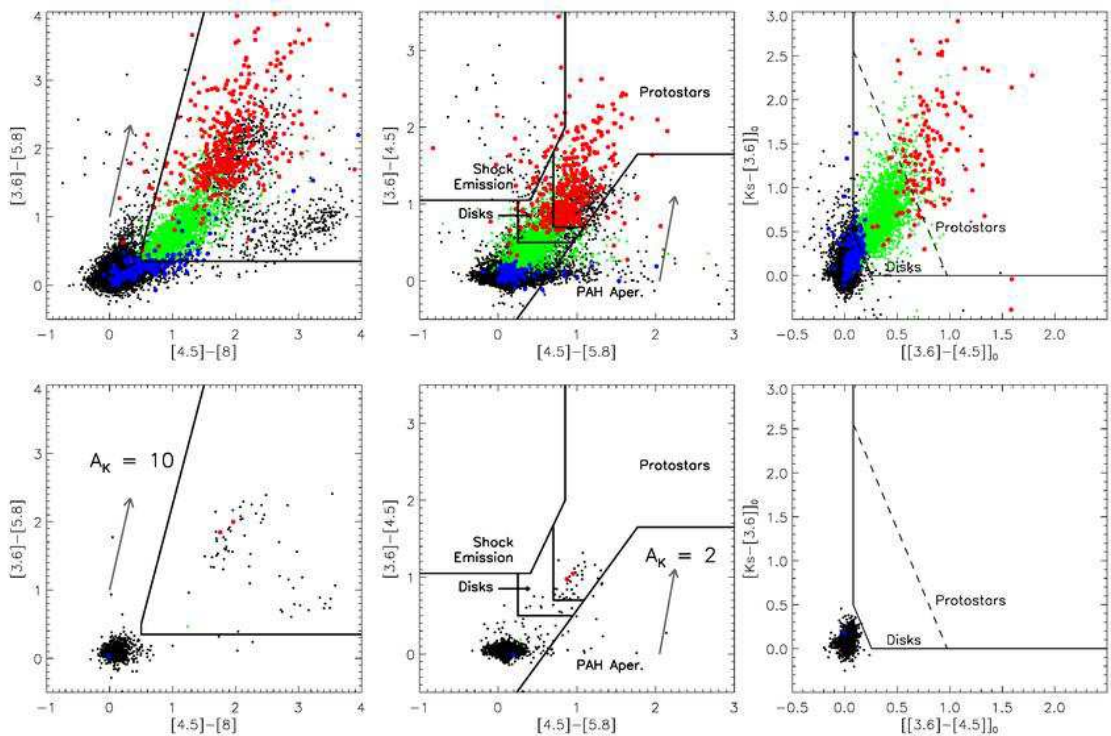}
\caption{The IRAC and IRAC-2MASS color-color diagrams used
to identify young stellar objects for
the star forming clouds ({\bf top}) and reference fields ({\bf bottom}).  The colored sources are those identified as likely YSOs: the green markers show sources classified as young stars with disks (Class II),
the red markers show the sources classified as protostars (Class~0, I and flat spectrum), and the blue markers show sources identified as transition disks on the basis of their 24~$\mu$m photometry.  Note that the protostars include the faint candidate protostars and the red candidate protostars detected at 24~$\mu$m but not at 4.5, 5.8 or 8~$\mu$m. The photometry of the $Ks-[3.6]$ vs $[3.6]-[4.5]$ plots has been dereddened as described in \citet{2008ApJ...674..336G} and \citet{2009ApJS..184...18G}.  The criteria used to identify the YSOs are delineated by the lines shown in each panel.  In addition, the criteria used to distinguish protostars from young stars with disks are shown. These criteria to distinguish protostars are only used four sources without 24~$\mu$m detections; the protostars that do not follow the criteria have been identified by other criteria using the 24~$\mu$m band.  The middle panel shows the criteria used to reject outflow knots (labeled {\it outflow emission}) and to reject sources where the aperture photometry was contaminated by PAH emission (labeled {\it PAH aper.}).  Some of the protostars have colors similar to shock emission: these were classified using their 24~$\mu$m magnitudes and cannot be outflow shocks (which produce weak 24~$\mu$m emission).  We have grouped the IRAC data for the Orion~B  filament field with the off-cloud reference fields given the lack of dusty YSOs detected in that region.  The two YSOs found in the reference fields may be misidentified galaxies; the small number of YSOs in these fields demonstrates that our criteria are successfully rejecting most contaminating galaxies.}  
\label{fig:cc_select}
\end{figure}

The \citet{2009ApJS..184...18G} methodology is divided into multiple phases
designed to classify sources with the most sensitive diagram for the
available photometric bands.  We first applied the Phase I criteria to
sources with detections in all four IRAC bands, 3.6, 4.5, 5.8 and
8~$\mu$m, as described in A.1. of \citet{2009ApJS..184...18G}. These criteria
isolate sources with excess emission in the Spitzer bands relative to
a pure stellar photosphere, and they reject probable
AGNs, galaxies with bright PAH emission, sources contaminated by
structured PAH emission, or knots of shock heated emission.  The
selection process is displayed in Fig~\ref{fig:cc_select}.  Sources
lacking detections in all four IRAC bands, but with magnitudes in the $H$, $K_s$,
3.6~$\mu$m and 4.5~$\mu$m bands were then applied to the Phase II
criteria, as shown see A.2 in \citet{2009ApJS..184...18G}. Again, we excluded
all sources with colors and magnitudes consistent with galaxies,
bright PAH emission, AGN, or knots of shock-heated emission from the
Phase II sample using the criteria in  \citet{2009ApJS..184...18G}.  
Note that in contrast to \citet{2009ApJS..184...18G}, we
group all of the Class~0/I, flat spectrum sources  and Class~II sources 
(i.e. all protostars and pre-main sequence stars with disks) 
together as YSOs with IR-excesses. The classification of  the
YSOs is performed later during the protostar selection (see Sec.~\ref{sec:protoid}).

We identified additional YSOs  by applying criteria designed to identify YSOs
that are not detected in the wavelength bands combinations required by the criteria of
\citet{2009ApJS..184...18G}.  The first of these criteria examines stars detected only in
the $J$, $H$ and 4.5~$\mu$m band.  We followed the approach of
\citet{2004ApJS..154..374G} and \citet{2007ApJ...669..493W} and identified as YSOs sources
satisfying the  inequality:

\begin{equation}
H-[4.5]+\sigma_{H-[4.5]} \ge 0.921 (J-H-0.55+\sigma_{J-H})
\end{equation}

\begin{figure}
\plotone{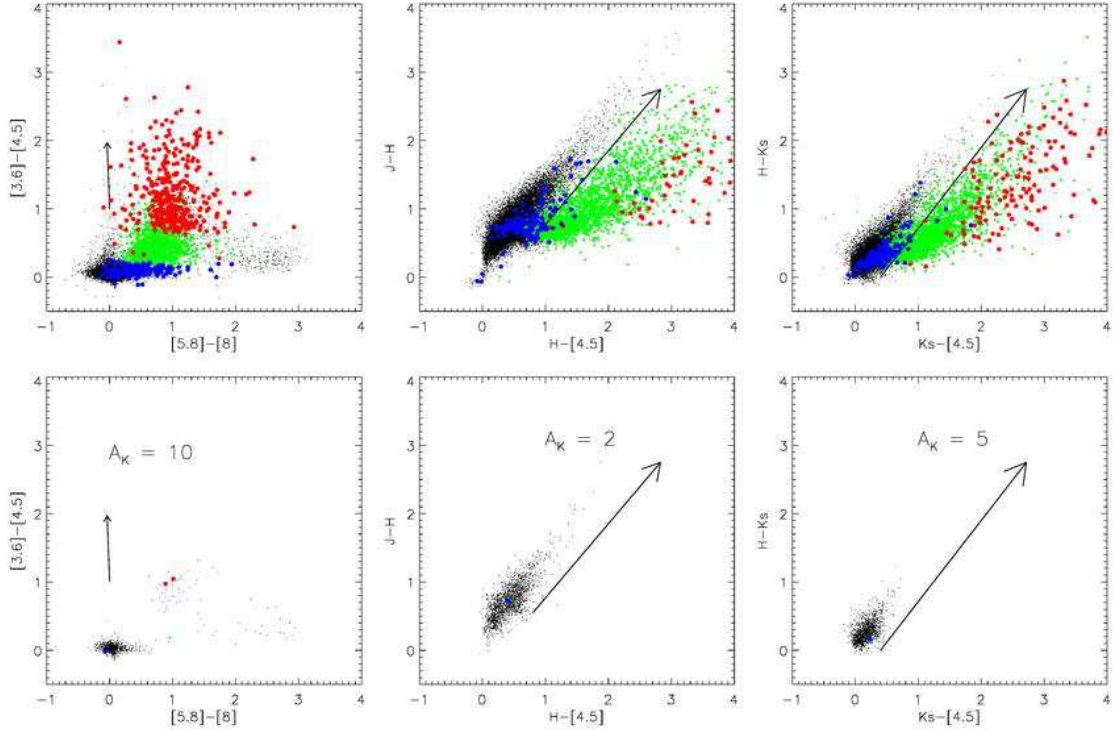}
\caption{Three IRAC/2MASS color-color diagrams commonly used to
identify infrared excesses in the literature.  The color scheme is the same as in 
Fig.~\ref{fig:cc_select}.  We show the star forming clouds in the top row and the reference fields in the bottom row.  The reddening vector displayed in the $J-H$ vs $H-[4.5]$ ({\bf middle panel}) and $H-K$ vs $K-4.5$ diagrams ({\bf right panel})  shows the division between excess and non-excess sources adopted by \citet{2007ApJ...669..493W}}. 
\label{fig:cc_classic}
\end{figure}

\noindent
This selection criterion is illustrated in Fig~\ref{fig:cc_classic}.
In addition, for sources with detections in the 3.6, 4.5, and 5.8~$\mu$m
bands that are missing  8~$\mu$m band detections or detections in the  $H$ and $K$-bands,
we identified YSOs which satisfy the inequalities:

\begin{equation}
[3.6]-[4.5] \ge 0.5 + \sigma_{[3.6]-[4.5]},~[4.5]-[5.8] \ge 0.25 + \sigma_{[4.5]-[5.8]}
\end{equation}

\noindent
These criteria are displayed in Fig~\ref{fig:cc_select}.  To reduce the contamination from 
background galaxies, we further required that YSOs identified by the above two criteria 
satisfy the inequality:

\begin{equation}
[4.5]-A_{4.5} \le 14
\label{eqn:le14}
\end{equation}

\noindent
where $A_{4.5}$ is the total cloud extinction determined from the $A_V$ map.

\begin{figure}
\plotone{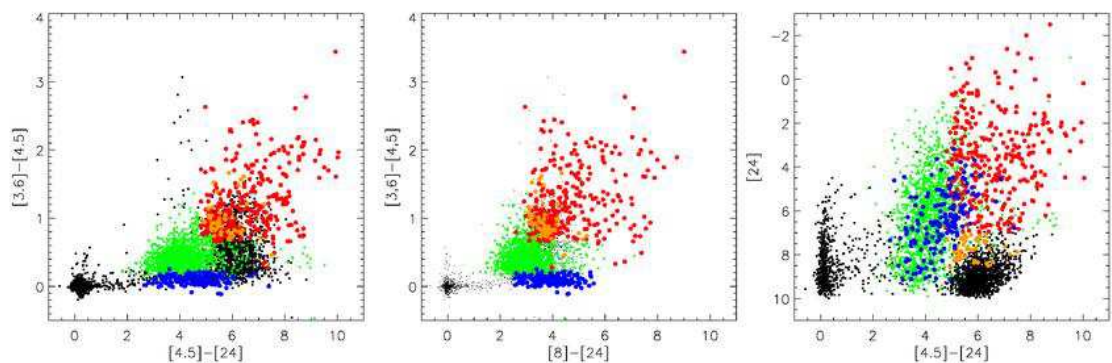}
\caption{The IRAC-MIPS color-color and color-magnitude diagrams
 used to identify YSOs and reject galaxies.  The {\bf left} panel shows the [3.6]-[4.5]
 vs. [4.5]-[24] diagram used to identify protostars.  The {\bf middle} panel shows the
 [3.6]-[4.5] vs. [8]-[24] diagram used to find transition disk sources with weak IRAC excesses due
 to holes in the inner disk.  The {\bf right} panel is the [4.5]-[24] vs [24] diagram used to
 eliminate galaxies from the protostar sample.  The green dots are stars with disks (Class II) 
 excepting the transition disks, which are marked with blue dots.
The red dots are the protostars (Class 0, I and flat spectrum )
 and the orange dots are  the candidate protostars (i.e. they satisfy the IRAC-only criteria used to distinguish
 YSOs from galaxies and AGN, but they are fainter than the MIPS 24~$\mu$m cutoff used to further 
 eliminate galaxies).}
\label{fig:cm_mips}
\end{figure}

The final phase in \citet{2009ApJS..184...18G} is the identification of
IR-excess sources through their MIPS photometry.  At this step, we
diverged from the \citet{2009ApJS..184...18G} criteria, and hence we describe
our criteria explicitly.  We began by identifying sources with weak
excesses in the IRAC bands, but a strong excess in the MIPS 24~$\mu$m
band.  Such objects  include transition disks with large
inner disk holes \citep{2010ApJ...708.1107M}. We took all sources
that satisfied our uncertainty limits in the four IRAC bands, but which
did not show an IR-excess in the IRAC bands.  We then identified as
YSOs sources that satisfied the condition:

\begin{equation}
[8]-[24] > 2.5 + \sigma_{[8]-[24]},
\label{eqn:transition}
\end{equation}

\noindent
where the 2.5~mag threshold comes from \citet{2009ApJS..184...18G}. The identified sources can be seen in Figs.~\ref{fig:cc_select}, \ref{fig:cc_classic}, and \ref{fig:cm_mips}. Note that we only mark sources as transition disk in these figures if they
have detections in all four IRAC bands, have no apparent IR-excess in the IRAC bands, and their dereddened [3.6]-[4.5] colors are less than 0.2~mag.   Next, we identified 
sources that showed strong excesses at 24~$\mu$m but were not identified in all 
four IRAC bands.  These sources satisfied one of the following four inequalities:

\begin{equation}
[3.6]-[24] \ge 2.5~{\rm or}~[4.5]-[24] \ge 2.5~{\rm or}~[5.8]-[24] \ge 2.5~{\rm or}~[8]-[24] \ge 2.5.
\label{eqn:yso24}
\end{equation}

\noindent
Again, the 2.5~mag thresholds come from \citet{2009ApJS..184...18G}. For a magnitude to be 
used in one of the above inequalities, it must satisfy the uncertainty limits
for that band.  In addition, we require that the sources satisfy the criterion:

\begin{equation}
[24] \le 7
\label{eqn:yso24limit}
\end{equation}

\noindent
This criterion rejects the large population of faint, red 24~$\mu$m sources  apparent in Fig.~\ref{fig:cm_mips}.  Unlike the YSOs, which are
concentrated within the boundaries of the cloud, these faint sources are evenly dispersed throughout the mapped regions and are most
likely extragalactic sources \citep{2012AJ....144...31K,2009AJ....137.4072M}. 
Note that the $[24] \le 7$ inequality is not required for the transition disk criterion
in Eqn.~\ref{eqn:transition}.  The reason is that the transition disk criterion
is applied only to the sources with detections in all four IRAC bands that have
already been subject to criteria designed to eliminate extragalactic sources.

After applying all the additional criteria described above, we further reduced contamination from 
galaxies by eliminating all sources with colors and magnitudes consistent with galaxies,
bright PAH emission, AGN, or knots of shock-heated emission using the criteria of  \citet{2009ApJS..184...18G}.

\subsubsection{Protostar Identification}
\label{sec:protoid}

The protostar identification serves two purposes: finding protostellar 
YSOs that were not found using the criteria overviewed in the 
previous section, and classifying the likely protostars among
the entire sample of YSOs.  We implemented the criteria of \citet{2012AJ....144...31K},
who used colors constructed from the IRAC and MIPS bands to find sources with 
flat or rising spectra between the 4.5, 5.8 or 8~$\mu$m IRAC bands and the 
24~$\mu$m MIPS band.  These criteria were derived from the spectral
index criteria of  \citet{1994ApJ...434..614G}, with the adopted color
thresholds equivalent to a spectral index criterion $\alpha \ge -0.3$,
where $\alpha = d\lambda F_{\lambda}/d\lambda$.  
To minimize extragalactic contamination, we adopted the magnitude cutoff
that \citet{2012AJ....144...31K} determined from the Orion cloud such that
all protostars must satisfy $[24] \le 7$.

In regions showing bright nebulosity or saturation in the 24~$\mu$m
bands, we must use alternative criteria for identifying protostars.
Following \citet{2009ApJS..184...18G}, we use two criteria based on the IRAC
and 2MASS photometry.  First, we identify sources with detections in
the 3.6, 4.5 and 5.8~$\mu$m bands guided by the criteria of
\citep{2009ApJS..184...18G}:

\begin{equation}
[3.6]-[4.5] \ge 0.7 + \sigma_{[3.6]-[4.5]},~[4.5]-[5.8] \ge 0.7 + \sigma_{[4.5]-[5.8]}
\end{equation}

To minimize contamination to the protostar sample, all sources
identified as AGN, galaxies with PAH emission, outflow shock knots,
and stars contaminated by PAH emission (as described in A.1 from
\citet{2009ApJS..184...18G}) are eliminated. Furthermore, we also require that
the protostars satisfy the criterion in Eqn.~\ref{eqn:le14}.
 For sources without 5.8~$\mu$m and 24~$\mu$m detections,
we use the Phase~II criteria from \citep{2009ApJS..184...18G} to find
protostars.  These criteria require photometry in the $H$, $Ks$,
3.6~$\mu$m and 4.5~$\mu$m bands.

Once the potential protostars without 24~$\mu$m detections are
identified, we reject sources where the upper limit to the [4.5]-[24]
is $< 4.761$~mag and the lower limit of the 24~$\mu$m magnitude is
$> 7$~mag.  We determine these limits using the 24~$\mu$m
magnitude at which 90\% of the sources are recovered for the local
value of $RMEDSQ$ (Fig.~\ref{fig:medvfrac24}).  In areas where the
24~$\mu$m data are saturated, we do not reject any sources.

As shown in Fig.~\ref{fig:cm_mips}, there are sources that clearly exhibit the red colors of protostars and
satisfy all the Phase~I criteria of \citet{2009ApJS..184...18G}; these sources should be
 considered dusty YSOs and are unlikely to be contamination.  Although they 
satisfy the color criteria for protostars, they ultimately fail the \citet{2012AJ....144...31K} critera
because they have $[24] > 7$.  We classify these YSOs as faint candidate protostars.

A number of objects are also detected that show MIPS detections with $[24] < 7$, but no detections 
in the 4.5, 5.8 or 8~$\mu$m bands.   To distinguish between protostars and sources such as 
asteroids or galaxies, these extremely red sources were visually inspected (asteroids would not 
be detected at the same position by IRAC due to their orbital motion). 
If they appeared as a distinct point source in a region with high extinction, if they were found in
a region of active star formation, or if they were associated with jets apparent in the IRAC images, they
were classified as red candidate protostars.  A total of 13 such objects were added to the catalog.
Far-IR observations with the PACS instrument onboard Herschel will be used to confirm 
their protostellar nature (Stutz et al., in prep.).

\subsubsection{Distinguishing YSO Candidates from Outflow Knots}

Some identified YSOs may be outflow knots which are unresolved in the Spitzer and 2MASS bands \citep{2008ApJ...674..336G}.  
To eliminate these, the position of each YSO in the catalog was examined in ground based $I$, $H\alpha$ and $[SII]$-band
images (J. Bally, private communication). If a source appeared to be a resolved  outflow knot, it was removed from the catalog.  
Only 6 sources were removed as likely outflow knots.

\subsection{YSO Classification, Statistics and Photometry}

We classified the YSOs as either protostars, faint candidate protostars, red candidate protostars, or 
pre-main sequence star with disks.  The protostars and candidate protostars were identified by the criteria in
Sec.~\ref{sec:protoid}.  Any YSO with an IR-excess that was not classified as a  protostar or candidate protostar 
was classified as a pre-main sequence star with a disk.  We have not  distinguished between transition disks and  
disks with optically thick inner regions in our catalog; these can be identified using the criteria described in Sec.~\ref{sec:ysoid} or with
criteria proposed in the literature  \citep[e.g.][]{2010ApJ...708.1107M,2010ApJ...712..925C,2010ApJ...718.1200M}.

Highly reddened stars with disks may be misidentified as protostars; this can occur when a star with a disk is 
observed through a dense region of the molecular cloud or when a  star is observed through a nearly  edge-on 
disk. \citet{2012AJ....144...31K} applied the approach of \citet{2009ApJS..184...18G} to put upper limits on the 
number of young stars misidentified as protostars due to their nearly edge-on disks, and they used Monte-Carlo 
simulations to estimate the number of  young stars with disks that are misidentified as protostars due to extinction from the molecular cloud. 
They estimated that at most 24\%  of the protostars are actually misclassified stars with disks.
The ongoing Herschel Orion Protostar survey \citep[][Fischer et al. in press]{2010A&A...518L.122F} will provide more stringent limits on this percentage by using
Herschel 70 and 160~$\mu$m observations to confirm the detection of infalling envelopes around Spitzer identified protostars.

The total statistics of the number of sources, their classification, and the method 
used to identify them, is given in Table~\ref{table:number_YSO}. Note that red candidate protostars
are not tabulated as IR-excess sources; since they are only detected in 1 to 2 bands, we cannot
establish whether they have excesses using our criteria. In the table, the number of
additional sources are given in each column, starting with the leftmost column
and progressing to the right.  

\begin{table}[h]
%\begin{center}
\caption{Dusty YSOs: Statistics on Detection and Classification}
\begin{tabular}{lccccccccc}
\tableline\tableline
Type & Total$^1$ & 3.6-8$^2$ & H, K$^2$ & 3.6,4.5$^2$ & J, H$^2$ & 3.6-8$^2$ &4.5,24$^2$ &3.6, 5.8$^2$ & 24$^2$ \\
&  & & 3.6,4.5 & 5.8 & 4.5 &  24  &  & or 8, 24 &  \\
\tableline
 all$^3$       & 306162$^4$ & 17337 & 23956 & 27934 & 27016 & 3296 & 4506 & 3365 & -  \\
ir-ex$^5$          & 3469 & 2004    & 704 & 117 & 5 & 183   & 411 & 45 & - \\
disks$^6$       & 2991 &  2004  & 689 & 61    & 5 & 183 & 26 & 23 & -    \\
proto$^7$ & 428   & -          & 15   & 56    & - & -      & 335 & 22 & -   \\
faint$^8$  & 50     &  -         & -      & -       & - & -      & 50   & - & -   \\
red$^9$   &   10    &  -         & -      & -      & - & -       & -   & - & 10 \\
\tableline
\tablenotetext{1} {Total number of sources for a given type.}
\tablenotetext{2} {Wavelength bands used to find and classify source.}
\tablenotetext{3} {Number of sources with detections in the wavelength bands listed for a given column.}
\tablenotetext{4} {Number of sources in the point source catalog.}
\tablenotetext{5} {Number of infrared excess sources; this includes stars with disks, protostars and faint candidate protostars.}
\tablenotetext{6} {Source classified as pre-main sequence stars with disks.}
\tablenotetext{7} {Sources classified as protostars.}
\tablenotetext{8}{Faint candidate protostars: dusty YSOs classified as candidate protostars which are fainter than the 24~$\mu$m threshold for protostars.}
\tablenotetext{9}{Red candidate protostars: tentative dusty YSOs that are only detected in the 24~$\mu$m band and occasionally, the 3.6~$\mu$m band. The implied red colors of these sources and their association with dark clouds, star forming regions, or jets indicate they are likely to be protostars.  These are not included as IR-excess sources.}
\end{tabular}
%\end{center}
\label{table:number_YSO}
\end{table}

The final YSO  catalog is given in Table~\ref{table:photometry}. This catalog contains all of the identified
{\it Spitzer} dusty YSOs, their 2MASS photometry, their photometry in all four IRAC
bands, their photometry in the MIPS 24~$\mu$m band, and their classification.  The uncertainties
given are the formal uncertainties from the extraction of the photometry.  In addition
to these formal uncertainties, the IRAC and MIPS instruments have absolute calibration uncertainties
of  0.05~magnitudes\footnote{http://irsa.ipac.caltech.edu/data/SPITZER/docs/irac/iracinstrumenthandbook/2/,
http://irsa.ipac.caltech.edu/data/SPITZER/docs/mips/mipsinstrumenthandbook/}.
Since the absolute uncertainties affect all the sources equally, we tabulate only the formal
uncertainties which varies from source to source.  For consistency with the tables in young cluster
survey of \citet{2009ApJS..184...18G}, we also include the extinction toward the source and the spectral index of the source
over the four IRAC bands.  The determination of those values are described in \citet{2009ApJS..184...18G}.

\begin{deluxetable}{cccccccccccccc}
\tabletypesize{\scriptsize}
\rotate
\tablecaption{{\it Spitzer}-identified YSOs: IRAC, MIPS, and 2MASS Magnitudes}
\tablewidth{0pt}
\tablehead{\colhead{Index} & \colhead{R.A.\tablenotemark{a}} & \colhead{Decl.\tablenotemark{a}} & \colhead{$J$} & \colhead{$H$} & \colhead{$K_S$} & \colhead{$[3.6]$} & \colhead{$[4.5]$} & \colhead{$[5.8]$} & \colhead{$[8.0]$} & \colhead{$[24]$} & \colhead{$A_K$\tablenotemark{b}} & \colhead{$\alpha_{IRAC}$\tablenotemark{c}} & \colhead{Class}\tablenotemark{d} }
\startdata
1 & 05:42:42.68 & -10:25:09.3 & $14.38 \pm 0.04$ & $12.06 \pm 0.03$ & $10.50 \pm 0.02$ & $8.53 \pm 0.01$ & $7.57 \pm 0.01$ & $6.71 \pm 0.01$ & $5.85 \pm 0.01$ & $1.97 \pm 0.01$ & 1.42 & 0.22 & P \\
2 & 05:42:30.55 & -10:10:47.8 & ... $\pm$ ... & ... $\pm$ ... & ... $\pm$ ... & $14.48 \pm 0.01$ & $13.97 \pm 0.01$ & $13.52 \pm 0.05$ & $12.95 \pm 0.05$ & ... $\pm$ ... & ... & -1.08 & D \\
3 & 05:43:01.59 & -10:07:50.3 & ... $\pm$ ... & ... $\pm$ ... & ... $\pm$ ... & $14.58 \pm 0.01$ & $13.91 \pm 0.01$ & $13.35 \pm 0.04$ & $12.95 \pm 0.05$ & ... $\pm$ ... & ... & -0.97 & D \\
4 & 05:42:13.07 & -10:05:34.9 & ... $\pm$ ... & $15.18 \pm 0.06$ & $13.60 \pm 0.04$ & $11.55 \pm 0.01$ & $10.57 \pm 0.01$ & $9.69 \pm 0.01$ & $8.84 \pm 0.01$ & $5.48 \pm 0.01$ & 0.35 & 0.26 & P \\
5 & 05:42:31.77 & -10:05:26.2 & $14.21 \pm 0.03$ & $12.75 \pm 0.02$ & $11.75 \pm 0.02$ & $10.45 \pm 0.01$ & $9.92 \pm 0.01$ & $9.48 \pm 0.01$ & $8.87 \pm 0.01$ & $6.22 \pm 0.02$ & 0.56 & -1.04 & D \\
6 & 05:42:38.03 & -10:03:43.6 & $14.05 \pm 0.03$ & $12.77 \pm 0.03$ & $12.14 \pm 0.03$ & $11.57 \pm 0.01$ & $11.30 \pm 0.01$ & $10.92 \pm 0.01$ & $10.55 \pm 0.02$ & $7.68 \pm 0.04$ & 0.63 & -1.64 & D \\
7 & 05:42:59.93 & -10:03:40.0 & $12.82 \pm 0.02$ & $11.84 \pm 0.03$ & $11.44 \pm 0.03$ & $10.98 \pm 0.01$ & $10.78 \pm 0.01$ & $10.33 \pm 0.01$ & $9.29 \pm 0.01$ & $5.79 \pm 0.01$ & 0.38 & -0.86 & D \\
8 & 05:42:59.99 & -10:03:35.4 & ... $\pm$ ... & $14.09 \pm 0.08$ & $13.49 \pm 0.06$ & $12.70 \pm 0.01$ & $12.44 \pm 0.01$ & $12.03 \pm 0.02$ & ... $\pm$ ... & ... $\pm$ ... & 0.27 & ... & D \\
9 & 05:42:57.84 & -10:03:35.3 & $16.16 \pm 0.08$ & $15.18 \pm 0.07$ & $14.48 \pm 0.10$ & $14.06 \pm 0.01$ & $13.75 \pm 0.01$ & $13.23 \pm 0.04$ & $12.83 \pm 0.05$ & ... $\pm$ ... & 0.09 & -1.38 & D \\
10 & 05:42:37.10 & -10:03:29.6 & $12.83 \pm 0.03$ & $11.64 \pm 0.03$ & $11.07 \pm 0.02$ & $10.45 \pm 0.01$ & $10.15 \pm 0.01$ & $9.75 \pm 0.01$ & $8.99 \pm 0.01$ & $6.21 \pm 0.02$ & 0.53 & -1.15 & D \\
11 & 05:42:44.44 & -10:03:18.4 & $13.50 \pm 0.03$ & $11.92 \pm 0.02$ & $11.14 \pm 0.02$ & $10.42 \pm 0.01$ & $10.19 \pm 0.01$ & $9.87 \pm 0.01$ & $9.17 \pm 0.01$ & $5.14 \pm 0.02$ & 0.95 & -1.39 & D \\
12 & 05:42:52.82 & -10:02:41.3 & $15.76 \pm 0.07$ & $14.63 \pm 0.06$ & $14.05 \pm 0.07$ & $13.41 \pm 0.01$ & $13.13 \pm 0.01$ & $12.80 \pm 0.03$ & $12.13 \pm 0.02$ & ... $\pm$ ... & 0.43 & -1.36 & D \\
13 & 05:42:52.59 & -10:02:30.4 & $14.70 \pm 0.03$ & $13.60 \pm 0.04$ & $13.00 \pm 0.03$ & $12.46 \pm 0.01$ & $12.28 \pm 0.01$ & $12.10 \pm 0.02$ & $12.03 \pm 0.02$ & $8.57 \pm 0.09$ & 0.37 & -2.33 & D \\
14 & 05:42:06.84 & -10:02:28.2 & ... $\pm$ ... & $15.92 \pm 0.14$ & $14.62 \pm 0.11$ & $13.93 \pm 0.01$ & $13.23 \pm 0.01$ & $12.38 \pm 0.02$ & $11.25 \pm 0.02$ & $7.31 \pm 0.03$ & 0.59 & 0.25 & FP \\
15 & 05:42:12.41 & -10:02:11.1 & $13.31 \pm 0.02$ & $12.27 \pm 0.03$ & $11.78 \pm 0.02$ & $11.27 \pm 0.01$ & $10.99 \pm 0.01$ & $10.54 \pm 0.01$ & $9.63 \pm 0.01$ & $6.82 \pm 0.01$ & 0.37 & -0.94 & D \\
16 & 05:42:34.41 & -10:01:51.9 & ... $\pm$ ... & ... $\pm$ ... & $10.15 \pm 0.03$ & $9.65 \pm 0.01$ & $9.45 \pm 0.01$ & $8.26 \pm 0.01$ & $6.54 \pm 0.02$ & $3.93 \pm 0.01$ & ... & 0.89 & D \\
\enddata
\tablenotetext{a}{J2000 coordinates}
\tablenotetext{b}{Only provided for sources with valid $JHK_S$ or $HK[3.6][4.5]$ photometry.}
\tablenotetext{c}{Extinction is not accounted for in these values.  High extinction can bias $\alpha_{IRAC}$ to higher values.}
\tablenotetext{d}{D: disks, P: protostars, FP: faint candidate protostar, RP: red candidate protostars (24~$\mu$m detection without 4.5, 5.8 or 8~$\mu$m detection).}
\label{table:photometry}
\end{deluxetable}

\subsection{Contamination of the YSO Catalog by Galaxies}
\label{sec:yso_galcont}

Background galaxies, and in particular AGN, often show colors similar to those of YSOs 
\citep{2005ApJ...631..163S,2008ApJ...674..336G}.   Although the YSO criteria described 
above are designed to  minimize contamination due to galaxies, a residual contamination is expected. 
To assess the residual contamination of galaxies, we take two approaches. First we use data from 
SWIRE {\it Spitzer} Legacy program,   a wide-field survey of galaxies executed with the 
IRAC and MIPS instruments \citep{2003PASP..115..897L}.  Second, we use the reference fields observed as
part of the Orion molecular cloud survey.

%ir exccess contamination=      14.5000+/-      2.69258
%protostars contamination =      2.50000+/-      1.11803
%cand protostar contamination =      4.00000+/-      1.41421

To estimate the contamination from the SWIRE survey, we used the IRAC and MIPS photometry from a 
circular region in the Elias N2 field with a total area of 2~sq.~deg.  We merged these data with the 
$J$, $H$ and $Ks$-band photometry from the 2MASS point source catalog.
To facilitate a comparison between the Orion survey and the deeper SWIRE data, we first ``degrade'' the 
uncertainties of the SWIRE data so that they are comparable to those achieved with the Orion survey.  Using the
Orion photometry, an empirical curve of uncertainty vs. magnitude is computed by calculating the median 
uncertainty as a function of magnitude for each band.  For each point source in the SWIRE survey,
the tabulated uncertainty was replaced by the median Orion uncertainty for the equivalent magnitude and 
wavelength band.  We then applied  our methodology for identifying YSOs to the SWIRE data. From this 
analysis, we find a total of $14.5 \pm 2.7$  galaxies are misidentified as YSOs per square~degree.  Of these, 
$2.5 \pm 1.1$ are misidentified as protostars and $4.0 \pm 1.4$ are misidentified as faint candidate protostars.

%ir exccess contamination=      10.3819+/-      5.99397
%protostars contamination =      3.46062+/-      3.46062
%cand protostar contamination =      0.00000+/-      0.00000
%ir exccess contamination=      2.75691+/-      2.75691
%protostars contamination =      0.00000+/-      0.00000
%cand protostar contamination =      2.75691+/-      2.75691
%ir exccess contamination=      6.13788+/-      3.06894
%protostars contamination =      1.53447+/-      1.53447
%cand protostar contamination =      2.75691+/-      1.53447

\begin{figure}
\plottwo{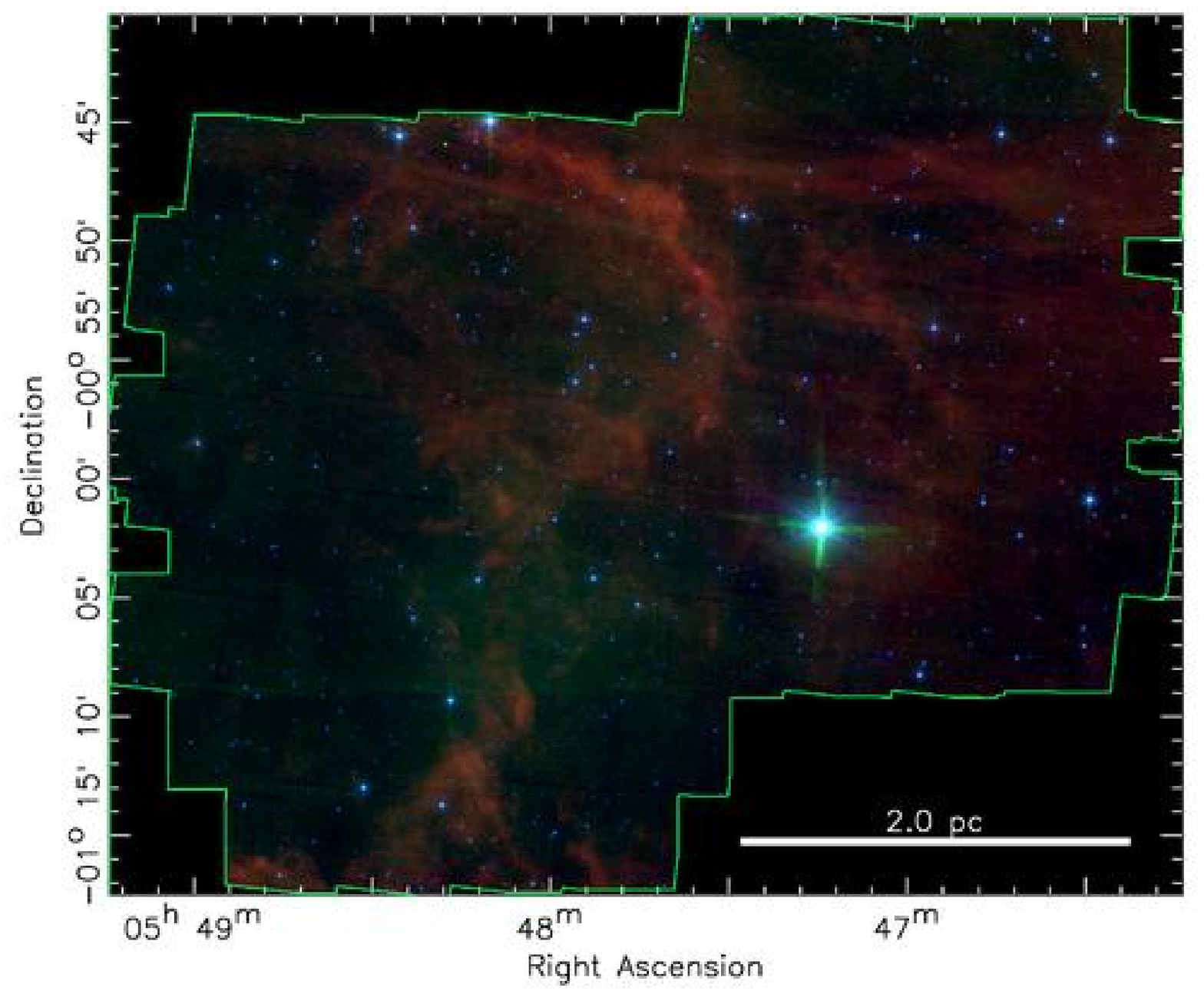}{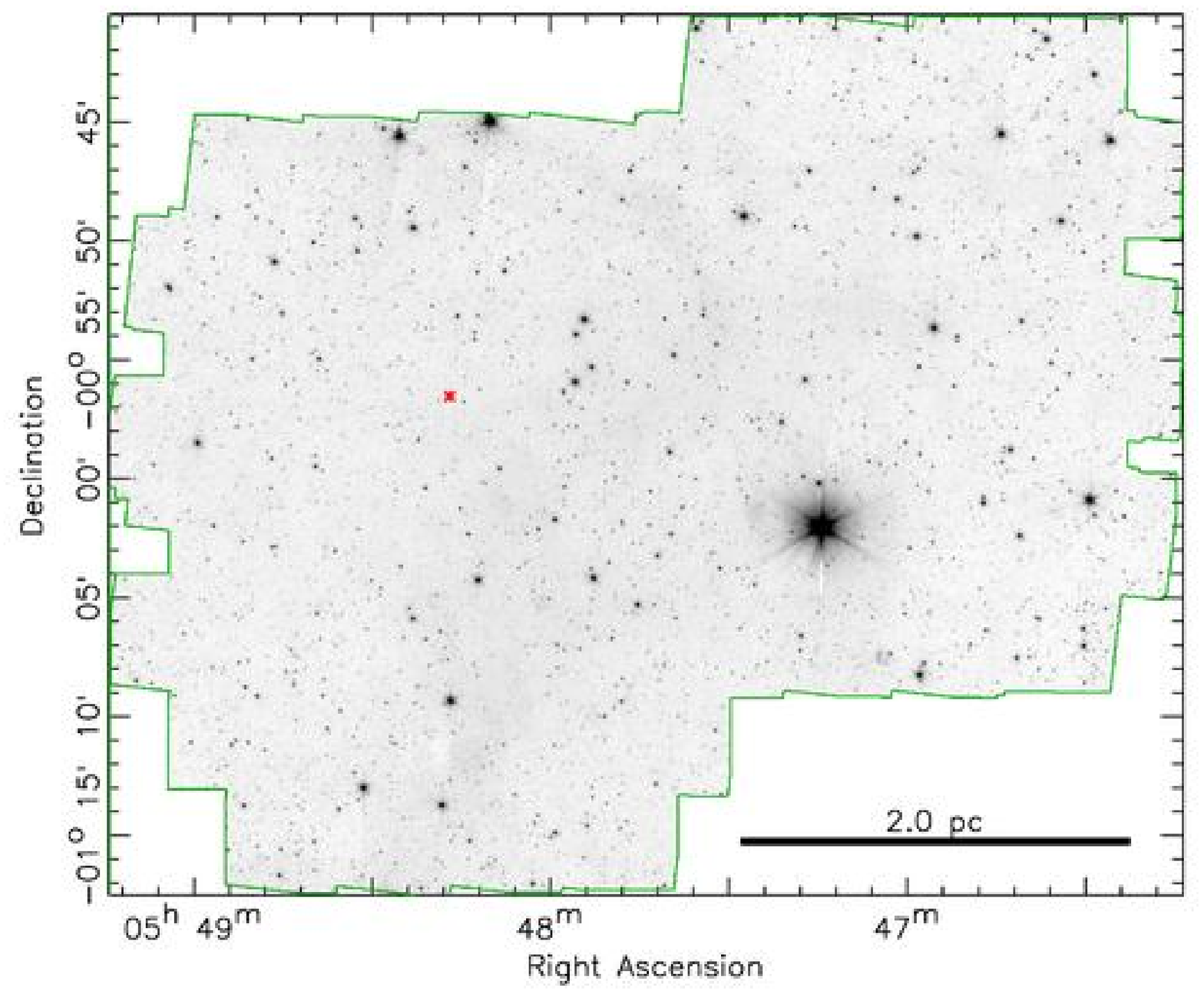}
\epsscale{1.}
\caption{{\bf Left:} Mosaic of the Orion~B filament field.  Blue is 4.5~$\mu$m, green is 5.8~$\mu$m and red is 24~$\mu$m. {\bf Right:} 4.5~$\mu$m image with the positions of dusty YSOs superimposed.  The red asterisk shows the only YSO identified in this field; a faint candidate protostars. In both panels, the green line outlines the surveyed field. }
\label{fig:fil}
\end{figure}

We also imaged eight off-cloud reference fields with the IRAC instrument; the overlapping regions
of 3.6 and 4.5~$\mu$m cover a total of 0.29~deg$^2$ (Fig.~\ref{fig:coverage}, Table~\ref{tab:irac_coverage}).
In these fields, we find a total of three IR-excess objects, for a density of $10.4 \pm 6.0$~objects per square~degree. In addition to these fields, 
there is one cloud field,  the Orion B "Filament", in which we detected only a single YSO (Fig.~\ref{fig:fil}). We include this 0.36~deg$^2$. map as a reference field since it was 
covered by MIPS and IRAC and since it provides a measurement of the contamination toward the cloud with its higher extinction.  If we assume that all the 
IR-excess sources in the combined reference  fields are objects misidentified as dusty YSOs,  we find the total contamination is $6.1 \pm 3.1$  
objects per square~degree, of which $1.5 \pm 1.5$ are misidentified as  protostars and $2.8 \pm 1.5$ are misidentified as faint candidate protostars.  Note,  that the 
density of IR-excess sources in the off-cloud fields is consistent with the density of galaxies misidentified as YSOs in the SWIRE field.  However,
when we include the reference field toward the Orion~B cloud, the contamination drops significantly; this may result
from the higher extinction through the on-cloud region.   We adopt the value for the combined
reference fields, $6.1 \pm 3.1$  objects per sq.~deg.~misidentified as YSOs, as the typical source contamination since
it is more representative of on-cloud regions targeted in this survey.
We note this is similar to the contamination of 8~objects~per~sq.~deg. quoted by \citet{2009ApJS..184...18G}.

%The SWIRE field may have a higher density of galaxies.  Alternatively this may
%result from differences in the extinction toward the two fields; the galactic latitude of the SWIRE field is $42^o$ compared to  the median
%galactic latitude of 
%$-15^o$ for the reference fields.  We find this explanation unlikely since the difference in the extinction between the two positions
%is expected to be only $A_V = 4.3$~mag. \citep{1998ApJ...500..525S}.    We adopt the reference fields as the typical source contamination 
%(3.4 objects per sq.~deg.).  We note this is similar to the contamination of 8~objects~per~sq.~deg. quoted by \citet{2009ApJS..184...18G}.

\subsection{The Effect of Nebulosity on Photometry and the YSO Catalog}
\label{sec:yso_nebcont}

Another form of contaminants are stars mis-classified as dusty YSOs due to their coincidence with bright nebulosity.    Aperture photometry of point sources superimposed on structured nebulosity may result in some of the nebulosity  contributing to the observed magnitudes and creating a fake IR-excess.  To assess the number of such misidentified stars, we first identified every star in the survey field with sufficient photometry to detect an IR-excess but without the detection of an IR-excess.  From this sample of stars we excluded likely extragalactic contamination leaving a total of 27339 sources.

%PHOT_REF        DOUBLE    = Array[18, 27339]

We then added an artificial star $10''$ east of every one of the identified stars. The artificial stars closely reproduce the spatial distribution of stars without IR-excesses, which are primarily field stars, and thus eliminate potential biases induced by the adopted spatial distribution.  For example, this experiment will correctly
account for the fact that most of the field stars are found in dark regions without bright nebulosity.  The colors and magnitudes of the artificial stars were taken from the set of stars in the reference fields which also had sufficient photometry to detect an excess, but were not identified as IR-excess sources or galaxies.  We ignore the MIPS data due to the small number of additional YSOs identified solely with MIPS and since the MIPS photometry was extracted with PSF fitting that may reduce, but not eliminate, the effect of nebulosity. Furthermore, all  but one of the adopted reference fields used to supply the colors and magnitudes of the artificial stars  lacked MIPS photometry. 

% LOADCT: Loading table B-W LINEAR
%   0.19414707 5-8
%      0.14345513 5-8 ref
 %   0.066811992 3-4
 % 0.054514543 3-4 ref

\begin{figure}
\plotone{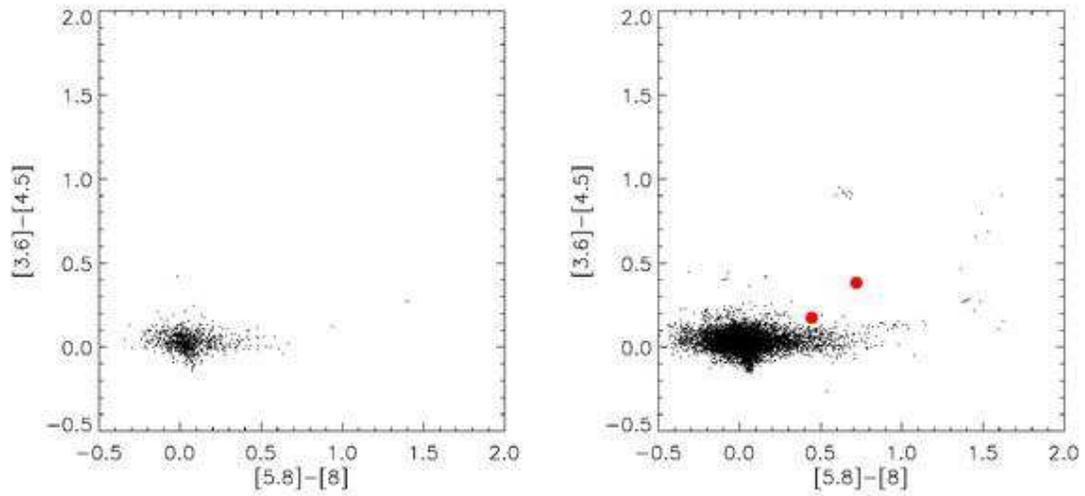}
\caption{{\bf Left panel:} The IRAC color-color diagram of the stars added to the Orion survey to test the effect of nebulosity on YSO identification.  A total 27339 stars, none of which had 
IR-excesses, were added to the image.  Displayed are the input colors  of the 11372 artificial stars with photometry in the four IRAC bands.   {\bf Right panel:} IRAC color-color diagram of the 8852 artificial stars  recovered in all four IRAC band.  Out of the 27339 added stars, only 12 showed IR-excesses in the recovered data.  Of the 12 IR-excess sources, only two had 
photometry in all four IRAC-bands (red dots).}
\label{fig:colorless}
\end{figure}

We next recovered the photometry of the artificial stars. The input and recovered photometry is shown in Figure~\ref{fig:colorless}. The IRAC color-color diagrams show a significantly larger spread in the $[5.8]-[8.0]$ color of the recovered data due to the nebulosity in the image.  In comparison, the increase in the spread of the $[3.6]-[4.5]$ color is negligible.   This can be shown quantitatively by examining the standard deviation; while the standard deviation of the $[3.6]-[4.5]$ color changed from 0.05~mag to 0.07 mag, the standard deviation  of the $[5.8]-[8]$ color changed from 0.14~mag to 0.19~mag. Furthermore, a  number of objects are scattered to $[5.8]-[8]$ colors
exceeding 0.5~mag but maintain small $[3.6]-[4.5]$ colors. This diagram demonstrates the necessity of rejecting candidate YSOs with IR-excesses only in the  
$[5.8]-[8.0]$ color. Out of the 27399 artificial stars, only 13 showed infrared excess using  our criteria for the IRAC color-color diagram.  We note that 11 of those 13 stars have recovered 4.5~$\mu$m magnitudes that are more than 0.5~mag brighter than the input magnitudes; these may be artificial sources that have been added on top of actual sources.  We conclude that the contribution to our YSO sample by stars with spurious IR-excesses due to confusion with nebulosity and stars is negligible.  

\section{The Spatial Distribution and Density of  Dusty YSOs}
\label{sec:spatial}

\begin{figure}
\plotone{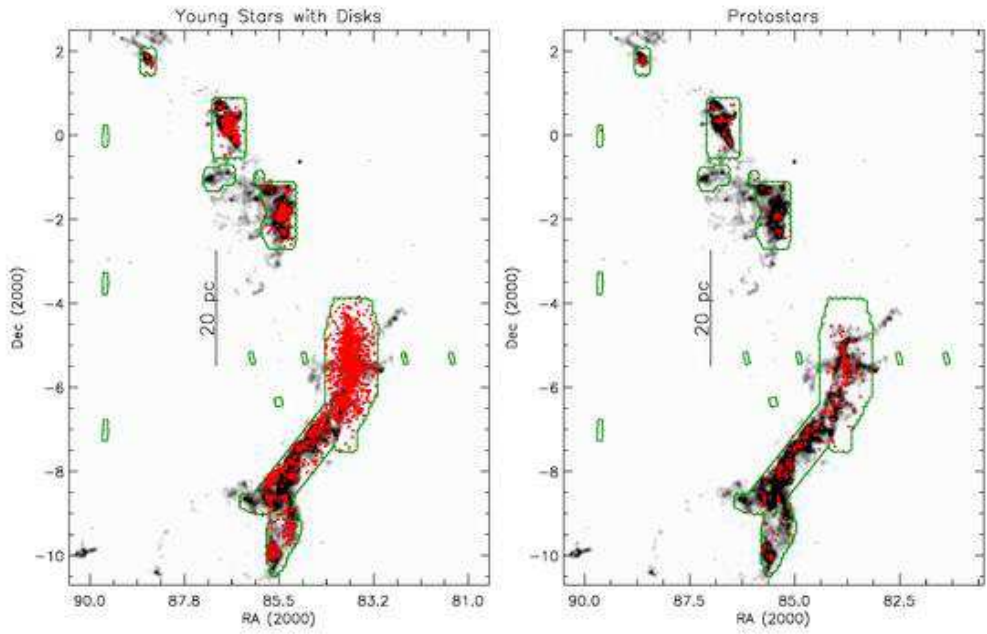}
\epsscale{1.}
\caption{The red dots show the spatial distribution of pre-main sequence stars with disks ({\bf left}) and protostars
({\bf right}) overlaid on the extinction map of the Orion clouds (grayscale).  The green line outlines the surveyed field.} 
\label{fig:yso_dist}
\end{figure}

The spatial distribution of all identified YSOs are displayed in Fig.~\ref{fig:yso_dist}. The
pre-main sequence stars with disks (hereafter: disks) and protostars are displayed separately.
Note that the reference fields, as well the ''Filament'' region of the Orion~B cloud are almost
devoid of YSOs; this demonstrates the reliability of our identification methodology.  This figure 
shows that the YSOs follow the  general structure of the cloud and that there are few parts
of the cloud that are completely devoid of YSOs \citep[also see ][]{2011ApJ...739...84G}).  The disks 
are more spread out  than the protostars, which more closely follow the distribution of the high $A_V$ gas.  

\begin{figure}
\plottwo{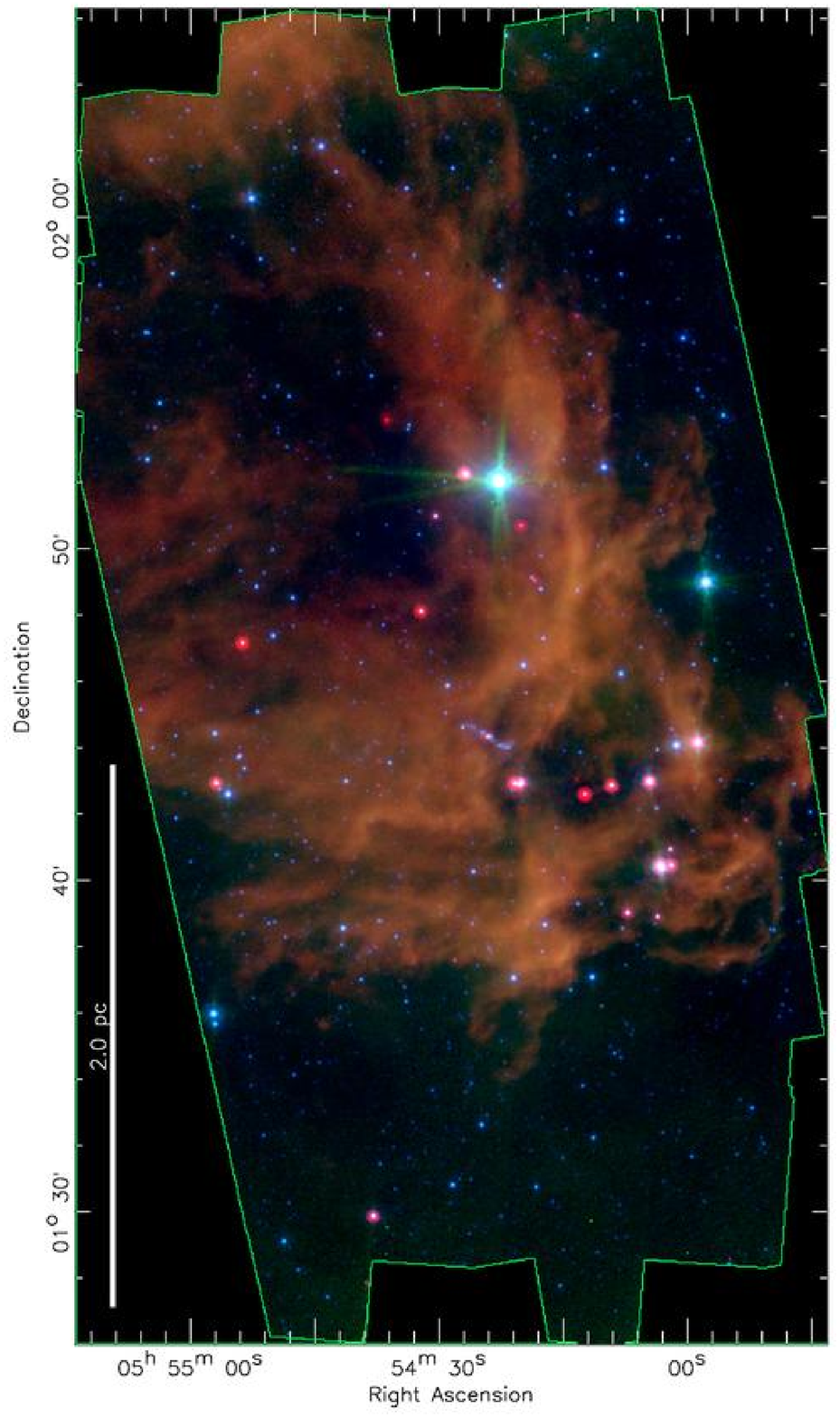}{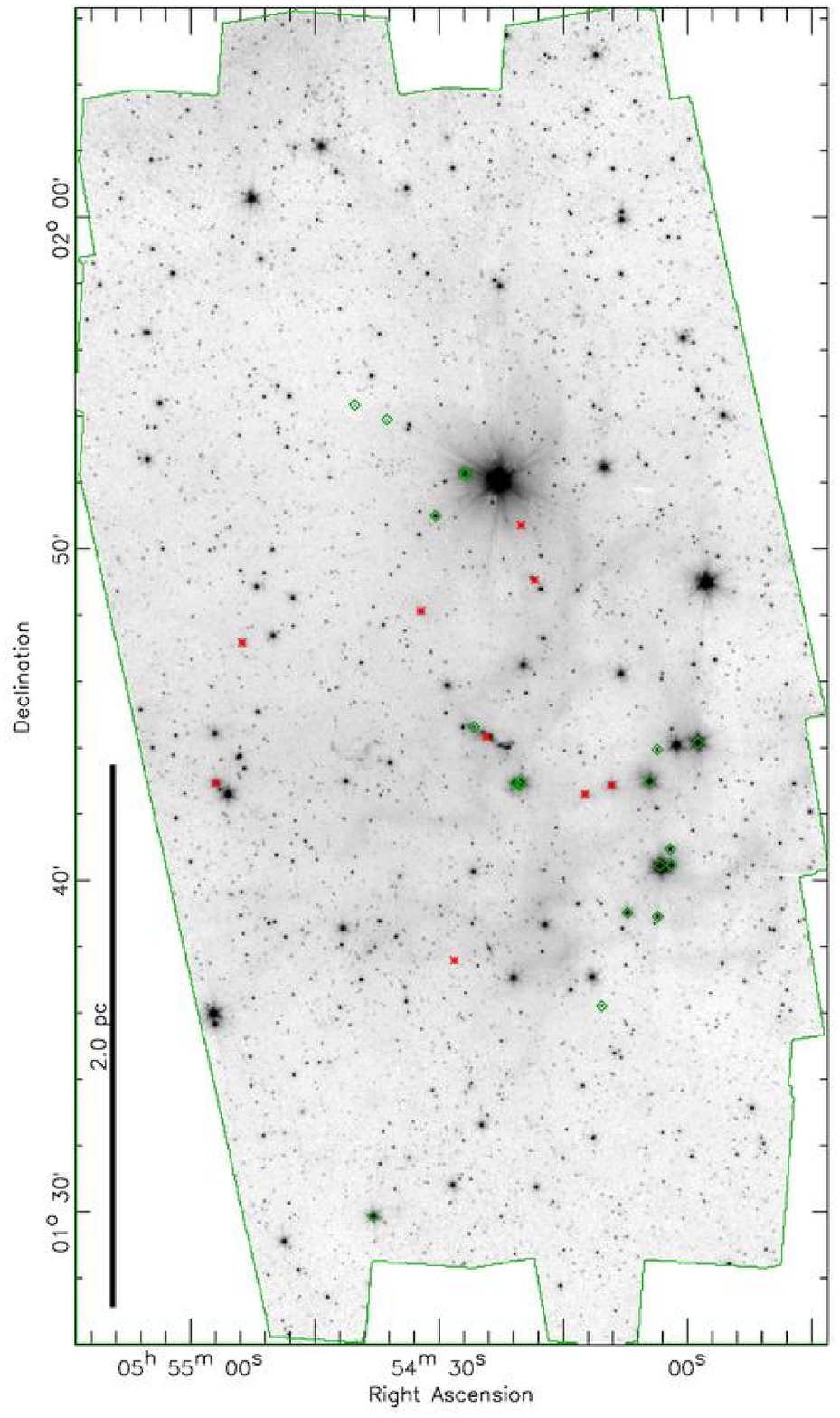}
\epsscale{1.}
\caption{{\bf Left:} Mosaic of the Lynds 1622 field.  Blue is 4.5~$\mu$m, green is 5.8~$\mu$m and red is 24~$\mu$m. {\bf Right:} 4.5~$\mu$m image with the  positions of dusty YSOs superimposed.  Green diamonds are young stars
with disks, red asterisks are protostars (including the faint candidate protostars and the 10 red candidate  protostars detected at 24~$\mu$m but not at 4.5, 5.8 and 8~$\mu$m). In both panels, the green line outlines the surveyed field.}
\label{fig:1622}
\end{figure}

\begin{figure}
\plottwo{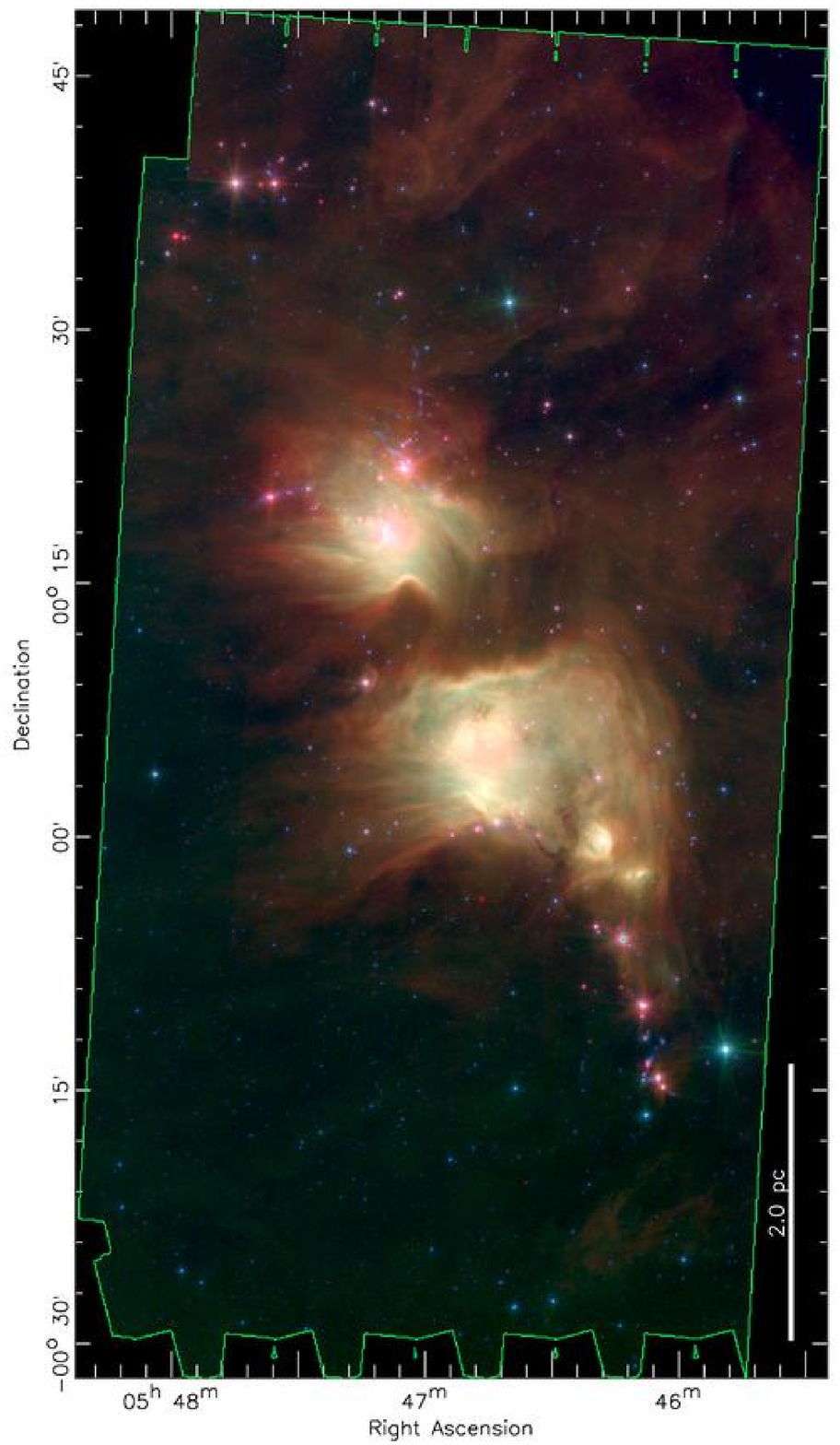}{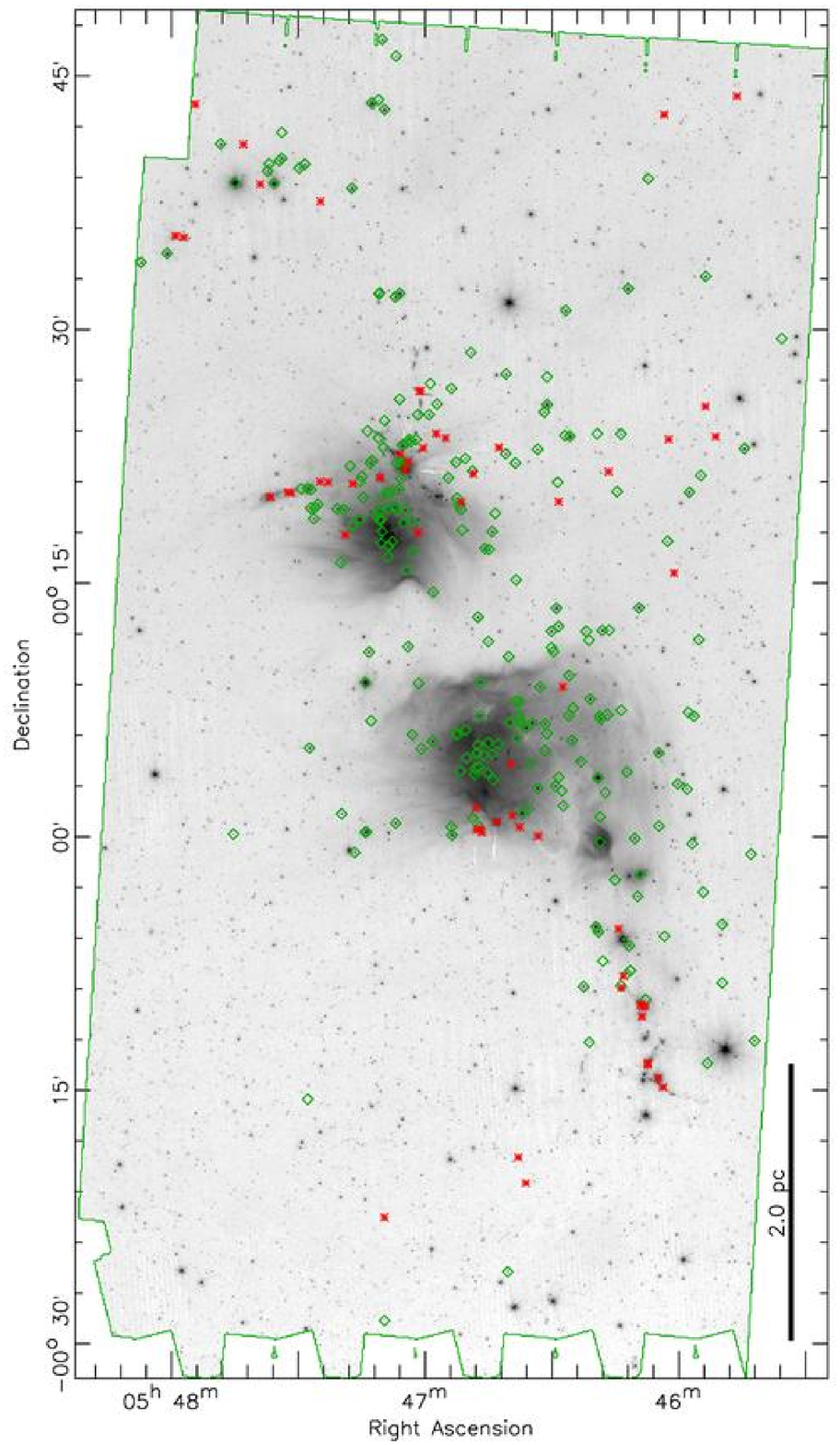}
\epsscale{1.}
\caption{{\bf Left:} Mosaic of the  NGC~2068/NGC~2071 field.  Blue is 4.5~$\mu$m, green is 5.8~$\mu$m and red is 24~$\mu$m. {\bf Right:} 4.5~$\mu$m image with the positions of dusty YSOs superimposed.  Green diamonds are young stars
with disks, red asterisks are protostars (including the  faint candidate protostars and the 10 red candidate protostars detected at 24~$\mu$m but not at 4.5, 5.8 and 8~$\mu$m). In both panels, the green line outlines the surveyed field. The upper reflection nebula is NGC~2071 while the lower nebula is NGC~2068. }
\label{fig:n2068}
\end{figure}

\begin{figure}
\plottwo{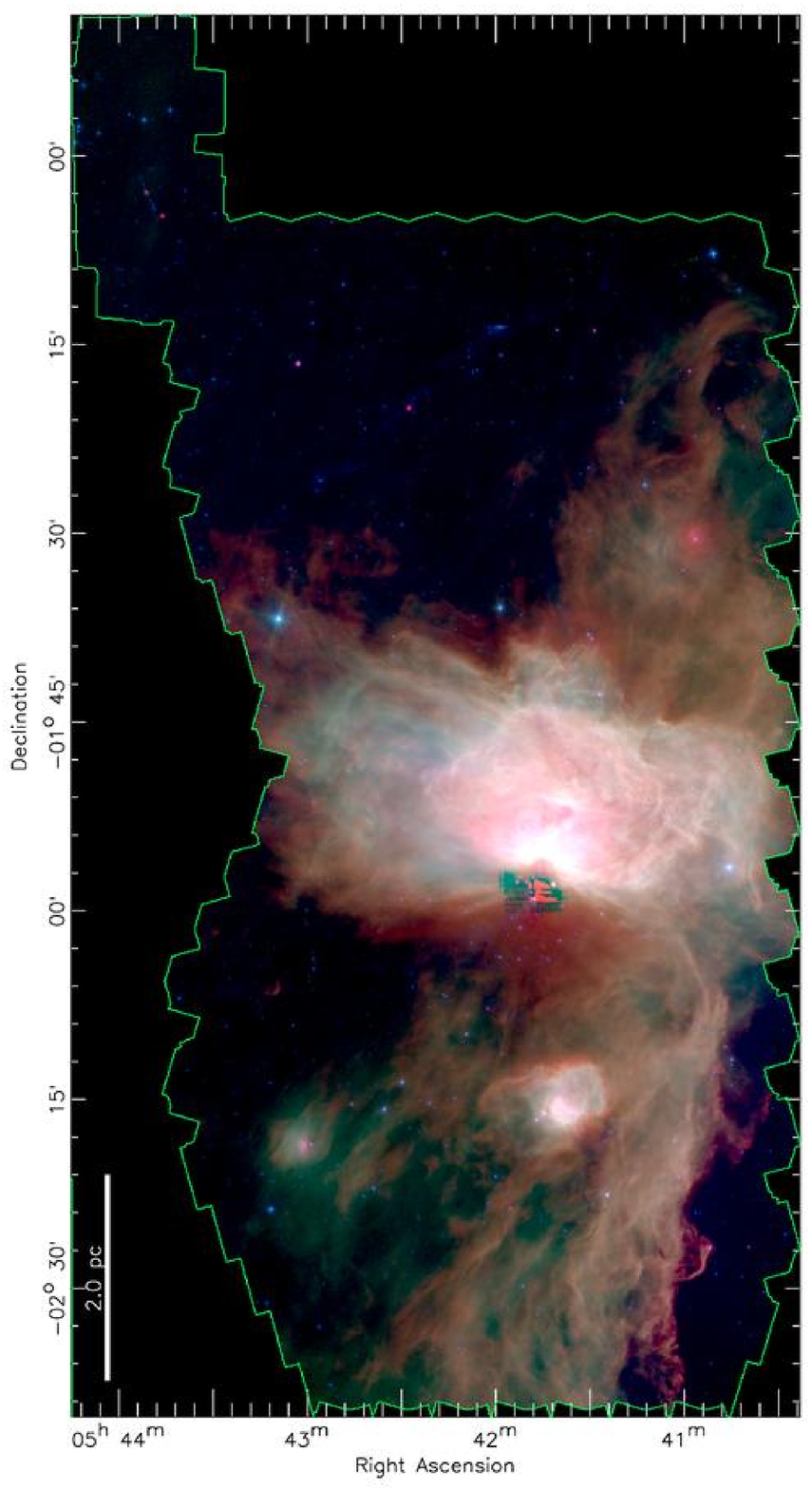}{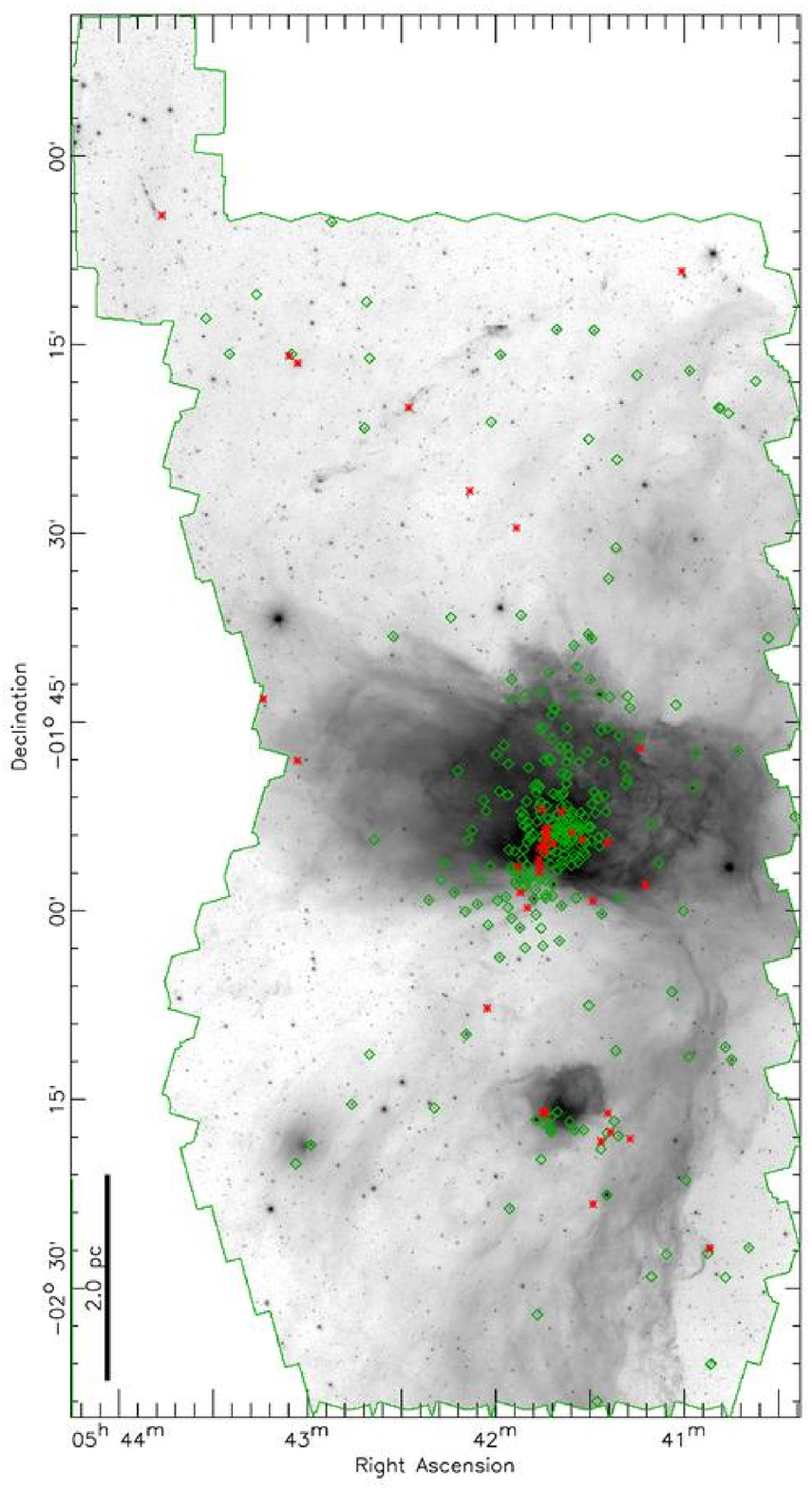}
\epsscale{1.}
\caption{{\bf Left:} Mosaic of the NGC~2024/NGC~2023 field.  Blue is 4.5~$\mu$m, green is 5.8~$\mu$m and red is 24~$\mu$m. \  {\bf Right:} 4.5~$\mu$m image with the positions of dusty YSOs superimposed. Green diamonds are young stars with disks, red asterisks are protostars (including the faint candidate protostars and the 10 red candidate protostars detected at 24~$\mu$m but not at 4.5, 5.8 and 8~$\mu$m). In both panels, the green line outlines the surveyed field.  NGC~2024 is the bright HII region in the center of the map while NGC~2023 is the smaller reflection nebula $20'$ to the south of NGC~2024.  The central region of NGC~2024 is saturated at 24~$\mu$m, and the region just south of the NGC~2024 nebula shows a pronounced artifact due to the steep gradient in intensity. The bright rimmed cloud which forms the Horsehead nebula is apparent in emission on the southwestern edge of the cloud; a protostar is detected at the apex of this cloud \citep{2009AJ....137.3685B}.}
\label{fig:n2024}
\end{figure}

\begin{figure}
\plottwo{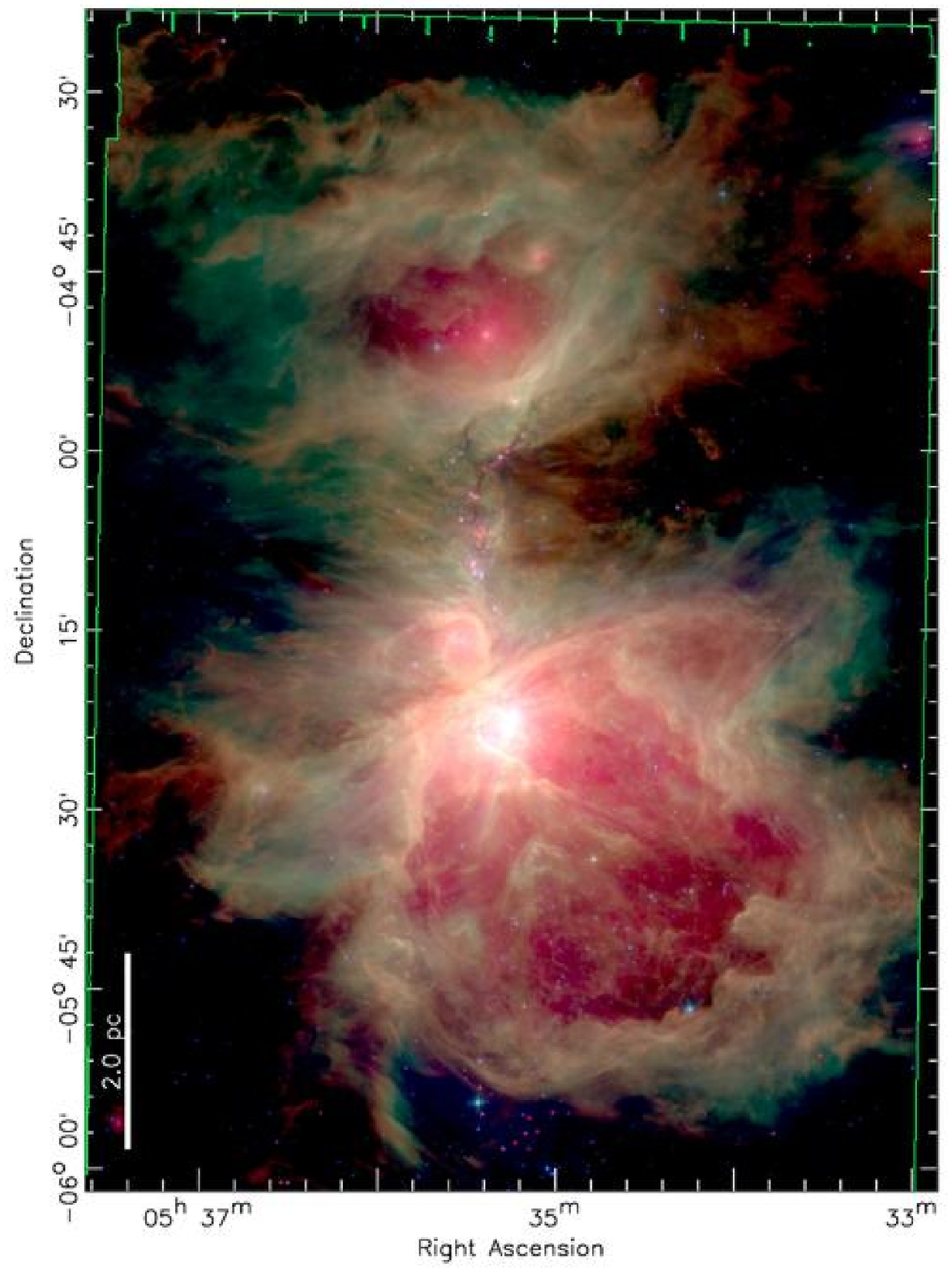}{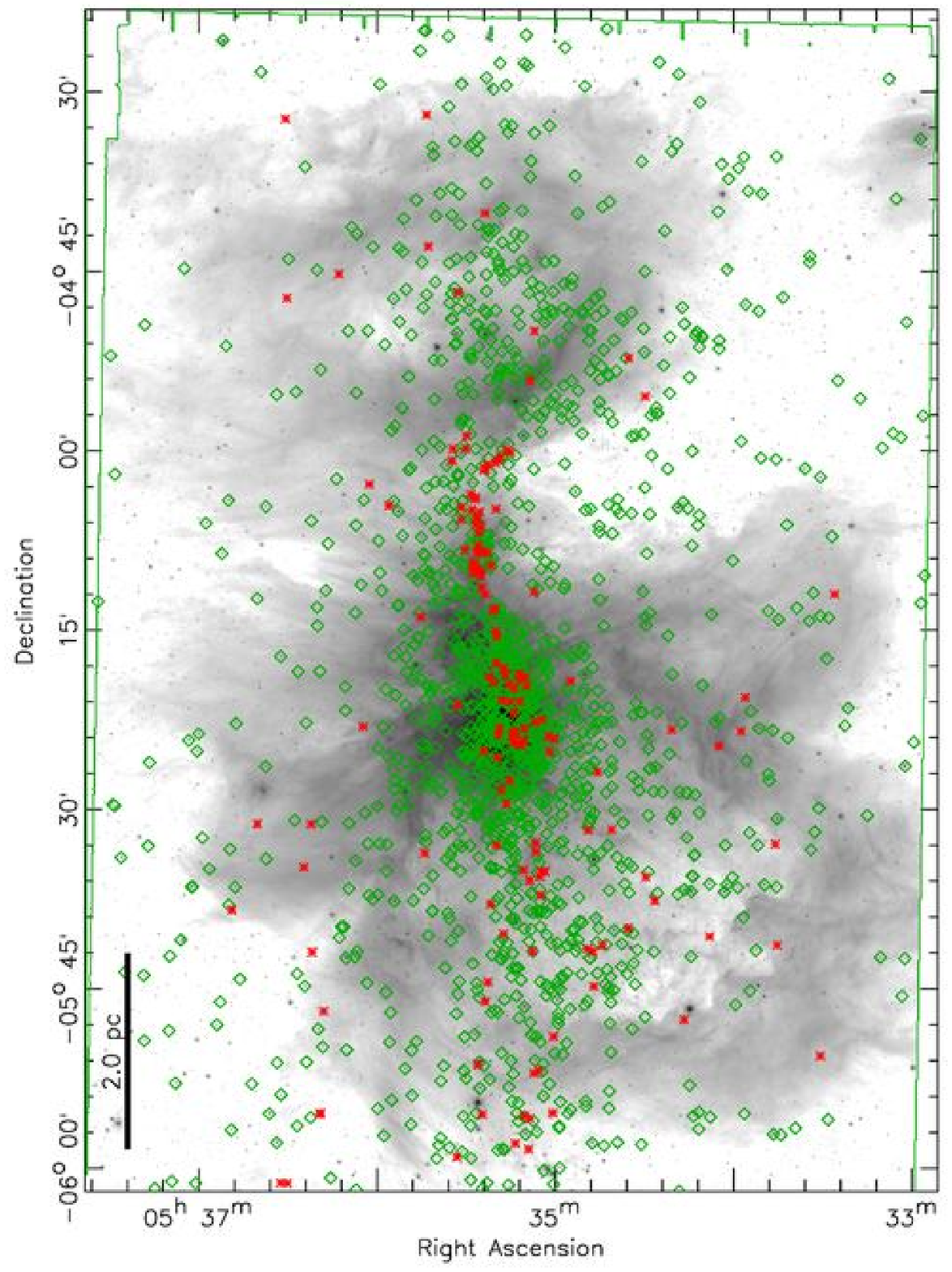}
\epsscale{1.}
\caption{{\bf Left:} Mosaic of the ONC field.  Blue is 4.5~$\mu$m, green is 5.8~$\mu$m and red is 24~$\mu$m.  {\bf Right:} 4.5~$\mu$m image with the positions of dusty YSOs superimposed.   Green diamonds are young stars with disks, red asterisks are protostars (including the faint candidate protostars and the 10 red candidate protostars detected at 24~$\mu$m but not at 4.5, 5.8 and 8~$\mu$m).  In both panels, the green line outlines the surveyed field.  The Orion Nebula is the extremely bright region just south of the center of the mosaic. The central region of this nebula is saturated in the 24~$\mu$m band.  The extended reflection nebula to the north of the Orion Nebula is NGC~1977.  Between the Orion Nebula and NGC~1977 is a filament rich in protostars known as the OMC 2/3 region. The large bubble to the southwest of the Orion Nebula is the extended Orion Nebula \citep{2008Sci...319..309G}.}
\label{fig:onc}
\end{figure}

\begin{figure}
\plottwo{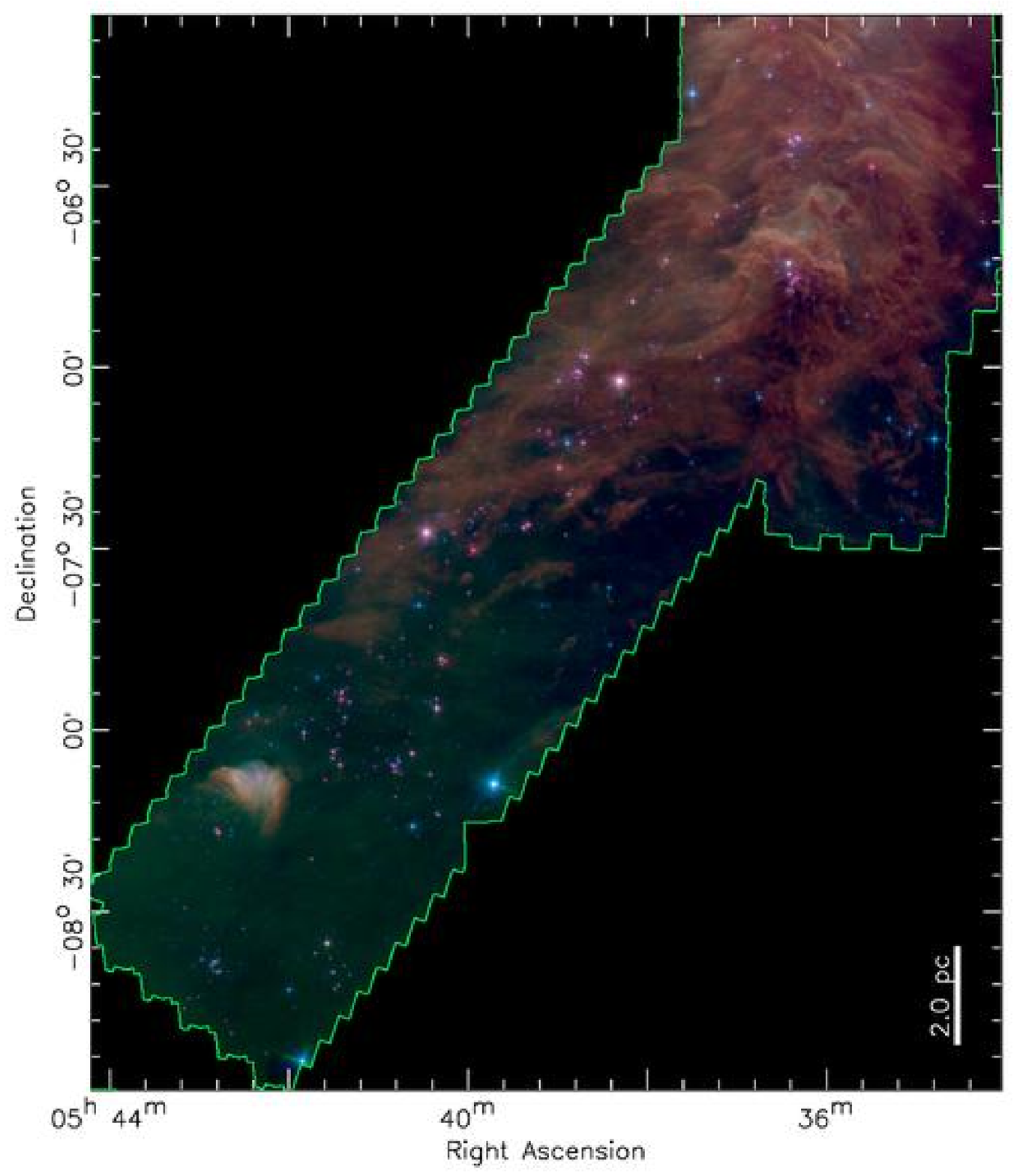}{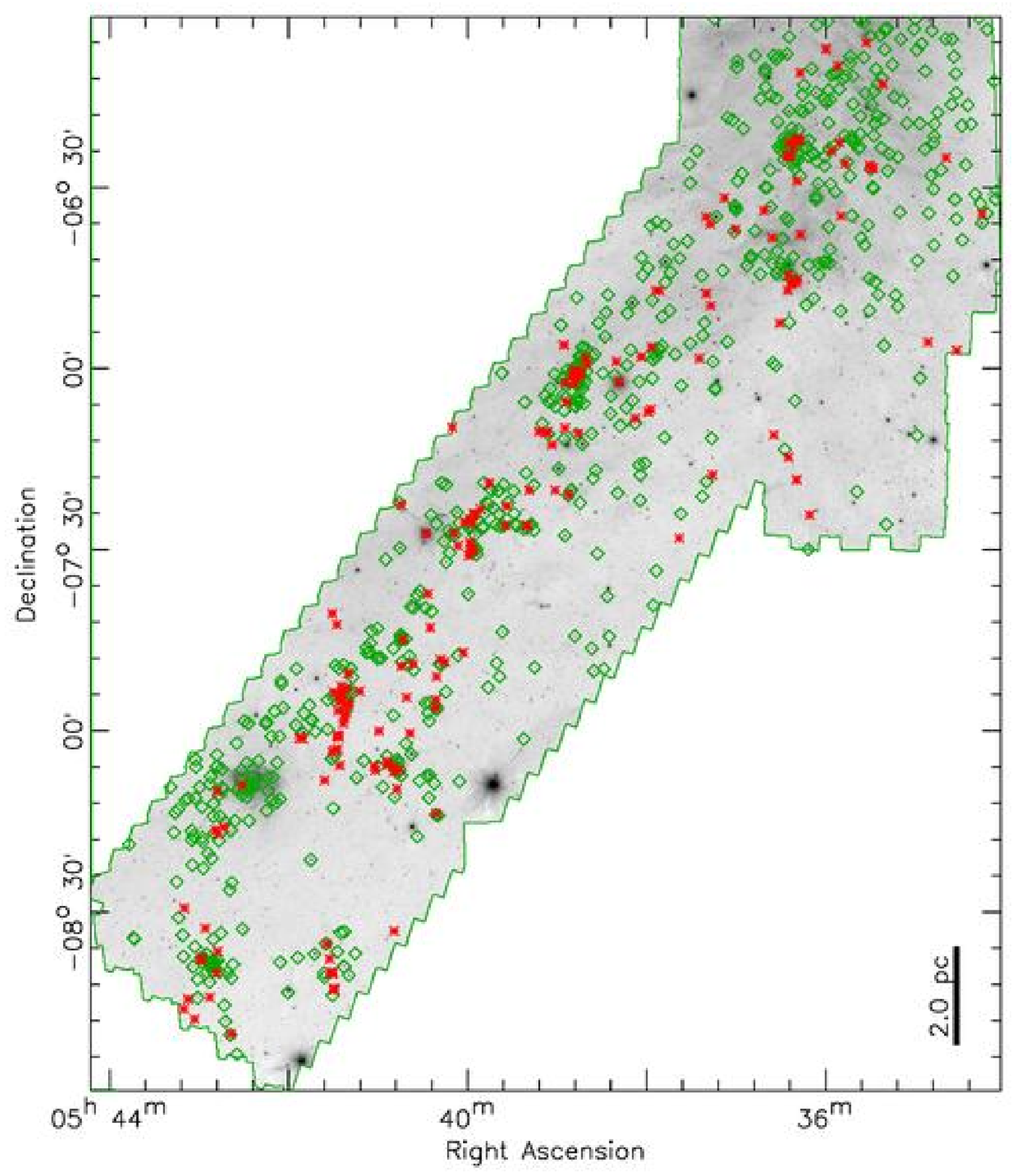}
\epsscale{1.}
\caption{{\bf Left:} Mosaic of the Lynds 1641 field.  Blue is 4.5~$\mu$m, green is 5.8~$\mu$m and red is 24~$\mu$m. {\bf Right:} 4.5~$\mu$m image with the positions of dusty YSOs superimposed.  Green diamonds are young stars
with disks, red asterisks are protostars (including the faint candidate protostars and the 10 red candidate protostars detected at 24~$\mu$m but not at 4.5, 5.8 and 8~$\mu$m). In both panels, the green line outlines the surveyed field.}
\label{fig:l1641}
\end{figure}

\begin{figure}
\plottwo{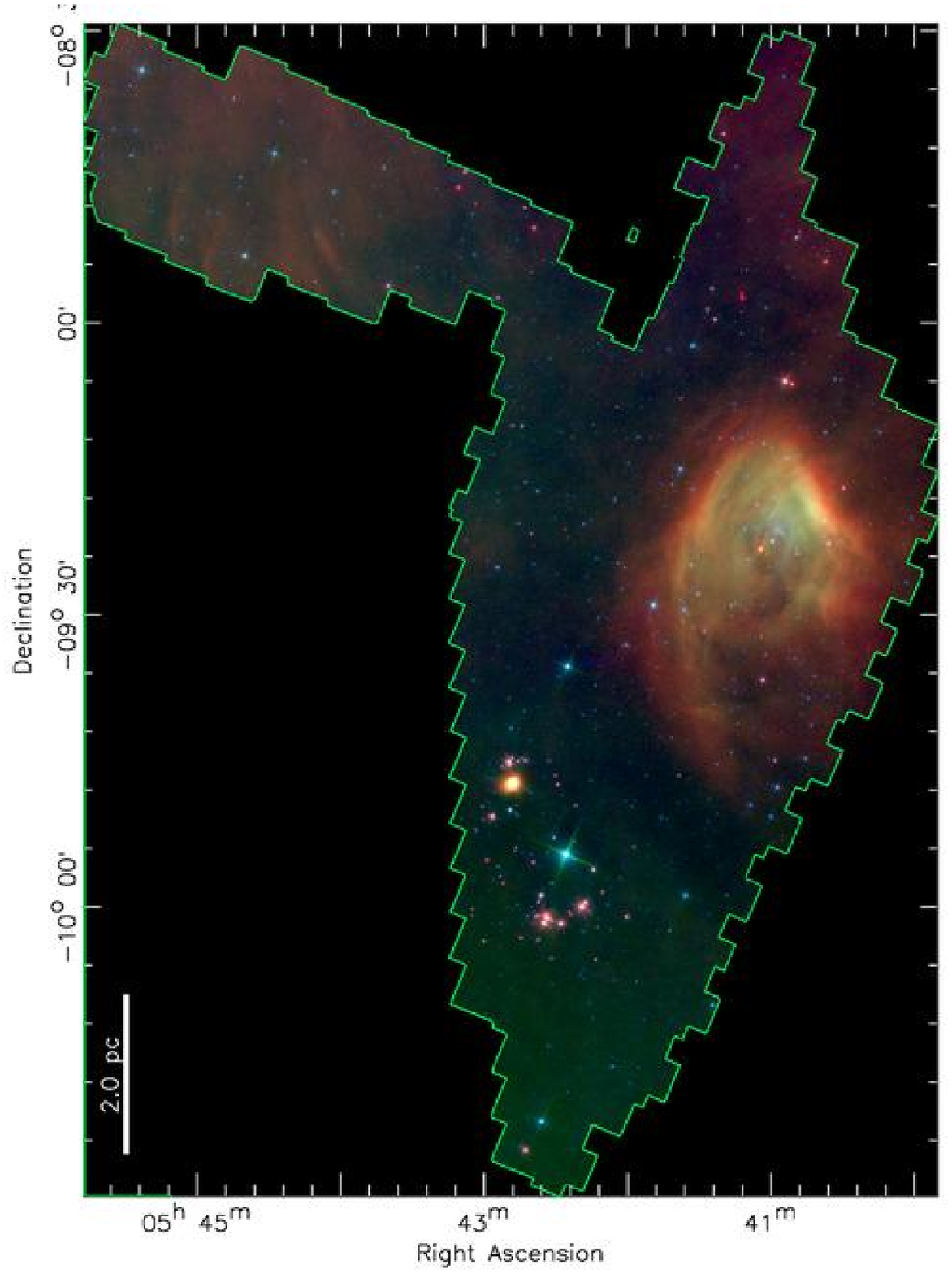}{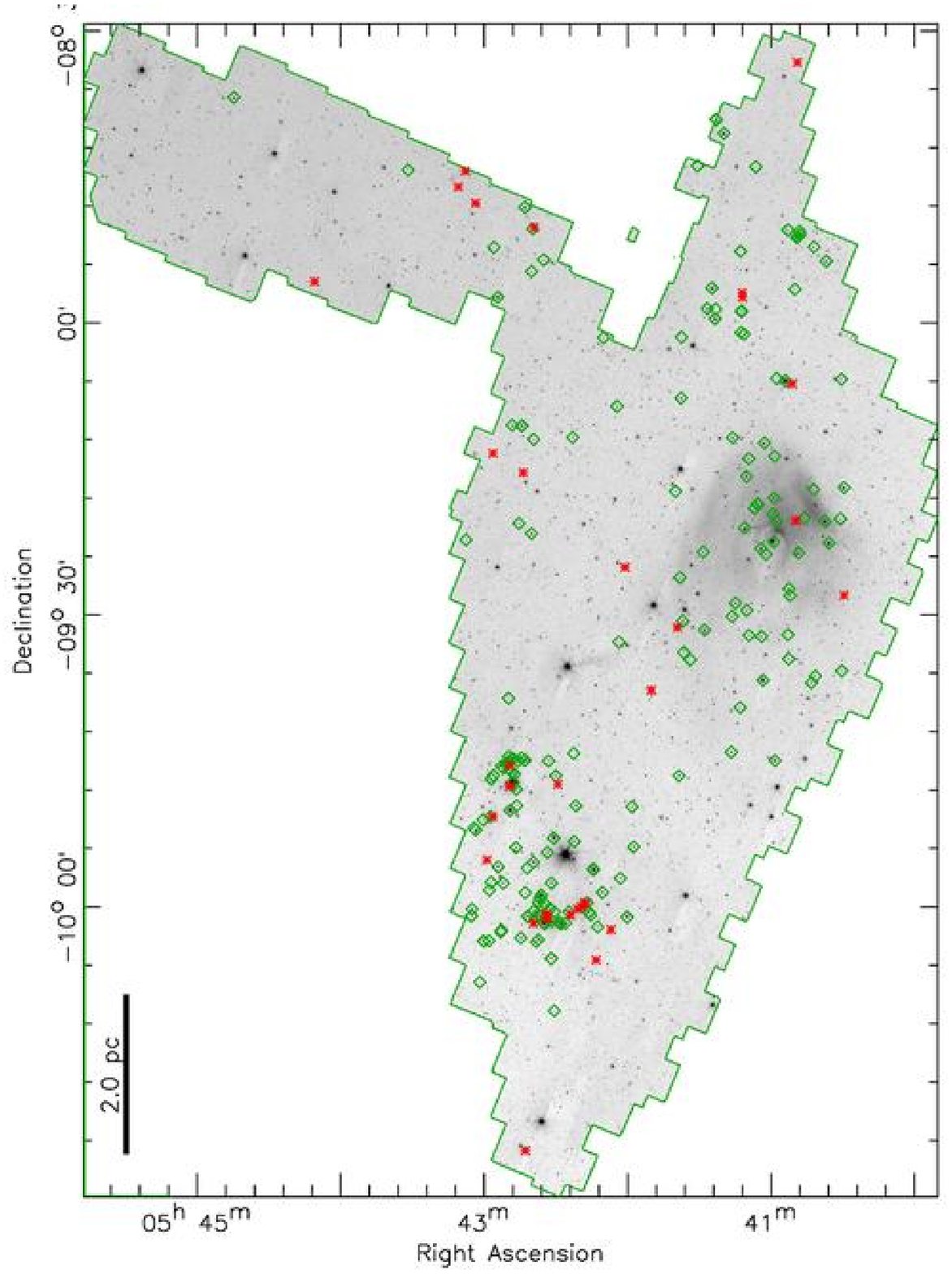}
\epsscale{1.}
\caption{{\bf Left:} Mosaic of the $\kappa$~Ori field.  Blue is 4.5~$\mu$m, green is 5.8~$\mu$m and red is 24~$\mu$m. {\bf Right:} 4.5~$\mu$m image with the positions of dusty YSOs superimposed.  Green diamonds are young stars
with disks, red asterisks are protostars (including the faint candidate protostars and the 10 red  candidate protostars detected at 24~$\mu$m but not at 4.5, 5.8 and 8~$\mu$m). In both panels, the green line outlines the surveyed field.}
\label{fig:kappa}
\end{figure}

 In Figs.~\ref{fig:1622} - \ref{fig:kappa}, we show color images constructed from the combined 4.5~$\mu$m, 5.8~$\mu$m and 24~$\mu$m mosaics, as well as the positions of the protostars and disks overlaid on the 4.5~$\mu$m mosaics.  These include the Lynds 1622  dark cloud \citep[Fig.~\ref{fig:1622}, ][]{2008hsf1.book..782R}, the field containing the reflection nebulae  NGC 2068 and 2071 region \citep[Fig.~\ref{fig:n2068}, ][]{2008hsf1.book..693G}, the field containing the HII region NGC~2024, the reflection nebula NGC~2023 and the Horsehead nebula \citep[Fig.~\ref{fig:n2024}, ][]{2008hsf1.book..662M}, the field containing the Orion Nebula, NGC~1977, and  OMC-2/3   \citep[Fig.~\ref{fig:onc}, ][]{2008hsf1.book..590P,2008hsf1.book..544O,2008hsf1.book..483M},  the Lynds~1641 region \citep[Fig.~\ref{fig:l1641}, ][]{2008hsf1.book..621A} and the part of the cloud closest to the star $\kappa$ Ori (Fig.~\ref{fig:kappa}, it is not clear if $\kappa$~Ori is in physical proximity to the cloud or just seen near the cloud in projection). Detailed descriptions of these region are given in the above references.  Rich clusters, small groups and relatively  isolated YSOs are clearly evident in these images. Multi-parsec long chains of protostars are also discernible in several of the mosaics.  In total, these figures illustrate the complex distribution of dusty YSOs within the Orion molecular clouds. 

The advantage of using {\it Spitzer} to trace recent and ongoing star formation is that we can reliably
identify young stars in regions with a low density of dusty YSOs.  This ensures that we get a census
of both clustered and (relatively) isolated star formation in Orion.  In contrast, previous studies
have relied on star counts that  are primarily sensitive to star formation in clustered regions \citep{1991ApJ...371..171L,1992ApJ...393L..25L,2000AJ....120.3139C}.  However, young clusters contain bright 
mid-IR nebulosity and crowded star fields which can lower the sensitivity of {\it Spitzer} to YSOs
(Figs.~\ref{fig:n2068}, \ref{fig:n2024}, and \ref{fig:onc}). Consequently, an unbiased census of dusty YSOs toward the Orion clouds requires an estimate of the 
completeness, particularly in regions with dense clusters and bright nebulosity, and a correction
for spatially varying incompleteness over the surveyed regions.  In a 2nd paper, we will
develop and apply approaches to correct for the spatially varying incompleteness and perform
a detailed analysis of the spatial distribution of YSOs in the Orion clouds.

\section{Mid-IR Variability in the Orion Sample}
\label{sec:var}

The variability of YSOs in the IRAC bands has now been established in
multi-epoch {\it Spitzer} observations of several star forming regions
\citep{2009AJ....138.1116S,2009ApJ...704L..15M,2011ApJ...733...50M}.
These results motivate an examination of the variability in the Orion
data set.  The dithered 10.4 second observations at most positions in the
Orion field were obtained in 2-3 distinct epochs.  By separating the
data into these epochs, we can use these data to characterize
variability between the observed epochs for a large sample of young stellar
objects. 

\subsection{Stetson Analysis of the Orion YSOs}

Since the Orion survey observations were not designed to measure
variability, a search for variable sources requires the extraction of
photometry from one or two individual data frames with no redundancy
to eliminate cosmic rays.  To reliably identify variable sources, we
use the Stetson index which measures correlated variability in the
3.6 and 4.5~$\mu$m bands of IRAC \citep{1996PASP..108..851S}.  These
bands were chosen as they are the least affected by  nebulosity and
typically show the highest signal to noise.  The requirement of
correlated variability minimizes the chance of mistaking a cosmic ray
strike during one of the epochs as variability.

We have adapted the Stetson analysis outlined in the 2MASS variability
study of \citet{2001AJ....121.3160C} to the IRAC 3.6 and 4.5~$\mu$m bands.
The Stetson index $S$ was calculated using the equation:

\begin{equation}
S = \frac{1}{N_{epoch}}\sum^{N_{epoch}}_{k=1} sgn(P_k)\sqrt{|P_k|}
\end{equation}

\noindent
where $N_{epoch}$ is the number of independent epochs and 

\begin{equation}
P_k = \delta_k(3.6~\mu m)\delta_k(4.5~\mu m)
\end{equation}

\noindent
The variable $\delta_K$ is calculated for each epoch from for the 3.6 and 4.5~$\mu$m magnitudes  using the equation:

\begin{equation}
\delta_k(\lambda) = \sqrt{\frac{n_{epoch}}{n_{epoch}-1}}\left( \frac{m_{\lambda}-\bar{m}_{\lambda}}{\sigma_\lambda} \right)
\end{equation}

\noindent
where $\bar{m}_{\lambda}$ is the average of $m_{\lambda}$ over the
epochs.  In the determination of $S$, we only used magnitudes with
uncertainties less than 0.1~mag and we required that this uncertainty
limit was met in at least 2 epochs. To increase the signal to noise
and reliability of the photomtery, we average observations which were
taken within 2 days of each other. This
will smooth out any short period variability over the 2 day interval. The number of overlapping
observations can range from 1 frame (in the case of L1641 where there
are separate 2nd and 3rd epochs), 2 frames (in the case of most of the
other fields), to as high as 6 frames (in the interstices of our
mapping grid where frames from 3 separate positions overlap). 
In cases where the peak signal of a source in the combined mosaics exceeded the
saturation limit of the 10.4 second frame or the magnitude of the short
0.4~second frame was brighter than the 10.4 second frame saturation
magnitude, we used the photometry for the 0.4~second frame.  A noise floor
was set by adopting a minimum value of 0.05~mag for $\sigma_\lambda$.

\begin{figure}
\epsscale{1.}
\plotone{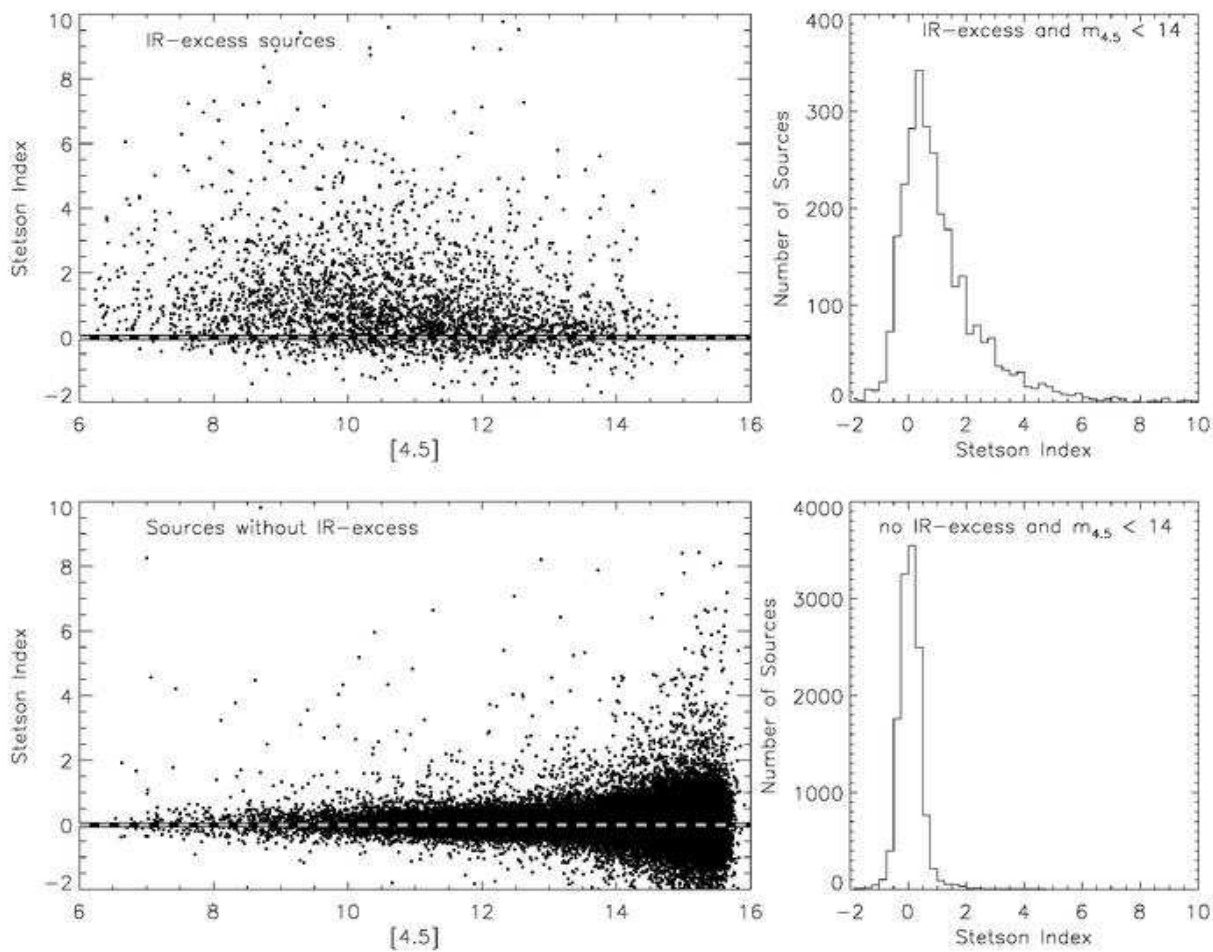}
\caption{{\bf Upper left:} Stetson index vs 4.5~$\mu$m magnitude for
  the sample of all YSOs with IR-excesses. {\bf Lower left:} Stetson
  index for the sample of all stars without IR-excesses, i.e. the pure
  photosphere sources.  This sample defines the range of Stetson
  indices for a population of objects which are not suspected to
  exhibit significant variablity.  {\bf Upper right:} distribution of
  Stetson indices for YSOs with IR-excesses.  {\bf Lower right:}
  distribution of the Stetson indices for pure photospheres. In both
  histograms, we require that the all the sources satisfy the criterion
  $[4.5] \le 14$.}
\label{fig:figure_stetson}
\end{figure}

The distribution of Stetson indices for the entire cloud sample are
displayed in Fig.~\ref{fig:figure_stetson}.  We compare the
distribution of young stellar objects with IR-excesses with that for
stars that show no evidence of IR-excesses (hereafter: pure
photospheres). The pure photospheres are a mixture of field stars in
the line of sight as well as diskless pre-main sequence stars.  The
difference between these two distributions is clearly apparent.  While
the IR-excess sources show a strong bias toward positive values of
$S$, the pure photospheres are mostly confined to small values of $S$
centered on $0$.  We show the histograms for sources with $m_{4.5} \le
14$; this magnitude threshold eliminates the sharp increase of $S$
with fainter magnitudes apparent for the pure photospheres.  We assume
the distribution of $S$ for the photospheres is dominated by noise and
systematic uncertainties in the photometry and not by intrinsic
variability in the sources. Given the that the distribution is
symmetric with $S=0$, this appears to be a reasonable assumption.  We
can then compare this distribution to that for the YSOs with
IR-excesses to assess the frequency of variability in both pre-main
sequence stars with disks and protostars.  This comparison shows that
the YSOs with IR-excesses are clearly biased to positive values of $S$,
indicating the presence of significant variability.

\begin{figure}
\epsscale{1.}
\plotone{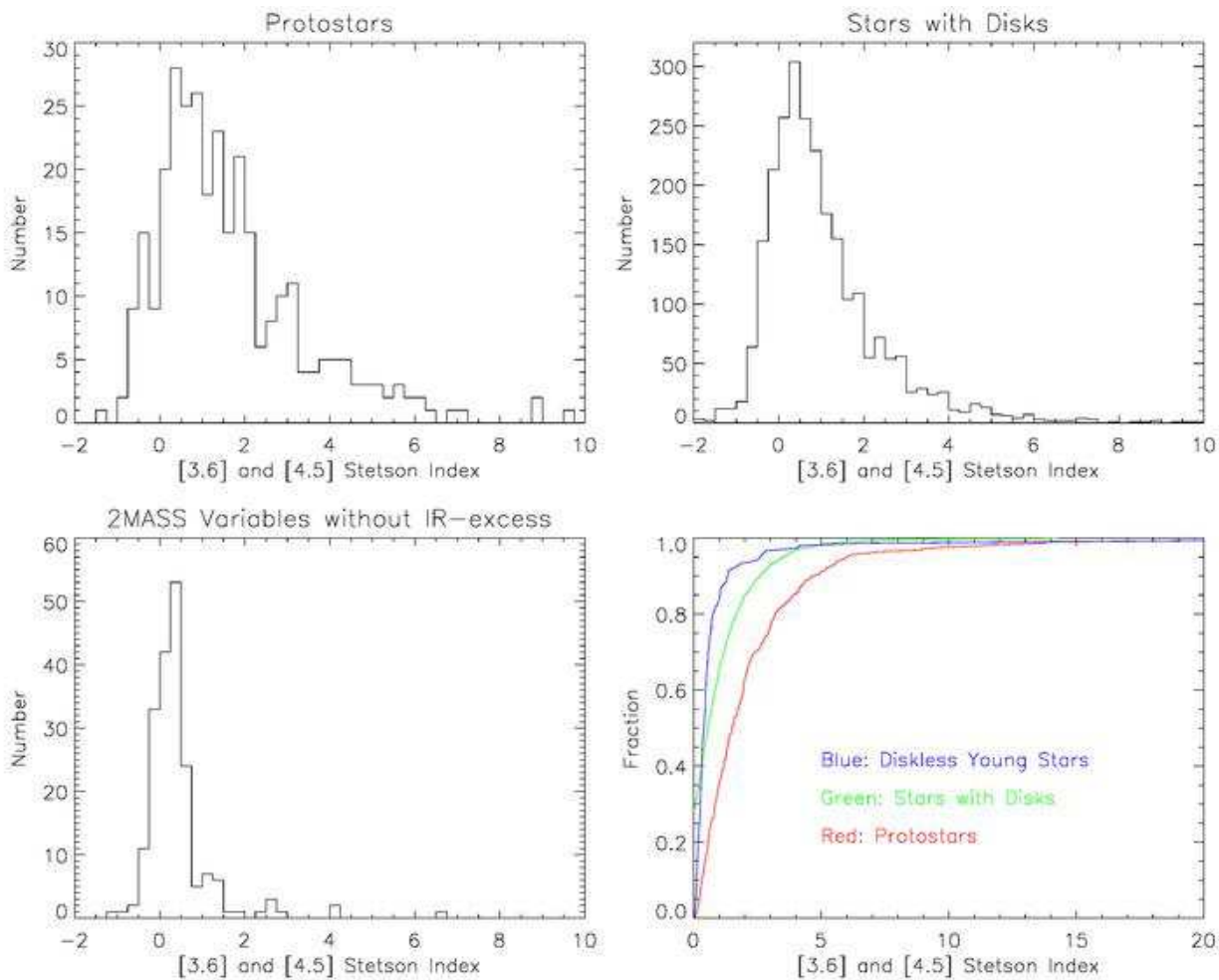}
\caption{A comparison of variability as a function of evolutionary
  class for the sample of YSOs in the Orion survey. We show the
  distributions for the protostars ({\bf upper left}), young stars
  with disks ({\bf upper right}) and 2MASS variables without
  IR-excesses ({\bf lower left}).  The 2MASS variables were selected
  as likely diskless pre-main sequence stars. While the protostars and
  stars with disks were taken from the entire Orion survey, the 2MASS
  variables were taken from the \citet{2001AJ....121.3160C}
  variability survey of the ONC and its immediate surroundings. The {\bf bottom left} panel
  shows the cumulative distributions for all three evolutionary classes. These
  show a growing incidence of variability (and larger Stetson
  indices) with progessively earlier evolutionary stages.  In all the
  distributions, we restricted the sample to sources with $[4.5] \le
  14$.}
\label{fig:figure_class_stetson}
\end{figure}

In Fig.~\ref{fig:figure_class_stetson}, we compare the variability as
a function of evolution.  (note: protostars do not include faint candidate protostars
in this analysis). To identify a sample of diskless pre-main 
sequence stars, we use the 2MASS variability survey to find sources
that show variability in the near-IR bands but show no evidence
for a disk in the {\it Spitzer} wavelength bands.  These objects are 
though to be pre-main sequence stars with large star spots that 
rotate in and out of view \citep{2001AJ....121.3160C}.  We
use all sources which are not found to show an IR-excess in
the {\it Spitzer} bands (despite having photometry in a sufficient 
number of wavelength bands to detect an excess) and which
have near-IR Stetson indices $\ge 0.55$ in the observations
of \citet{2001AJ....121.3160C}.   

The histograms show a clear 
progression in the distribution of $S$ with evolution, with the protostars 
showing the highest values of $S$, the young stars with disks exhibiting intermediate
values, and the diskless young stars showing typically very small
values.  We note that the IR-excess sources exhibit much higher values
of $S$ even though the diskless pre-main sequence stars were 
selected by their inherent variability.  Nevertheless, the diskless
pre-main sequence stars show a higher incidence of variability than the total
sample of pure photospheric sources that include both field stars
and diskless pre-main sequence objects.

To better quantify the fraction of YSOs that exhibit variability, we
tabulate  the fraction of sources with $S > n \sigma_{S(photo)}$, where
$\sigma_{S(photo)}$ is the standard devation in the $S$ for the
photospheres (Table~\ref{table:stetson}). We show that the
IR-excesss sources have systematically larger values than the
photospheres.  If we use the photospheres as a reference distribution
for non-variable sources (i.e. the spread in the values of $S$ result
from random and systematic errors in the uncertainties), we estimate
that $\sim 50\%$ of the young stellar objects with IR-excesses exhibit
variability.

\begin{table}[h]
%\begin{center}
\caption{Fraction of Young Stellar Objects Exhibiting Variability}
\vskip 0.1 in
\begin{tabular}{lccccc}
\tableline\tableline
    &  \multicolumn{5}{c}{Fraction of souces with $s > n \sigma_{photo}$}  \\
n  & Photospheres & YSOs with & Diskless & YSOs with & Protostars \\
  &             & IR-excesses & YSOs     & Disks &  \\
\tableline
 1 & 0.09 & 0.60 & 0.28 & 0.59 & 0.73 \\
 2 & 0.02 & 0.41 & 0.12 & 0.40 & 0.57 \\
 3 & 0.01 & 0.28 & 0.06 & 0.26 & 0.45 \\
\tableline
\end{tabular}
%\end{center}
\label{table:stetson}
\end{table}

\subsection{Characterizing the Changes in Magnitude and Color}

\begin{figure}
\plotone{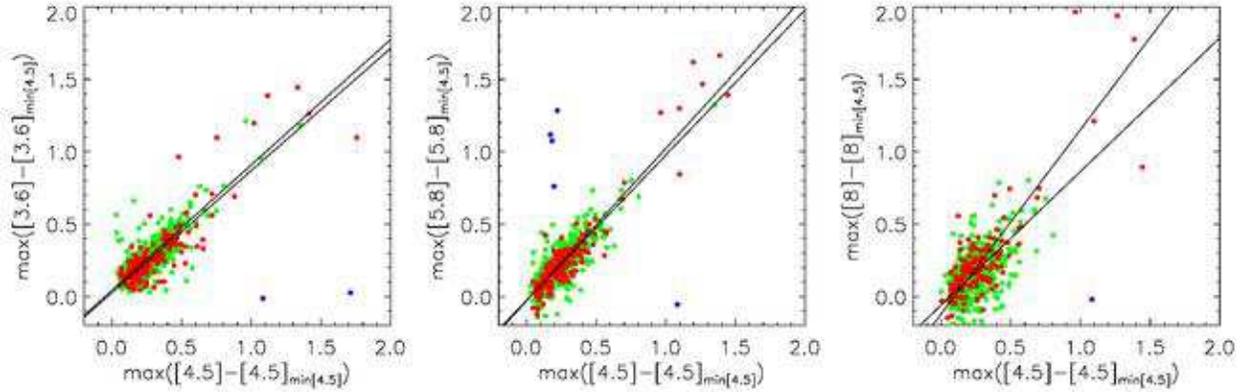}
\epsscale{1.}
\caption{The observed variation in magnitude for the variable
  sources. In all three panels, we chose sources where $S > 2\sigma_{S(photo)}$.  
  In each panel, we identify the epochs that show the minimum and 
  maximum 4.5~$\mu$m magnitude. We then plot
  the magnitude variation between those two epochs for the 3.6, 5.8
  and 8~$\mu$m bands.  The green points are stars with disks,
  the red points are protostars. The blue points show strong
  variability in only one band, and may result from cosmic ray
  strikes; they have been eliminated from further analysis.  The lines
  show the fits to the points: the lower lines show the fits to all the green and red
  points while the upper lines show the fits excluding points with deviations less than
  0.25~mag.}
\label{fig:figure_deltamag_variable}
\end{figure}

We now characterize the changes in magnitude during the variations.
For the IR-excess sources where $S > 2 \sigma_{S(photo)}$,
Fig.~\ref{fig:figure_deltamag_variable} displays the change in the
magnitude in each of the IRAC bands.  In each case, we identified the
epoch that showed the minimum and maximum 4.5~$\mu$m magnitudes.  
We then plotted the difference in magnitude for each IRAC band between
those two epochs.  In each case, we plot the difference in magnitude
relative to that observed in the 4.5~$\mu$m band.  We find that the
distribution of maximum deviations are strongly correlated and
approximately follow a diagonal in each plot.  This indicates that the
change in magnitude is approximately the same in all four wavebands.  Note
that there are several points where variability is seen in one band
but not the other.  In these cases, the photometry may have been
affected by cosmic ray strikes and are removed from the analysis.
Nevertheless, there is a strong correlation between the
independent bands: the Pearson's r correlation coefficients are given
in Table~\ref{table:deltamag_slope}.  These show strong correlations in
all three cases which can be further strengthened by limiting the analysis
to sources showing changes $\ge 0.25$~mag.  

We fit a line to each correlation using the functional form
$[\lambda]-[\lambda]_{4.5_{min}} = A + B \times
([4.5]-[4.5]_{4.5_{min}})$ after eliminating all the points that are
likely cosmic ray strikes from the fit
(Fig.~\ref{fig:figure_deltamag_variable}). We try fits that include all the datapoints (after
rejecting the possible cosmic ray strikes) and fits limited to those
datapoints showing changes in both bands greater than 0.25~mag. By ignoring sources
that show small variations, we reduce the sensitivity of the fits to  spatially varying nebulosity.
The results of the fits are given in Table~\ref{table:deltamag_slope}.
In three out of  four of the observed correlations, the slopes are within 2$\sigma$ of unity, 
indicating that the variability does not change the color of the sources.  
The only exception is the slope for the $\Delta$3.6~$\mu$m vs
$\Delta$4.5~$\mu$m correlation, which shows a somewhat lower slope. 

% If, on the other hand,we only consider the sources where the change is less than 1~mag, the slope for 
% $\Delta$5.8~$\mu$m vs $\Delta$4.5~$\mu$m  and  $\Delta$8~$\mu$m vs $\Delta$4.5~$\mu$m 
% decrease significantly.  However, examination of Fig~\ref{fig:figure_deltamag_variable} shows 
% that these are poor fits to the observed data for changes between 0.5 and 1~mag and do not reproduce the observed variability well.  

\begin{table}[h]
%\begin{center}
\caption{Slope of $\Delta m_{\lambda}$ Variations in the IRAC Bands}
\vskip 0.1 in
\begin{tabular}{lccc}
\tableline\tableline
Bands  & r & A  & B \\
\tableline
$\Delta$[3.6] vs $\Delta[4.5]$  & 0.85 & $ 0.025 \pm 0.004$ & $0.844 \pm 0.016$ \\
$\Delta$[3.6] vs $\Delta[4.5]^1$  & 0.87 & $ 0.043 \pm 0.016$ & $0.865 \pm 0.034$ \\
%$\Delta$[3.6] vs $\Delta[4.5]^2$  & 0.79 & $ 0.037 \pm 0.004$ & $0.782 \pm 0.019$ \\
$\Delta$[5.8] vs $\Delta[4.5]$  & 0.88 & $-0.027 \pm 0.005$ & $1.003 \pm 0.019$ \\
$\Delta$[5.8] vs $\Delta[4.5]^1$   & 0.92 & $-0.031 \pm 0.018$ & $1.055 \pm 0.036$ \\
%$\Delta$[5.8] vs $\Delta[4.5]^2$   & 0.80 & $-0.006 \pm 0.006$ & $0.897 \pm 0.023$ \\
$\Delta$[8] vs $\Delta[4.5]$   & 0.69   & $-0.059 \pm 0.010$ & $0.922 \pm 0.037$\\
$\Delta$[8] vs $\Delta[4.5]^1$ & 0.83   & $-0.121 \pm 0.045$ & $1.273 \pm 0.091$ \\
%$\Delta$[8] vs $\Delta[4.5]^2$ & 0.56   & $-0.018 \pm 0.010$ & $0.713 \pm 0.040$ \\

\tableline
\tablenotetext{1} {Only for sources which vary by more than 0.25~mag in both wavebands.}

\end{tabular}
%\end{center}
\label{table:deltamag_slope}
\end{table}

\begin{figure}
\epsscale{1.}
\plotone{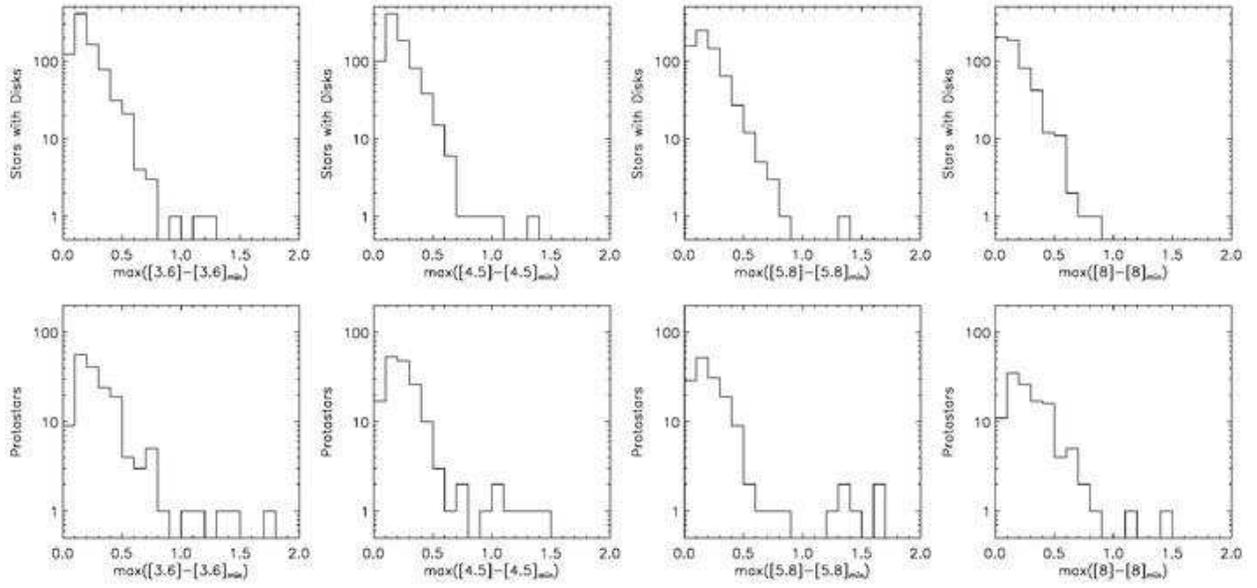}
\caption{The distributions of the maximum change in magnitude for
  variable sources.  In all cases, we chose sources where $S >
  2\sigma_{S(photo)}$. We show the maximum deviation in
magnitude over the 2-3 epochs for all four IRAC bands, dividing the variable
sources into stars with disks ({\bf upper row}) and protostars ({\bf lower row}).}
\label{fig:figure_deltamag_hist}
\end{figure}

The distribution of the magnitude variations are shown for all four IRAC bands in
Fig.~\ref{fig:figure_deltamag_hist}.  We plot the maximum
variation in magnitude for each source  where $S > 2 \sigma_{S(photo)}$.  Although the 
variations can be as large as 1.76~mag, most of the variations are smaller than 0.5~mag.
The maximum variations in each band as a function
of percentiles are given in Table.~\ref{table:deltamag_hist}.  In addition
to having a higher incidence of variability,  the protostars exhibit larger 
variations in magnitudes than the young stars with disks.

\begin{table}[h]
%\begin{center}
\caption{Percentiles for $\Delta m(\lambda)$}
\vskip 0.1 in
\begin{tabular}{lcccccc}
\tableline\tableline
Wavelength & Evolutionary & Number & \multicolumn{4}{c}{$\Delta$mag by percentile} \\
Band     &  Class &  & 25\% & 50\% & 75\% & 10\% \\
\tableline
3.6~$\mu$m & disks & 838 & 0.12 & 0.17 & 0.25 & 0.36 \\
4.5~$\mu$m & disks & 838 & 0.13 & 0.18 & 0.26 & 0.37 \\
5.8~$\mu$m & disks & 665 & 0.10 & 0.17 & 0.25 & 0.36 \\
8~$\mu$m   & disks & 542 & 0.07 & 0.13 & 0.21 & 0.32 \\
3.6~$\mu$m & protostars & 167 & 0.16 & 0.24 & 0.38 & 0.53 \\
4.5~$\mu$m & protostars & 167 & 0.15 & 0.23 & 0.33 & 0.45 \\
5.8~$\mu$m & protostars & 151 & 0.12 & 0.20 & 0.31 & 0.44 \\
8~$\mu$m   & protostars & 119 & 0.16 & 0.24 & 0.40 & 0.53 \\
\tableline
\end{tabular}
%\end{center}
\label{table:deltamag_hist}
\end{table}

\subsection{Constraints on the Mechanism for Variability}

We find evidence of variability for $50\%$ of the YSOs with IR-excesses. The magnitude changes are approximately the same in
all four IRAC bands and are typically on the order of 0.2~mag. Both the fraction of sources with variability and the changes in magnitude are largest for protostars.
Since diskless objects do not exhibit similar variability, it is likely that the observed variability is due to the changes in the
emission from the inner disk region where dust emits strongly in the IRAC bands.  This emission
is thought to originate primarily in the inner rim of the disk; at this inner rim the central stars heats the dust to the sublimation temperature of
silicate  grains. Dust within this radius will be sublimated, thus eliminating the primary source of opacity in the disk.
Since the disk is vertically extended, the emission in the inner rim originates come from an ``opacity wall'' facing the central
star.  The infrared emission from the disk traced by the IRAC bands arises mainly in the rim, but it includes contributions from disk radii 
up to 1 AU, particularly in the 8~$\mu$m band \citep{2006ApJ...638..314D}.

The variations in overall disk flux could be due to changes in the heating of the disk, fluctuations in its physical structure, or some combination of the two.
The wavelength dependence of the variability, as measured by the slope of $\Delta m_{\lambda}$ versus $\Delta$[4.5], and the timescale of the variations provide 
constraints on the mechanism for the infrared variability.    A physical model for the variability must explain how material over the inner 1 AU could uniformly change its flux over the course of six months. 
%A change in the accretion rate through the disk, and the corresponding change in surface density of the disk, would uniformly affect the disk flux, but the viscous timescale at which this operates is on the order of $10^5$ years. Vertical waves created by thermal instabilities and traveling inward from the outer disk could also influence large areas of the disk, but they operate on the thermal timescale of the midplane, which is on the order of decades for this region of the disk \citep{2008ApJ...672.1183W}. 
Variations on timescales of less than one year could be due to fluctuations and rotating structures in the inner 1 AU of the disk where the dynamical timescale ranges from months to days with decreasing radius.   These variations may result from a variety of reasons such as a rotating disk warp or MRI instabilities that launch dust out of the midplane  \citep{2010ApJ...708..188T}.  Previous observational studies have invoked scale height perturbations to the dust at the very inner edge of the disk to explain the wavelength dependence and timescale of the variability for evolved transition disks where the inner disk has been cleared. Scale height perturbations at the very inner edge of a disk result in a 'seesaw' behavior for the infrared SED, in which the short wavelength flux increases while the long-wavelength flux  decreases with the entire SED appearing to pivot at 8~$\mu$m \citep{2011ApJ...732...83F,2011ApJ...728...49E}.
The observed variability could also result from changes in the heating of the inner disk by the central star.  The disk is heated by both
the intrinsic luminosity of the star and the accretion luminosity generated by gas funneled down onto the star.  Changes in the heating of a disk could arise from fluctuations of the accretion rate onto the central star and the resultant change in accretion luminosity.   The accretion luminosity is thought
to be concentrated in accretion hot spots; the rotation of these hot spots may also result in variations in the heating of the portions of the inner disk 
most visible from the Earth  \citep{1989A&A...211...99B,2009ApJ...702.1507M}.
 
To examine how changes in the structure and heating of the inner disk would affect the observed magnitudes in the IRAC bands, we employ the CGPlus models \citep{1997ApJ...490..368C,2001ApJ...560..957D}. These models include a two-layer, flared disk with a puffed-up inner wall. While more complicated combinations of grain growth and settling are often needed to fit detailed SEDs, we can use these models to estimate the wavelength dependence due to changes in the disk and star. Our base model consists of a 0.5M$_{\odot}$, 2L$_{\odot}$, T$_{eff}$=3800 K star surrounded by a 0.01M$_{\odot}$ disk that extends from the dust sublimation radius, defined as the point where T$_{dust}$=1500 K, out to 100 AU. From these models we can directly estimate magnitudes, and by varying different model parameters we can calculate the slope of $\Delta m_{\lambda}$ versus $\Delta$[4.5] in the same manner as was done with the actual data. First we consider variations due to a   puffed up inner rim by increasing the scale height of the dust and gas of the inner rim. In this case, we find slopes of 1.014, 0.96, 0.60 for [3.6], [5.8] and [8.0], respectively (Table~\ref{var_model}). While the radiative transfer in the shadowed region behind the inner rim is not fully understood, changing the scale height in this way  produces a 'seesaw' behavior similar to what has been observed in transition disks and thus serves as a useful approximation. The calculated slopes are inconsistent with the observations, suggesting that in the majority of the YSOs, the variability is not due to a change in the scale height of the inner disk rim. We next vary the radial location of the inner rim of the disk by lowering the temperature at which the disk becomes truncated. In this situation, we find slopes of 0.53, 1.468, 1.499 which are also highly inconsistent with the data (Table~\ref{var_model}). This implies that, unlike in the evolved disks, we are not observing a change in the structure of the dust at the very inner edge of the disk.  We cannot rule out synchronized changes in the disk scale height over the entire inner 1 AU which would affect all the IRAC bands equally; however, it is not clear what mechanism could alter the disk in this manner.   

Next, we consider changing the stellar flux in the CGPlus models without varying the flux of the disk.  This produces slopes of 1.09, 0.861, 0.687 for [3.6], [5.8] and [8.0], respectively; these values show the decreasing dominance of the stellar photosphere with increasing wavelength. The lack of agreement with the observed slopes confirm the need for a change in the disk flux. If we vary the stellar luminosity while allowing the disk structure to respond to this change in the impinging flux, we find slopes of 0.985, 1.012, 1.013.  If we vary the disk heating without varying the infrared emission from the photosphere, which would be the case if the disk was heated by UV radiation from the accretion of gas onto the star, the resulting slopes of 0.922, 1.053,  1.124 provide the best match to the data (Table~\ref{var_model}).   In this case, the reduced slope at [3.6] is due to the significant contribution of the photosphere in that band.  The thermal timescale of the upper layers of the disk that are responsible for the observed infrared emission are very short \citep{1997ApJ...490..368C} and the disk will respond almost instantaneously to changes in the illumination. This suggests that we are observing a change in the overall flux level of the disk by an amount that is relatively constant with wavelength.  Variable heating may come from
rapid changes in the rate of accretion of gas onto the central stars; the accretion luminosity contributes significantly to the total luminosity of protostars and young stars with disks.
Alternatively, \citet{2009ApJ...702.1507M} constructed models of the variable heating in a protostar due to a rotating accretion hot spot. In this case,
the variability is not due to a change in total luminosity, but due to the region of the inner disk rim illuminated by the hot spot rotating in and out of view.  These models also produce a wavelength dependence that is similar to what has been observed.  We note that the \citet{2009ApJ...702.1507M} models which add a disk warp rotating in sync with the hot spot predicts a weaker variability at [8.0]; this is inconsistent with our observations.

We conclude that the time variable heating of an extended portion of the disk driven by rotating accretion hot spots or rapid accretion variations is the most likely explanation for the observed variations in the mid-IR magnitudes. This is in distinct contrast to transition disks for which structural perturbations of the puffed-up wall at the dust sublimation radius can explain the observed mid-IR variability \citep{2011ApJ...732...83F,2011ApJ...728...49E}.   Hence, the mechanism for  mid-IR variability may change as young stars objects and their disks evolve.  A more definitive assessment of the origin of the variability will require coordinated observations designed to study the time dependence of mid-IR emission, stellar luminosity and accretion (e.g. Flaherty et al. in press).

% We also consider changing the overall strength of the disk flux, while holding the stellar flux constant, and we derive slopes of 0.922, 1.053, 1.124 (Table~\ref{var_model}), with a 30\%\ increase in disk luminosity resulting in a 0.3 mag increase at [4.5]. These results are also consistent with the actual data, including the slope of slightly below 1 for [3.6] and slightly above 1 for the longer wavelengths. 

\begin{deluxetable}{cccc}
\tablecaption{Slope of Model $\Delta m_{\lambda}$ Variations\label{var_model}}
\tablehead{\colhead{Model}&\colhead{[3.6]}&\colhead{[5.8]}&\colhead{[8.0]}}
\startdata
Observed & 0.865 $\pm$ 0.034 & 1.055 $\pm$ 0.036 & 1.273 $\pm$ 0.091\\
Inner disk scale height & 1.014 & 0.96 & 0.60\\
Inner disk radius & 0.53 & 1.468 & 1.499\\
Vary star, not disk & 1.09 & 0.861 & 0.687\\
Vary star and disk & 0.985 & 1.012 & 1.013\\
Vary disk, not star & 0.922 & 1.053 & 1.124\\
\enddata
\end{deluxetable}

\section{Summary}

We present a survey of dusty YSOs identified in the Orion~A and B clouds using the IRAC and  MIPS
instruments onboard the {\it Spitzer} Space Telescope and the 2MASS point source catalog. This survey 
covered 8.8 sq.~degrees of the Orion clouds in the four IRAC bandpasses and 16.4 sq.~degrees with the 
MIPS instrument, with a combined total coverage of 8.8~deg$^2$.  In addition, eight reference fields
covering 0.29~sq.~degrees were observed with IRAC.  In this paper, we present the catalog of point
sources and young stellar objects, with the following results:

\noindent
{\bf 1:} Within this field, 306162 point sources are identified; a total of 34176 are in a sufficient number of wavebands
that they can be examined for infrared excesses.   We characterize the completeness of the 
point source catalog as a function of magnitude in the four IRAC and one MIPS band.  The 
completeness is spatially varying and is strongly dependent on the amount of nebulosity in a given field.  

\noindent
{\bf 2:} A total of 3479 dusty YSOs are identified by the presence of mid-infrared emission in excess
of that expected for a reddened photosphere.   This includes 2991 young stars with disks, 428 protostars and
50 faint candidate protostars.  In addition, another 10 sources detected only in the MIPS 24~$\mu$m band are candidate
protostars.  The identified dusty YSOs are 
found distributed throughout the Orion A and B  clouds molecular clouds;  a detailed  analysis of the spatial distribution 
will be presented in a subsequent paper.
We employed several criteria for rejecting contaminating galaxies; the residual contamination to the 
YSO sample from background galaxies is estimated to be $6.1 \pm 3.1$ per deg$^2$.  Using a Monte-Carlo simulation
of the background contamination, we estimate that an additional 13 sources may not be YSOs,
but instead show YSO-like colors due to a blending with nebulosity and other point sources. 

\noindent
{\bf 3:} The YSOs were examined for variability using three epochs for the L1641 cloud and two epochs for
the remainder of the survey.
The dusty YSOs show a high degree of correlated variability in the 3.6 and 4.5~$\mu$m bands relative to pure photospheric sources without IR-excesses, with
50\% of the dusty YSOs exhibiting significant variability.    The median variability is 0.2~mag, with variations up to 1.76 magnitudes detected. 
The protostars show both a higher incidence of variability and larger variations in their mid-IR magnitudes than young stars
with disks.  The variability is correlated in all four IRAC bands with the changes in the 5.8 and 8~$\mu$m 
bands approximately equal to those at 4.5~$\mu$m, and the 3.6~$\mu$m band showing somewhat
weaker changes.   Using models of the heating of an inner, puffed up disk by a central
source, we find that models that invoke changes in the heating of the inner disk best reproduce the observed
variability.  These changes may arise from fluctuations in the accretion rate or from rotating accretion hot
spots.

\acknowledgements
This publication makes use of data products from the Two Micron All Sky Survey, which is a joint project of the University of Massachusetts and the Infrared Processing and Analysis. Center/California Institute of Technology, funded by the National Aeronautics and Space Administration and the National Science Foundation.  This work is based in part on observations made with the Spitzer Space Telescope, which is operated by the Jet Propulsion Laboratory, California Institute of Technology under a contract  with NASA.  This work received support through that provided to the IRAC and MIPS instruments by NASA through contracts 960541 and 960785, respectively, issued by JPL.  Support for this work was also provided by NASA through awards  issued to STM and JLP by JPL/Caltech.

\facility{Facilities: Spitzer}

\section{Appendix A: Measuring  the Completeness to Point Sources in each Wavelength Band}

\begin{table}[ht]
%\begin{center}
\caption{Fits to Single Band Completeness Curves}
\begin{tabular}{lcccccccccccc}
\tableline\tableline
& \multicolumn{2}{c}{3.6~$\mu$m}&\multicolumn{2}{c}{4.5~$\mu$m} & & \multicolumn{2}{c} { 5.8~$\mu$m}& \multicolumn{2 }{c}{8~$\mu$m} & &\multicolumn{2}{c}{24~$\mu$m} \\
Mag & a$^1$ & b$^1$ & a$^1$ & b$^1$ &  Mag & a$^1$ & b$^1$ & a$^1$ & b$^1$ & Mag  & a $^1$ & b$^1$ \\
\tableline
   8.0 &    4.37 &    0.30 &    3.84 &    0.15 &    6.0 &    4.01 &    0.22 &    4.14 &    0.25 &   2.00 &    4.09 &    0.50 \\
    8.5 &    4.06 &    0.27 &    3.75 &    0.17 &    6.5 &    3.83 &    0.32 &    3.94 &    0.26 &   2.50 &    3.93 &    0.50 \\
    9.0 &    3.74 &    0.24 &    3.49 &    0.28 &    7.0 &    3.61 &    0.26 &    3.72 &    0.26 &   3.00 &    3.72 &    0.53 \\
    9.5 &    3.65 &    0.30 &    3.28 &    0.23 &    7.5 &    3.44 &    0.30 &    3.54 &    0.26 &   3.50 &    3.34 &    0.44 \\
   10.0 &    3.41 &    0.28 &    3.04 &    0.18 &    8.0 &    3.19 &    0.27 &    3.34 &    0.28 &   4.00 &    3.15 &    0.40 \\
   10.5 &    3.20 &    0.30 &    2.87 &    0.23 &    8.5 &    3.01 &    0.26 &    3.12 &    0.23 &   4.50 &    2.94 &    0.39 \\
   11.0 &    3.01 &    0.25 &    2.69 &    0.23 &    9.0 &    2.82 &    0.28 &    2.96 &    0.25 &   5.00 &    2.78 &    0.41 \\
   11.5 &    2.81 &    0.28 &    2.48 &    0.23 &    9.5 &    2.59 &    0.24 &    2.75 &    0.28 &   5.50 &    2.63 &    0.46 \\
   12.0 &    2.60 &    0.27 &    2.25 &    0.22 &   10.0 &    2.40 &    0.27 &    2.54 &    0.23 &   6.00 &    2.42 &    0.44 \\
   12.5 &    2.41 &    0.28 &    2.04 &    0.21 &   10.5 &    2.19 &    0.24 &    2.31 &    0.28 &   6.50 &    2.21 &    0.43 \\
   13.0 &    2.19 &    0.27 &    1.84 &    0.21 &   11.0 &    1.99 &    0.26 &    2.08 &    0.26 &   7.00 &    1.98 &    0.43 \\
   13.5 &    1.99 &    0.25 &    1.65 &    0.21 &   11.5 &    1.79 &    0.28 &    1.87 &    0.28 &   7.50 &    1.80 &    0.44 \\
   14.0 &    1.81 &    0.29 &    1.44 &    0.20 &   12.0 &    1.58 &    0.24 &    1.65 &    0.25 &   8.00 &    1.58 &    0.46 \\
   14.5 &    1.58 &    0.24 &    1.23 &    0.20 &   12.5 &    1.37 &    0.27 &    1.44 &    0.26 &   8.50 &    1.32 &    0.44 \\
   15.0 &    1.37 &    0.27 &    1.00 &    0.20 &   13.0 &    1.13 &    0.25 &    1.20 &    0.25 &   9.00 &    0.89 &    0.54 \\
   15.5 &    1.13 &    0.26 &    0.76 &    0.20 &   13.5 &    0.89 &    0.27 &    0.96 &    0.25 &   9.50 &    0.21 &   0.07 \\
   16.0 &    0.89 &    0.26$^2$ &    0.49 &    0.19 &   14.0 &    0.64 &    0.24 &    0.65 &    0.26 &  10.00 &    0.16 &    0.05 \\
   16.5 &    0.64 &    0.24$^3$ &    0.21 &    0.15 &   14.5 &    - &    - &    - &    - &  - &   - &    - \\
\tableline
\tablenotetext{1} {Uncertainties for $a$ and $b$ are $\le 0.01$ unless noted.}
\tablenotetext{2} {Uncertainty is $0.11$.}
\tablenotetext{3} {Uncertainty is $0.15$.}
\end{tabular}
%\end{center}
\label{tab:completeness}
\end{table}

To assess the completeness of the data, artificial stars were added to the
ONC field in the four IRAC bands and the MIPS 24~$\mu$m bands. 
As described in Sec.~\ref{sec:psc}, the Orion data exhibit spatially varying 
completeness primarily due to confusion with the 
bright and highly structured nebulosity detected in all wavelength bands.  Instead
of measuring the completeness for each position in the IRAC and MIPS maps, we determine 
completeness as a function of RMEDSQ, as defined in Sec.~\ref{sec:psc}.  
To do this, we used the ONC field since this field shows the largest range in RMEDSQ
values.  

For the IRAC artificial stars, we used a PRF that was 
reconstructed from the observations of an IRAC standard star\footnote{http://irsa.ipac.caltech.edu/data/SPITZER/docs/irac/calibrationfiles/psfprf/}.  
To account  for the rotation of the spacecraft during the two epochs of the 
observation, the artificial PRF was the sum of two rotated PRFs.  The 
artificial star  analysis was done separately for ``faint'', ``bright'' and
``ultra-bright'' stars.  The input magnitudes of the faint stars were
distributed in 0.5 mag increments between 12.5 to 16.5~mag for the 3.6
and 4.5~$\mu$m bands, and between 10.5 to 14.5~mag for the 5.8
and 8~$\mu$m bands.   For the bright stars, the stars were added in
0.5~mag increments between 8 and 12~mag for the 3.6 and 4.5~$\mu$m
bands, and between 6 and 10~mag for the 5.8 and 8~$\mu$m bands.  Finally,
the ultra-bright stars covered 3.5 to 7.5~mag in 0.5~mag
increments for the 3.6 and 4.5~$\mu$m bands and 1.5 to 5.5~mag in
0.5~mag increments for the 5.8 and 8~$\mu$m bands.  In all three cases,
the stars were distributed in a rectangular grid in which each added star 
was separated by 12.5'' from the adjacent artificial stars. 

\begin{figure}
\plotone{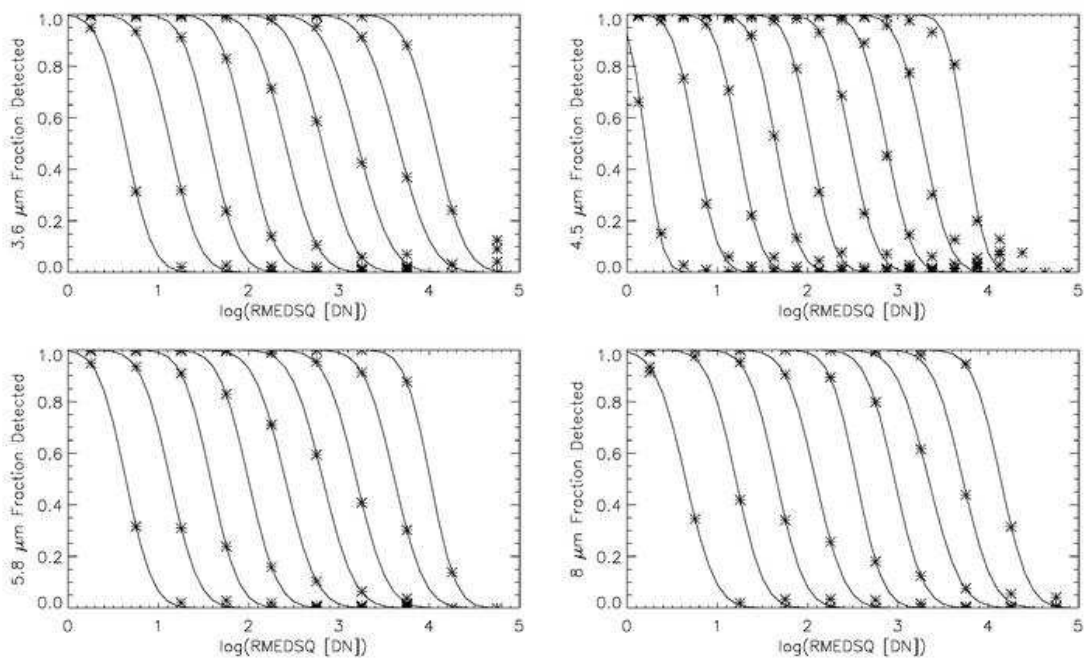}
\caption{The fraction of recovered point sources as a function of
  $log(RMEDSQ)$ for the four IRAC bands. The asterisks show the
  measured fractions, while the curves passing through/near the asterisks
  display the fit to the measured values for a single magnitude. For the 3.6 and 4.5~$\mu$m
  bands, we plot left to right the curves for [3.6]
  or $[4.5] = 16.5$, 15.5, 14.5, 13.5, 12.5, 11.5, 10.5, 9.5, 8.5~mag.
  For the 5.8 and 8~$\mu$m bands, we show left to right the curves for
  [5.8] or $[8] = 14$, 13, 12, 11, 10, 9, 8, 7, 6~mag.}
  \label{fig:medvfrac}
\end{figure}

The added stars were then recovered with PhotVis using automated
source detection; any source within $1.5"$ of the input position, within 0.25~mag of the input 
source magnitude, and with uncertainties within the limits set for YSO
identification  (0.1~mag for the 3.6 and 4.5~$\mu$m bands and 0.15~mag
for the 5.8 and 8~$\mu$m bands) was considered recovered.  The PhotVis object detection
parameters were the same as those used for the point source catalog.
The recovered faint stars were binned by their log(RMEDSQ) values using
a  bin of 0.5. The reason for the logarithmic binning is that increasingly bright nebulous regions
cover progressively smaller portions of the image;  a logarithmic
binning thus provides better statistics for regions where $RMEDSQ >
1000$.  We then calculated the 
fraction of recovered stars ($n_{rec}$) to added stars ($n_{add}$)  using the
equation $ f = n_{rec}/n_{add}$.
The fractions as a function of input magnitude and log(RMEDSQ) are shown
for the IRAC bands  in Fig.~\ref{fig:medvfrac}.

For the purpose of the correction, it is of value to obtain for each
magnitude a continuous curve parameterized by the RMEDSQ.   
For each input magnitude of the artificial stars, we fit a function to
the fraction of recovered stars vs. log(RMEDSQ).  We fit the equation

\begin{equation}
x = log(RMEDSQ),~f = \frac{1}{2}(1-\erf(\frac{x-a}{\sqrt{2}b}))
\label{eqn:unc}
\end{equation}

\noindent
where a and b are the value of RMEDSQ at which the
completeness declines to 0.5 and the width of the decline, respectively.   This
equation has desirable features of having a fraction approaching one for small values of 
log(RMEDSQ) and a fraction approaching zero for high values of log(RMEDSQ).
When fitting the data, we adopted an uncertainty

\begin{equation}
\sigma_f^2 =   (f(1-f)/n_{add})^2+0.00001+0.000001(1-f)
\label{eqn:logerf}
\end{equation}

\noindent
which is the uncertainty from \citet{1989ApJ...341..168B} with a noise floor added to both
high and low values of $f$.  The noise floors were set by using trial and error to find values of $\sigma_f$ for which the fit would converge. The resulting fits are shown in Fig.~\ref{fig:medvfrac}.
The values of $a$ and $b$ for the selected magnitudes are shown in Table~\ref{tab:completeness}.

%\begin{equation}
%x = RMEDSQ
%f = (1-a^2)(0.5-erf(\frac{x-b_1}{2c_2})/2)+a^2(0.5-0.5 erf(\frac{x-b_2}{\sqrt{2}c_3}))
%\end{equation}

%\noindent
%where $a$, $b_1$, $c_1$, $c_2$ and $b_2$ are the parameters.
%For the logarithmic functions we fit as

\begin{figure}
\plotone{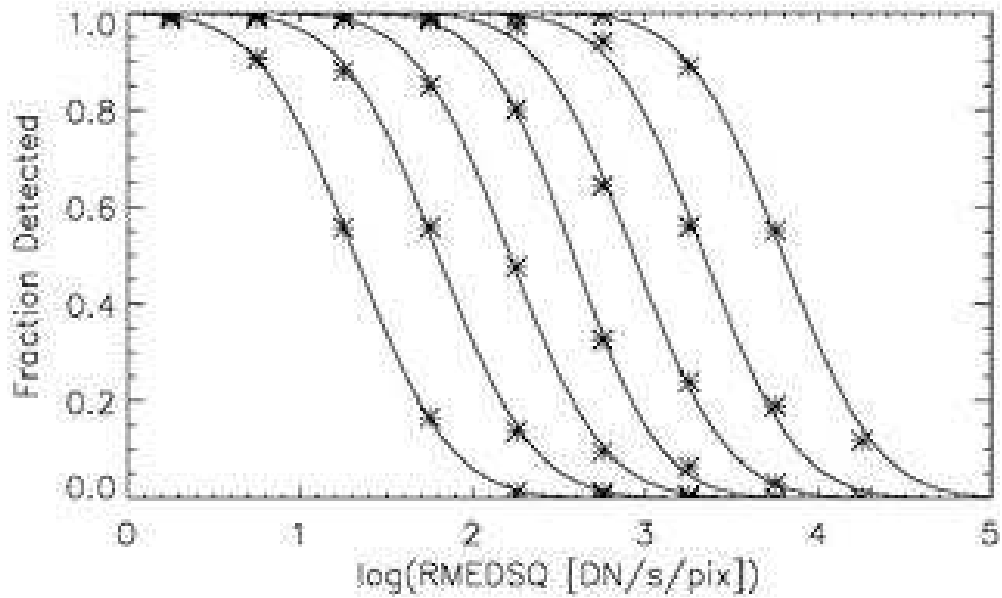}
\caption{The fraction of recovered point sources as a function of
  $log(RMEDSQ)$ for the MIPS 24~$\mu$m band. The points show the measured values while
  the curves show the fits to the measured values for a given magnitude.  Going from left to right,
  we plot the curves for sources with $[24]
  = 8.5$, 7.5, 6.5, 5.5, 4.5, 3.5 \& 2.5~mag.  The asterisks show the
  fractions; the curves passing through/near the points display the
  fit to the measured values.}
\label{fig:medvfrac24}
\end{figure}

The MIPS 24~$\mu$m completeness was assessed in a similar way.  A grid of point sources
were added to the images; the point source was generated by STinyTim for a 50~K source. 
This was done with two iterations: one for faint stars with magnitudes ranging from 6.5 to 10.5~mag
in increments of 0.5~mag and one for bright stars with magnitudes ranging from 2 to 6~mag in
increments of 0.5~mag.  The stars we separated by 12.5~pixels.
Point  sources detected by PhotVis with positions and magnitude within $1.5''$  of the input
position,  0.25~mag of 
the input magnitude and with an uncertainty $\le 0.25$~mag were considered recovered. The PhotVis object detection
parameters were the same as used for the point source catalog and the magnitudes
were determined using aperture photometry using the aperture radius, sky annulus, and calibration described in Sec.~\ref{sec:mipsred}.
The number of recovered and added stars were binned in intervals of log(RMEDSQ), and the resulting 
fraction was fit repeating the steps for the IRAC data (Fig.~\ref{fig:medvfrac24}).   The values of $a$ 
and $b$ for the MIPS data are shown in Table~\ref{tab:completeness}.

\input{abbrev}

\bibliography{ADS}

\clearpage

\clearpage

\clearpage 

\end{document}

%% file: abbrev.tex
\def\newblock{}
\def\aap{AA}
\def\aj{AJ}
\def\apj{ApJ}
\def\apjl{ApJL}
\def\apjs{ApJS}
\def\baas{BAAS}
\def\nat{Nature}
\def\mnras{MNRAS}
\def\apss{APSS}
\def\araa{ARAA}
\def\angst{\AA}